\documentclass[12pt]{article}
\usepackage[a4paper, margin=1.22in]{geometry}
\usepackage{mathrsfs}
\usepackage{color}
\usepackage{psfrag}
\usepackage{graphicx}
\usepackage{caption,subcaption}
\usepackage{enumerate}
\usepackage{amsmath,amssymb,calc,amsfonts}
\usepackage{latexsym}
\def\beq{\begin{eqnarray}}
\def\eeq{\end{eqnarray}}

\def\nn{\nonumber\\}

\def\={\stackrel{\Delta}{=}}

\DeclareMathOperator\erf{Erf}
\usepackage[utf8]{inputenc}

\title{Marginally Trapped Surfaces in Spherical Gravitational Collapse}
\author{Ayan Chatterjee\footnote{ayan.theory@gmail.com} \\
Department of Physics and Astronomical Science\\
Central University of Himachal Pradesh, Dharamshala-176206, India.\\
Amit Ghosh\footnote{amit.ghosh@saha.ac.in}\\
Theory Division, Saha Institute of Nuclear Physics,\\
1/AF, Bidhannagar, Kolkata -700064, India.\\
Suresh Jaryal~\footnote{suresh.fifthd@gmail.com}\\
Department of Physics and Astronomical Science\\
Central University of Himachal Pradesh, Dharamshala-176206, India.}

\usepackage{graphicx}
\begin{document}
\date{}
\maketitle

\begin{abstract}
This paper deals with a detail study of gravitational collapse of dust and viscous fluids
under the assumptions of spherical symmetry. Our main goal is to closely analyze the
horizons which arise during this gravitational phenomenon. To 
this end, we examine the formation and evolution of trapped surfaces 
in these spacetimes, with special attention to trapped regions and 
cylinders foliated by marginally trapped surfaces. The time evolution
of trapped surfaces, collapsing shell as well as the event horizon are identified
analytically as well as numerically. Using different density profiles of matter,
we analyze, how the nature of the marginally trapped surfaces 
modify as we change the energy momentum tensor. These studies reveal that
depending on the mass function and the mass profile, it is possible to envisage 
situations where dynamical horizons, timelike tubes or isolated
horizons may arise.

\end{abstract}

\section{Introduction}

The study of collapse of a self- gravitating isolated system is of great importance 
in general relativity. Not only this problem is of physical importance, particularly in
understanding the formation of black holes and large scale structures in 
the universe, but also raises fundamental queries related to 
formation of horizons, spacetime singularities and the cosmic censorship
conjecture \cite{Hawking_Ellis, Wald, Landau_Lifshitz, joshi, Penrose:1964wq,Penrose:1969pc}. 
The study of gravitational collapse, began with the independent work of Dutt \cite{SD} and
Oppenheimer and Snyder \cite{OS} (OSD). Although the OSD model was 
limited to collapse of a dust cloud of homogeneous 
density, it provided important clues to the nature and mode of 
formation of spacetime singularities out 
of a self gravitating matter system. In particular, it shows that 
within a finite proper time, the spherical ball of 
matter collapses to a proper radius smaller than its Schwarzschild radius. Eventually, 
the entire self gravitating matter collapses to a point of infinite density and 
curvature, usually called a spacetime singularity. Furthermore, once the matter
has crossed the Schwarzschild horizon, no light is able to escape to observers at asymptotic 
infinity and hence the singularity remains hidden to the outside world 
\cite{Hawking_Ellis, Wald,Landau_Lifshitz}. 
This is the well known scenario of black hole formation.
Though the OSD model is simple, the mathematical structure may be used to 
understand more complicated and real life examples of collapse of massive astrophysical 
systems. It has been argued, quite rightly, that the gravitational collapse of
real stars may not follow the idealised dynamical situation as 
described in the OSD model. The collapsing cloud of matter may be  inhomogeneous, 
have internal pressure, and even possess properties generic to fluids, like viscosity
and pressure anisotropies. For example, the well known Lemaitre- Tolman- Bondi (LTB) model
of collapse describes the inhomogeneous, pressureless gravitational collapse \cite{Lemaitre:1933gd,
Tolman:1934za, Bondi:1947fta}.
To understand these large variety of situations, general formalism to study dynamical evolution 
of collapse has been developed. In 
this method, under various regularity and energy
conditions, the initial data is provided in terms of the initial 
density, pressures, and velocity profiles, and the dynamical evolution is studied 
using the Einstein equations \cite{joshi, Misner_Sharp, Hernandez:1966zia}. It has been argued 
that under a large number of fairly regular initial data, both black holes and 
naked singularities may evolve (a nice review is given in \cite{joshi,Joshi:2012mk}. 
The OSD model for example, shows 
that once the collapse starts it eventually reaches an epoch 
(that of horizon formation at the Schwarzschild radius) when no light emitted from its 
surface can escape to faraway observers, and hence leading to a black hole. A naked singularity
is not covered by a horizon but is also interesting since an observer at infinity 
may communicate with it. Naked singularities have a large literature
and are discussed in \cite{Eardley:1978tr, 
Christodoulou_1984, Papapetrou, Newman:1985gt, Ori:1987hg, Newman_joshi, 
Dwivedi:1989pt,Ori:1989ps,
Lake:1991bff, Lake:1991qrk, Joshi:1993zg, Christodoulou_1994, Dwivedi:1994qs,Singh:1994tb,
Jhingan:1996jb,Harada:1998cq,Harada:1998wb,Maeda:1998hs,Joshi:1998su,Maeda:2006pm,Goswami:2006ph}. 
The singularity is said to be naked locally (or globally) 
if a non- spacelike geodesic emanates from it to 
reach the neighbourhood (or asymptotic infinity). Indeed, in many instances, regular initial data in
inhomogeneous, pressureless collapse develop 
into strong curvature naked singularities. Another class of singularity
which remains visible to observers at infinity is called the shell- crossing singularity
\cite{Yodzis:1973gha}. 
These indicate breakdown of the coordinate system and hence, are not genuine spacetime singularities.
Shell- crossing singularities are gravitationally weak \cite{Clarke},
but care must be taken to avoid them.

It is now well known that any general 
relativistic collapse of isolated gravitating matter satisfying regular initial data always 
results in a spacetime singularity in the form of geodesic incompleteness if a trapped surface forms, 
and certain reasonable energy conditions on matter and causal structure of the spacetime holds
\cite{Hawking_Ellis, Wald}. 
Additionally, the censorship conjecture, rules that 
the gravitational collapse of matter fields under generic conditions result in the formation 
of spacetime singularity which shall remain clothed from the outside world by a horizon. 
The statement of this conjecture are encoded in the weak as well as the strong versions of 
the conjecture and has been shown to hold true for simple systems
(see \cite{Christodoulou_1999,Dafermos:2008en} for review). So, if the censorship conjecture
is assumed correct, the collapse must be followed by an horizon. The most natural 
description of horizons is the Event Horizon (EH).
However, one fundamental objection against 
the event horizon formalism is that it is too global. 
Indeed, for this definition to work, one must have access to the global 
development of the spacetime, which may not be possible in all cases. For example, 
in a numerical study of a black hole spacetime, the location of the EH is only possible if the entire
evolution of the spacetime has been obtained, although the numerical evolution itself requires
the horizon to be located on each time slice. Such
inconsistencies have led to many local definitions of horizon
(a detailed overview is in \cite{Dreyer:2002mx,Ashtekar:2004cn,Booth:2005qc,Schnetter:2006yt}). 
Out of these, the marginally trapped surface (MTS) is quite useful. This is a closed $2$- dimensional
surface, such that the expansion scalar of the outgoing null normal vanishes $\theta_{(\ell)}=0$,
while that of the ingoing null normal is negative $\theta_{(n)}<0$. This definition has
proved to be quite useful, since the Marginally Trapped Tube (MTT) formed out of
stacking MTSs does not have any signature associated to it.
Indeed, a null MTT is an isolated horizon (IH) and hence 
describes a black hole horizon in equilibrium. When the  MTT is spacelike,
it is a dynamical horizon (DH), and describes a growing black hole.
If the MTT has a timelike signature, it is called a timelike tube, through
which matter may cross in either directions. Thus MTTs provide an unified
framework to study time evolution of black holes through different phases. 
The nature of MTTs and their behavior due to some dust models of collapse like the LTB
have been studied in \cite{Booth:2005ng}, and very recently in \cite{Sharif_maharaj, Sharif_maharaj_2}. 
However, our study focuses on both the analytical and numerical aspect of the evolution of the MTT.
In particular for the homogeneous dust models, we identify the beginning of formation of MTT, and 
trace it until it stops evolving,
eventually matching with the isolated horizon. Furthermore, we also include a
detail study of MTTs forming due to gravitational collapse of    
more general energy- momentum tensors including viscous effects.

The main motive of this paper is to discuss methods which 
will be useful to, (i) construct spherically symmetric models of spacetime for fluids with 
general energy- momentum tensors, (ii) study the collapse end state with
special emphasis on the formation of horizons, and in particular, track the marginally trapped tubes
in each of the cases, and (iii) identify, for the mass profiles considered here, the
regions of the parameter space where the MTT evolves as a DH (when matter fall through it),
where it might be timelike, and when it does become an IH.
In each of the examples, the exterior geometry will be assumed to be the Schwarzschild spacetime.
The study will include OSD/LTB models and the ones
obtained by dropping the assumptions of homogeneity and local anisotropy in the fluid 
energy- momentum tensor. 
Interest in these generalities in the equation of state stems from the fact that 
there is a growing attention in understanding the phenomenon of collapse of astrophysical 
systems with different equation of states and energy-momentum tensors 
\cite{Shapiro:1983du, Baumgarte:2010ndz, Rezzolla_ Zanotti, HL, LH}. 
As particular examples, the energy momentum tensors we consider below shall include 
locally anisotropic fluids, without heat flux, but with shear and bulk viscosity. 
Local anisotropy in the interior fluid have been argued to arise due to 
various causes including viscosity or local anisotropic velocity distributions. 
Such anisotropies would naturally break the conditions of fluid isotropy and hence, in presence 
of viscosity, radial and transverse pressures must be different (in section 
$5$, we present a proof). Furthermore, it has been shown that shearfree condition, particularly in 
the presence of anisotropy of the pressure and dissipation, leads to instability.  
These kind of inhomogeneities and anisotropies in the matter fields are expected to occur 
quite naturally in the astrophysical systems, particularly during the gravitational collapse, 
and are expected to play a major role in deciding the spacetime structure. In particular, it 
has been argued that shear may be responsible for the violation the cosmic censorship 
leading to a naked singularity in the spacetime \cite{joshi}. 
Thus, it is important that a detail study into 
these aspects must be made and indeed similar kind of models have 
received attention on the past (see for example \cite{Chan:1992zz, Chan:2000}). However, most of 
the metric configurations are either static or with restricted time 
dependence. On the other hand, we expect that, in presence of 
heat flux, viscosity or pressure terms, the interior spacetime would be highly dynamical and respond to any 
fluctuations in the energy momentum tensor. To incorporate these attributes, in this paper, we relax the 
assumptions of staticity and generalise these geometries to include time dependence, making these models 
closer to realistic dynamical systems. More specifically, including anisotropic pressure and the shear and 
bulk viscosity terms, we construct explicitly dynamical metric functions (we must however
ensure that the viscosity effects are not huge to destroy the 
spherical symmetry of the spacetime and in the examples, we have chosen the coefficients
in this manner). These metric functions are then 
used to study the end state of the spherically symmetric collapse, leading to a central singularity, and
trapped surface formation.

The paper is arranged as follows. In the section \ref{sec2}, we set up the definitions
of the different class of horizons which we shall use in this paper.
We discuss the inadequacies of the EH and the advantages of the MTT formalism.
In section \ref{sec3}, we set- up our conventions and the mathematical framework
for the Einstein equations as an initial value framework. We also spell out the boundary conditions,
along with those required for smooth gravitational collapse. For example,
we ensure that there is no shell crossing singularities and that the initial
spacelike surface does not have any trapped region. The section \ref{sec4} discusses
the pressureless collapse models including the OSD and the LTB models. While these models
are quite well known (the marginally bounded OSD models are described in 
\cite{Kanai:2010ae, blau} using the Painleve- Gullstrand coordinates,
and in \cite{Landau_Lifshitz, Rezzolla_ Zanotti} using the standard coordinates), we give 
detailed analytical calculations, along with the boundary conditions to show how
the collapse scenario proceeds.
In particular, we present analytical formula for (i) collapse of the shells, 
(ii) time development of event horizon, and (iii) time development of the marginally trapped tube
(or sometimes called the apparent horizon).
In several of the collapse scenarios, like the OSD models, the study of
formation and evolution of MTT can be carried out exactly through analytical methods. 
In section \ref{sec4}, the behavior of MTT
for each of the three subcases of the OSD model: marginally bounded, bounded and 
unbounded collapse respectively are given. 
The analytical results, followed by detail numerical models corroborate exactly. 
For inhomogenous dust collapse models like LTB, the 
simple mass profiles may be understood through analytical tools, but
for realistic mass profiles, we rely on numerical methods.
We consider several density profiles and, in each case, determine 
the evolution of MTT. Section \ref{sec5} studies models of spacetimes
due to viscous fluids. We begin with some generic properties these spacetimes must be endowed with.
We show that  if the fluid has shear and bulk viscosity, as well as pressure anisotropy, 
then \emph{generically} the spacetime will not admit isotropy, conformal flatness or spatially
uniform expansion scalar. We also take various cases to show that the effects 
of viscosity (again keeping the non- spherical effects small) on the formation of the MTT.
In particular, we show that the viscous effects may delay or advance the formation of MTT
depending on the coefficients of viscosity.  These effects of viscosity are exemplified through various
choices of parameters and mass functions. We conclude in
section \ref{sec6}.

\section{Horizons and Marginally Trapped Tubes}\label{sec2}

In the standard description, a black hole horizon is a future event horizon (EH), 
which, in an asymptotically flat spacetime, is defined as boundary of the past of future null infinity,
$\partial[J^{-}(\mathscr{I}^{+})]$ \cite{Hawking_Ellis, Wald}. This definition is powerful since it is 
an invariant construct based only on geometrical arguments and asymptotic structure of spacetime.
However, this definition is difficult to implement in many practical situations, although 
the situation simplifies  for equilibrium situations (see \cite{Ashtekar:2004cn,Booth:2005qc}
for a detailed review). In equilibrium, 
the spacetime is stationary and 
admits Killing vector fields one of which may be identified with the 
time translation generator at asymptotic infinity. 
Naturally, these Killing vector fields are tangential to the EH. 
Thus, in stationary spacetimes, the EH may be described as a Killing Horizon, 
generated by a null Killing vector field. A clear example is provided by 
the spherically symmetric Schwarzschild spacetime. 
This is a stationary spacetime which admits four global Killing vector fields, three of which
are spacelike, generating the $2$- sphere isometries and a time translation 
generator $(\partial/\partial t)$. This timelike Killing vector becomes null generator on 
the horizon $R=2M$. Note however, that if a spacetime is not in equilibrium (for example, the
Vaidya spacetime), Killing Horizon cannot be constructed and hence, one has to resort to
the abovementioned definition of EH to locate horizon for this spacetime. 

However, the EH, as a definition, is far from useful even in dynamical spacetimes. 
The notion is global since it requires knowledge of the entire future evolution of 
the spacetime (which may include a black hole region as well), to locate the future null infinity and hence,
ascertain the existence of a black hole horizon. Quite simply, the definition does not
work in practical situations, like those studied in numerical relativity where, one cannot even evolve the
spacetime if the horizon is not located on a given time slice (a comprehensive discussion on these
problems, and the need for quasilocal horizons in numerical relativity is discussed in 
\cite{Dreyer:2002mx,Ashtekar:2004cn,Booth:2005qc,Schnetter:2006yt}). A vivid description of the 
difficulties associated with this definition and its teleological nature
is captured through the Hartle- Hawking formula \cite{Chandrasekhar}.
This formula shows that the area of an event horizon of a dynamical black hole at 
the final time $t_{f}$ (when the matter has stopped falling) is  
dependent also on the beheviour of geometrical quantities at times $t> t_{f}$. More precisely, area 
of any dynamically evolved EH may only be determined if the full global evolution
of the spacetime along with precise knowledge of values of geometrical fields for all future times
is known. A further critique of EH was given in \cite{Hajicek:1986hn}, where it was argued 
that simple local modifications of the black hole region (which may be
quantum mechanical in nature, and hence beyond the scope of classical GR) may get rid of the
formulation of EH. These arguments underscore the need for a quasilocal description of 
black hole horizons which however, must capture all the essential physics details EH has given us.

The local notion of horizons are based on the definition
of trapped surfaces which, loosely speaking, characterize regions of spacetime
from which light rays cannot escape to infinity \cite{Penrose:1964wq}. In these regions,
null rays orthogonal to closed $2$- surfaces have negative expansion. More precisely,
if $\ell^{a}$ and $n^{a}$ are respectively the outgoing and
the ingoing null vectors orthogonal to a $2$- sphere, then this $2$-surface
is called trapped if their respective expansions, $\theta_{(\ell)}$ and $\theta_{(n)}$ 
are both negative. In general, the $2$-sphere is called untrapped, trapped or marginally trapped
depending on whether $\theta_{(\ell)}$ is greater, less or equal to zero respectively. 
Using these marginally trapped surfaces, one may formulate a definition of horizon.    
The notion of apparent horizon is one such local description which found several applications
in local black hole dynamics. However, since apparent horizon depends on the choice of foliation
of spacetime by spacelike hypersurfaces, several examples exist where the apparent horizon has been
difficult to locate. Even in simple cases like the Schwarzschild spacetime, where the existence of
horizon is unambiguous, the apparent horizon may be difficult to locate \cite{Wald:1991zz}. As a remedy,
a quasilocal formulation called Trapping Horizon (TH) were introduced in \cite{Hayward:1993wb}. 
They are defined as follows:
A trapping horizon, denoted here by $\Delta_{T}$ is a $4$-dimensional spacetime, 
foliated by $S^{2}$ such that the expansions of the null normals $\ell^{a}$ (outgoing) 
and $n^{a}$ (ingoing) orthogonal to the foliations have expansions $\theta_{(\ell)}|_{\Delta_{T}}= 0$,  
$\theta_{(n)}|_{\Delta_{T}}\neq 0$ such that $\pounds_{n}\theta_{(\ell)}|_{\Delta_{T}}\neq 0$. 
If $\theta_{(n)}|_{\Delta_{T}} <0$, the horizon is termed future, whereas it is past otherwise.  
Furthermore, a horizon is called an outer trapping horizon if 
$\pounds_{n}\theta_{(\ell)}|_{\Delta_{T}}< 0$ , whereas, it is termed inner if 
$\pounds_{n}\theta_{(\ell)}|_{\Delta_{T}} > 0$. In fact, Future Outer Trapping horizons (FOTH)
have found important applications in the proof of laws of black hole mechanics \cite{Hayward:1993wb}
as well as in developing local formulations 
of Hawking radiation mechanism \cite{Chatterjee:2012um}.

The formulations of Isolated Horizons (IH) 
\cite{Ashtekar:2000sz,Ashtekar:2000hw} and Dynamical Horizons (DH)
\cite{Ashtekar:2002ag,Ashtekar:2003hk}, which are closely 
related to THs, have led to crucial insights towards understanding classical and quantum behavior
of black hole horizons. They have also found important applications
in the development of numerical methods used to study horizon formation, their mergers
and gravitational waves. Indeed, the IH formalism which describes the equilibrium states
of black hole, have been used to develop Hamiltonian techniques for 
black hole mechanics. More specifically,
the space of solutions of Einstein's theory, with IH as an
inner boundary, admits a phase- space formulations along with 
a well defined symplectic structure and surprisingly, the first law of black hole
mechanics turns out to be the necessary and sufficient condition for
a consistent Hamiltonian evolution in the phase- space \cite{Ashtekar:2000sz,Ashtekar:2000hw}.
Additionally,
this framework also provides the boundary symplectic structure which 
allows identification of the boundary quantum states responsible for black hole entropy
\cite{Ashtekar:1997yu}.
The DH plays a crucial role in understanding smooth dynamical evolution of black hole
horizons. The first law (in terms of fluxes) for such a dynamical evolution provides
a useful theoretical framework to model black holes evolution 
\cite{Ashtekar:2002ag,Ashtekar:2003hk}.             
Further developments in these directions have been in the development of quasispherical
and perturbative approximations of the IH and the DH formalisms. In the dynamical
set-up it has been possible to construct a class of evolving horizons, called 
the Conformal Killing Horizons (CKH) which are null, but admit a well defined phase- space description
in the first order connection- tetrad variables. Indeed, it has also been possible, even in this dynamical
framework, to derive a differential version of the first law of black hole 
mechanics, arising due to influx of (scalar) matter terms \cite{Chatterjee:2014jda,Chatterjee:2015fsa}.

A unified quasilocal framework to describe horizons, called Marginally Trapped Tube (MTT)
was developed in \cite{Ashtekar:2005ez}.
Let $(\mathcal{M},g_{ab})$ be a $4$- dimensional spacetime with signature $(-,+,+,+)$. We 
shall use the Newmann- Penrose null basis $(\ell^{a}, n^{a}, m^{a}, \bar{m}^{a})$,
where $\ell\cdot n=-1$, $m\cdot\bar{m}=1$, while all other dot products vanish. Let $\Delta$
be a hypersurface in $\mathcal{M}$. In the following, we shall
not restrict the signature of $\Delta$ and hence, it may be spacelike,
timelike or even null. Let us assume that 
$\Delta$ is topologically $S^{2}\times \mathbb{R}$. Let $\ell^{a}$ and $n^{a}$
are respectively the outgoing and the ingoing vector fields orthogonal to 
the $2$- sphere cross-sections of $\Delta$. If $t^{a}$ is a vector field tangential 
to $\Delta$ and normal to foliations, $t^{a}$ may be written in terms of the 
ingoing and the outgoing null vector fields as: $t^{a}=\ell^{a}-Cn^{a}$ 
(the sign of $C$ is in accordance with the conventions in \cite{Booth:2005ng}).
Since $t\cdot t=2C$, the constant $C$ determines the signature of the MTT.

The hypersurface $\Delta$ 
will be called a  Marginally Trapped Tube (MTT) if the following conditions hold true on $\Delta$:
\begin{enumerate}
\item $\theta_{(\ell)}= 0$,
\item $\theta_{(n)} <0$.
\end{enumerate}
Several comments are in order regarding these boundary conditions.
First, MTT may be viewed as a unified formalism to describe
black hole horizons since $\Delta$ has no restriction on its signature.
When MTT is null it describes black holes in equilibrium (an IH),
a growing black holes (a DH) when it is spacelike, or simply a timelike membrane (when 
$\Delta$ is timelike), allowing matter to cross it. 
The advantage of the MTT formalism is that instead of looking at the evolution of horizons through 
various phases: dynamical horizons, isolated horizons, and timelike membranes (each phase multiple
number of times), one may view horizons as the time evolution of a single MTT.
Secondly, MTT admits much weaker set of conditions
than either the IH or the DH formalism. For example, no restiction
on $\pounds_{n}\theta_{(\ell)}$ is assumed. If $\pounds_{n}\theta_{(\ell)}<0$,
the MTT shall be called a FOTH. Thirdly, MTTs are foliated by marginally 
trapped $2$- spheres. Since $t^{a}$ is orthogonal to the foliations 
and tangential to $\Delta$, it generates a foliation preserving flow so that 
the following condition holds on $\Delta$
\begin{equation}\label{liet}
\pounds_{t}\,\theta_{(\ell)}\triangleq 0.
\end{equation}
Fourth, the constant $C$ also measures the evolution of MTT. To see this, note that
if $m^{a}$ and $\bar{m}^{a}$ are tangential to the $2$- sphere
cross-sections, the area element is given by ${}^{2}\mathbf{\epsilon}=im\wedge\bar{m}$.
Under the flow generated by $t^{a}$, the area element of MTT evolves as:
\begin{equation}
\pounds_{t}\,{}^{2}\mathbf{\epsilon}=-C\,\theta_{(n)}\,{}^{2}\mathbf{\epsilon}
\end{equation}
Naturally, the timelike MTT (for which $C<0$) contracts, 
null MTT ($C=0$) does not grow, whereas spacelike MTT (for which $C>0$) expands.
Furthermore note that no condition on the energy- momentum tensor is assumed on $\Delta$.
The Einstein equation $G_{ab}\equiv R_{ab}-(1/2)R\,g_{ab}=T_{ab}$\footnote{We use the 
units of $c=1$ and $8\pi G=1$, or equivalently, we scale the components of the 
energy- momentum tensor by $8\pi G$.}shall be assumed to hold on $\Delta$.

Several conclusions follow from these conditions, detail calculations of various
equations are given in \cite{booth_fairhurst_2005}. From \eqref{liet}, 
the constant $C$ is determined by the condition
\begin{equation}
C=\frac{\pounds_{\ell}\,\theta_{(\ell)}}{\pounds_{n}\,\theta_{(\ell)}}.
\end{equation}
To determine the value of the constant $C$, one uses the Newmann- Penrose equations for the MTT.
Using $\theta_{(\ell)}=0$ and the Einstein equation $G_{ab}=T_{ab}$,
we get the following two equations:
\begin{eqnarray}
\pounds_{\ell}\,\theta_{(\ell)}&=&-T_{ab}\ell^{a}\ell^{b},\\
\pounds_{n}\,\theta_{(\ell)}&=&-(\mathcal{R}/2) +T_{ab}\ell^{a}n^{b}.
\end{eqnarray}
Here, $\mathcal{R}$ is the scalar curvature of the round $2$- sphere and may be rewritten
as $\mathcal{R}=(8\pi/\mathcal{A})$, where $\mathcal{A}$ is area of $2$- sphere.
These equations imply that the constant $C$ which determines the nature of the MTT is given by:
\begin{equation}\label{value_of_c}
C=\frac{T_{ab}\ell^{a}\ell^{b}}{(4\pi/\mathcal{A})- T_{ab}\ell^{a}n^{b}}
\end{equation}

It follows from the discussion above that the signature of $\Delta$, determined by $C$,
is a quantity of utmost importance since it decides the nature and \emph{stability} of horizon
\cite{Andersson:2005gq,Andersson:2007fh}. 
From the above equation \eqref{value_of_c}, this value is controlled by 
the energy- momentum tensor and area of the marginally trapped
surfaces. In the following, we shall use several energy-momentum
tensors, including dust models and viscous fluids,
and evaluate $C$ in each case. However, as we shall see below,
the form of $T_{ab}$ is not the only criteria deciding the signature of MTT, the mass
profile and the equations of collapse are also important factors.  
Given these complicated constraints, the generic behavior of MTT is not 
known for arbitrary black hole evolution. Consider for example, Vaidya- type black holes
evolving under matter fields satisfying dominant energy conditions, the evolution of the 
MTT is described by the equation $R=2m(v)$, where $v$ is 
the advanced Eddington- Finkelstein coordinate. In this case, the MTT is spacelike (more precisely,
it is a DH) if $\dot{m}(v)>0$, where dot indicates derivative with 
respect to the advanced time coordinate \cite{Ashtekar:2002ag,Ashtekar:2003hk, Booth:2010eu}.
However, this conclusion does not hold true for any arbitrary collapse scenarios.
Indeed, even for simple situations like the OSD models of homogeneous
dust collapse, a timelike MTT (or a timelike membrane) appears just as the matter cloud reaches
the Schwarzschild radius. This timelike membrane, together with the matter cloud,
eventually collapses into the singularity at exactly the same time.
Examples of trapped surfaces are also
discussed in \cite{Schnetter:2006yt,Booth:2005ng, Kanai:2010ae,krasinski_hellaby, 
Bengtsson:2008jr,Bengtsson:2010tj,
 Bengtsson:2013hla,Booth:2012rm, Creelman:2016laj,Booth:2017fob,Schnetter:2005ea, Nielsen:2010wq}.
In more realistic
LTB inhomogenous collpase models, the matter cloud and the MTT
behave drastically differently: the cloud shells reach singularity at different times
and the MTT is not purely timelike. For large number of 
cases, in which the matter profile is smooth, the MTT begins as a spacelike
hypersurface from the center of the cloud and asymptotes
to the null event horizons as infall of matter is discontinued. 
For mass profiles with more complicated functional forms, time evolution of MTT shows strange behavior:
for example, turning timelike from being spacelike through an intermediate (expanding) null regions.     
These details are studied with large number of examples as well in the following sections.

\section{Spherical Symmetric Collapse formalism}\label{sec3}

Let us consider a general spherically symmetric ball of fluid 
with the line element
\begin{equation}
 ds^{2}=-e^{2\alpha(r,t)}dt^2 + e^{2\beta(r,t)}dr^2 + R(r,t)^2 d \theta^2+ 
 R(r,t)^2 \sin^2{\theta}\, d\phi^2,
 \label{1eq1}
\end{equation}
where $\alpha(r,t)$ and $\beta(r,t)$ are spacetime dependent functions, $\theta$
and $\phi$ are the angular variables on the sphere and $R(r,t)$ is the radius 
of the sphere. This is the standard frame, where the fluid velocity is 
$u^{a}=u\,(\partial/\partial t)^{a}$. This frame allows for a simpler integration of the Einstein
equation and the Bianchi identities.
The tetrad basis is suitable for obtaining the Einstein equations. The set of tetrad one forms obtained from the metric are as follows:
 \begin{equation}
 e^{0}=e^{\alpha(r,t)}\, dt, ~~ e^{1}=e^{\beta(r,t)}\, dr, ~~ e^{3}=R\,d\theta,
 ~~ e^{4}=R\sin\theta\, d\phi.
 \end{equation}
 The Riemannian spin- connection may be obtained from the torsion- free condition,
 $de^{I}+\omega^{I}_{J}\wedge e^{J}=0$, where $I,J$ are the internal flat basis,
 with $I,J, K, \dots =0,1,2,3$. The spin- connections are obtained to be:
 \begin{eqnarray}
 \omega^{0}{}_{1}=\left(\alpha^{\prime}\,e^{-\beta}\right)\,e^{0} +\left(\beta^{\prime}\,e^{-\alpha}\right)\,e^{1},~~~~&&\omega^{0}{}_{2}=(\dot{R}/R)\,e^{-\alpha}\,e^{2},\nn
 \omega^{1}{}_{2}=-(R^{\prime}/R)\,e^{-\beta}\,e^{2},~~~~&&\omega^{2}{}_{3}=-(\cot\theta/R)\,e^{3}.
 \end{eqnarray}
 The curvature two form, $\Omega_{IJ}$, are given by, 
 $\Omega_{IJ}=d\omega_{IJ}+\omega_{IK}\wedge\omega^{K}{}_{J}$ and 
 the non-zero ones are ($i,j=2,3$):
 \begin{eqnarray}
 \Omega^{0}{}_{1}&=&\left[\left\{e^{-\alpha}\,(e^{\beta})_{,t}\right\}_{,t}-\left\{e^{-\beta}\,(e^{\alpha})_{,r}\right\}_{,r}\right]e^{-(\alpha+\beta)}\,\,e^{0}\wedge e^{1},\nn
\Omega^{0}{}_{i}&=&\left[\frac{e^{-\alpha}}{R}\left\{(R)_{,t}\, \, e^{-\alpha}\right\}_{,t}-\frac{R_{,r}}{R}(e^{\alpha})_{,r}\,\,e^{-(\alpha+2\beta)}\right]\,e^{0}\wedge e^{i} \nn
 &&~~~~~~~~~~~~~~+ \left[\frac{e^{-\beta}}{R}\left\{(R)_{,t}\, \, e^{-\alpha}\right\}_{,r}-\frac{R_{,r}}{R}(e^{\beta})_{,t}\,\,e^{-(\alpha+2\beta)}\right]\,e^{1}\wedge e^{i}\nn
 \Omega^{1}{}_{i}&=&\left[-\frac{e^{-\alpha}}{R}\left\{(R)_{,r}\,e^{-\beta}\right\}_{,t}-\frac{R_{,t}}{R}(e^{\alpha})_{,r}\,\,e^{-(2\alpha+\beta)}\right]\,e^{0}\wedge e^{i}\nn &&
 ~~~~~~~~~~~~~~+ \left[-\frac{e^{-\beta}}{R}\left\{(R)_{,r}\,e^{-\beta}\right\}_{,r}-\frac{R_{,t}}{R}(e^{\beta})_{,t}\,\,e^{-(2\alpha+\beta)}\right]\,e^{1}\wedge e^{i}\nn
\Omega^{2}{}_{3}&=&\left[\left(\frac{R_{,t}}{e^{\alpha}R}\right)^{2}-\left(\frac{R_{,r}}{e^{\beta}R}\right)^{2}\right]\,e^{2}\wedge e^{3} +(1/R^{2})\, e^{2}\wedge e^{3}.
 \end{eqnarray}

The components of the curvature two torms may be extracted by using $\Omega_{IJ}=(1/2)\,\Omega_{IJKL}\, e^{K}\wedge e^{L}$ and the Ricci tensor components are given by $R_{IJ}=\Omega^{K}{}_{IKL}$. The Einstein tensor in the orthornormal frame is 
easily obtained by using the equation $G_{IJ}=R_{IJ}-(1/2)\eta_{IJ}R$, with $R=\eta_{IJ}R^{IJ}$.
In this tetrad basis, the four velocity is given by $u^{I}=(1,\vec{0}\,)$. Using the 
standard frame transformation rules, $e^{I}{}_{a}\,u^{a}=u^{I}$, the velocity vector 
in the coordinate basis becomes,  $u^{a}=e^{-\alpha}\, (\partial/\partial t)^{a}$. From now on, we shall
use the coordinate basis for our explicit calculations.


We envisage the solutions of the Einstein equation for the 
energy- momentum tensor of the spherical ball given by the following form:
\begin{eqnarray}
T_{ab}=(p_t+\rho)\, u_a u_b +p_t \, g_{ab}+(p_r-p_t)\,X_{a}\, X_{b}-2\eta\, \sigma_{ab}-
\zeta \theta h_{ab}, \label{tmn}
\end{eqnarray}
where  $\eta$ and $\zeta$ are the coefficients of shear 
and bulk viscosity, $X^{a}$ is a unit 
space-like vector tangential to the spacelike section orthogonal to
$u^{a}$ respectively satisfying $X_a X^a=1$.
The quantities $\sigma_{ab},\theta$, and $h_{ab}$ are shear, 
expansion and projection tensor
and, $\rho$, $p_{t}$ and $p_{r}$ are the energy density and tangential
and radial components of pressure respectively. The expressions for these quantities
are 
\begin{eqnarray}
\theta&=&\nabla_{a}u^{a}, ~~~~ h^{a}{}_{b}=(\delta^{a}{}_{b}+u^{a}u_{b}\,)\\
\sigma^{ab}&=&\frac{1}{2}\left(h^{ac}\,\nabla_{c}\,u^{b}+h^{bc}\,\nabla_{c}\,u^{a}\right)-
\frac{1}{3}\theta P^{ab},\\
 X^a&=&e^{-\beta(r,t)}\,(\partial/\partial r)^{a} \label{uX}
\end{eqnarray}
Their values for this metric is easily determined to be:
\begin{eqnarray}\label{theta_shear_def}
&&\theta=e^{-\alpha}(\dot{\beta}+2\dot{R}/R), \\
&&h_{ab}=e^{2\beta(r,t)}dr^2 + R(r,t)^2 d \theta^2+ R(r,t)^2 \sin^2{\theta}\, d\phi^2,\\
&& \sigma^{1}{}_{1}=(2/3)\,(\dot{\beta}-\dot{R}/{R})e^{-\alpha},\\
&& \sigma^{2}{}_{2}=\sigma^{3}{}_{3}=(-1/3)\,(\dot{\beta}-\dot{R}/{R})e^{-\alpha}.
\end{eqnarray}
Let us define a shear scalar, $\bar{\sigma}^{2}=\sigma_{ab}\,\sigma^{ab}$ and from the above expressions, we get 
\begin{equation}
\bar\sigma^{2}=(2/3)e^{-2\alpha}(\dot{\beta}-\dot{R}/{R})^{2}.
\end{equation}
In many cases, for simplification, we shall get rid of the $(2/3)$ factor and 
redefine $\sigma=e^{-\alpha}(\dot{\beta}-\dot{R}/{R})$. The non- zero components of energy-momentum tensor are given by the following quantities:
 \begin{eqnarray}
 T^0{}_0=-\rho,\,\, ~~ T^1{}_1=p_{r}-\frac{4}{3}\eta\sigma-\theta\zeta,\,\,~~ T^2{}_2=
 T^3{}_3=p_{t}+\frac{2}{3}\eta\sigma-\theta\zeta.
 \end{eqnarray}
Let us now have a look at the Bianchi identities, $\nabla_{a}\,T^{ab}=0$. This gives the following two equations. The first is the $t$- equation
\begin{eqnarray}\label{t-Bianchi}
&&\dot{\rho}e^{-\alpha}+(\rho+p_{t})\theta +(p_{r}-p_{t})\dot{\beta}e^{-\alpha}-(4/3)\,\eta\sigma^{2}-\zeta\theta^{2}=0,
\end{eqnarray}
and the second is the $r$- equation given by the following equation, 
\begin{eqnarray}\label{r-Bianchi}
&&(p_{t}-\zeta\theta)^{\prime}+\alpha^{\prime}(\rho+p_{t}-\zeta\theta)
-(4/3)\,\eta\sigma^{\prime}-(4/3)\eta\sigma\,(\alpha^{\prime}+3R^{\prime}/R)\nonumber\\
&& ~~~~~~~~~~~~~~~~~~~~~~~~~~~~~~~~~~+(p_{r}-p_{t})^{\prime}+(\alpha^{\prime}+2R^{\prime}/R )(p_{r}-p_{t})=0.
\end{eqnarray}
A simple rearrangement of equation $\eqref{t-Bianchi}$ leads to the following expression for the $\dot{\beta}$:
\begin{equation}\label{beta_dot}
\dot{\beta}=-\frac{\dot{\rho}}{\rho+p_{r}-(4/3)\,\eta\sigma}-\frac{2\dot{R}}{R}\,\frac{\rho+p_{t}+(2/3)\eta\sigma-\zeta\theta}{\rho+p_{r}-(4/3)\eta\sigma-\zeta\theta} \, .
\end{equation}
On rearranging, the equation \eqref{r-Bianchi} similarly leads to the following equation:
\begin{equation}\label{alpha_prime}
\alpha^{\prime}=\frac{2R^{\prime}}{R}\frac{p_{t}-p_{r}+2\eta\sigma}{\rho+p_{r}-(4/3)\eta\sigma-\zeta\theta}
-\frac{(p_{r}-4/3\,\eta\sigma-\zeta\theta)^{\prime}}{\rho+p_{r}-(4/3)\eta\sigma-\zeta\theta}\,\,.
\end{equation}
Let us now consider the $R_{01}$ component of the Einstein equation, which is given by:
\begin{equation}\label{R01equation}
\alpha'\dot{R}+\dot{\beta}R^{\prime}-{\dot{R}}^{\prime}=0.
\end{equation}
Using the expressions of $\alpha^{\prime}$ from equation \eqref{alpha_prime} and $\dot{\beta}$ from
\eqref{beta_dot} in the abovementioned equation \eqref{R01equation} and multiplying by $R^{2}$,
we get that:
\begin{equation}
[(p_{r}-4/3\,\eta\sigma-\zeta\theta)\,R^{2}\,\dot{R}\,]_{,r}+[\rho R^{2}R^{\prime}]_{,t}=0
\end{equation}
It is natural to interpret the exact differential to construct a function $F(r,t)$, such that
\begin{eqnarray}\label{F_eqn}
&&F^{\prime}\propto\rho\, R^{2}R^{\prime}, \\
&&\dot{F}\propto-(p_{r}-4/3\,\eta\sigma-\zeta\theta)\,R^{2}\,\dot{R},
\end{eqnarray}
with the same proportionality factors. We shall call this function $F(r,t)$ as the \emph{mass function}.
To detremine the exact form of the mass function $F$, let us look at the Einstein equations 
$G_{00}$ and $G_{11}$ respectively. They are given by the following two equations:
\begin{eqnarray}
\left(\frac{2R^{\prime}\beta^{\prime}}{R}-\frac{2R^{\prime\prime}}{R}-
\frac{R^{\prime 2}}{R^{2}}\right)\, e^{-2\beta}
+\left(\frac{2\dot{R}\dot{\beta}}{R}+\frac{\dot{R}^{2}}{R^{2}}\right)\, e^{-2\alpha}
+\frac{1}{R^{2}}=\rho,\\
\left(\frac{2\alpha'{R}'}{R}+\frac{{R'}^2}{R^2}\right)e^{-2\beta}
 +\left(\frac{2\dot{\alpha}\dot{R}}{R}-\frac{\dot{R}^2}{R^2}-\frac{2\ddot{R}}{R}\right)e^{-2\alpha}-\frac{1}{R^2}=p_{r}-\frac{4}{3}\eta \sigma-\zeta \theta.
\end{eqnarray}
Using the $G_{01}=R_{01}$ equation, and multiplying $G_{00}$ by $RR^{\prime}$ 
and the $G_{11}$ equation by $R\dot{R}$, the equations above reduce to:
\begin{eqnarray}\label{F_equations}
&&\left[R(1+\dot{R}^{2}e^{-2\alpha}-R^{\prime 2}e^{-2\beta})\right]_{,\,r}=\rho R^{2}R^{\prime},\nonumber\\
&& \left[R(1+\dot{R}^{2}e^{-2\alpha}-R^{\prime 2}e^{-2\beta})\right]_{,\,t}=-(p_{r}-\frac{4}{3}\eta \sigma-\zeta \theta) R^{2}\dot{R}.
\end{eqnarray}
So, comparing \eqref{F_eqn} and \eqref{F_equations}, the mass function is given by the following
\begin{equation}\label{mass_function}
F(r,t)=R(1+\dot{R}^{2}e^{-2\alpha}-R^{\prime 2}e^{-2\beta}).
\end{equation}
This is the same equation as the Misner- Sharp mass function for the spherical symmetry.
Let us now collect the equations for our problem of gravitational collapse of viscous fluids. Defining 
the two functions, $H=e^{-2\alpha(r, t)}\dot{R}^2$ and $ G=e^{-2\beta(r, t)}R'^2$,
the equations reduce to: 
\begin{eqnarray}
 &&\rho=\frac{F'}{R^2 R^{\prime}}\,\, ,\label{1eq1}\\
 &&{p}_{r}=-\frac{\dot{F}}{R^2 \dot{R}}+\frac{4}{3}\eta \sigma+\zeta \theta\,\, ,
 \label{1eq2}\\
 &&\alpha^{\prime}=\frac{2R'}{R}\frac{p_{t}-p_{r}+2\eta \sigma}{\rho+p_r-\frac{4}{3}\eta\sigma-\zeta\theta}- \frac{({p}_{r}-\frac{4}{3}\eta\sigma-\zeta\theta)^{\prime}}{\rho+p_r-\frac{4}{3}\eta\sigma-\zeta\theta}  \label{1eq3} \,\, ,\\
 && 2\dot{R}'=R'\frac{\dot{G}}{G}+\dot{R}\frac{{H}'}{H}\, \, ,  \label{1eq4}\\
 && F=R(1-G+H).  \label{1eq5}
 \end{eqnarray}
The first two are the $G_{00}$ and the $G_{11}$ equations, as described in the equations \eqref{F_equations}, the third is the Bianchi identity \eqref{r-Bianchi}, the fourth is the \eqref{R01equation} and the fifth is the mass function equation \eqref{mass_function}. These set of equations are the ones to be used to study the gravitational collapse of the fluid.

Let us now make some remarks on the domain of validity of various functions in the equations above. Note 
that the number of unknowns in the above equations are three metric variables $\alpha(r,t)$, $\beta(r,t)$
and $R(r,t)$, and the matter variables $p_{t},\,p_{r},\, \rho, \eta\sigma$ and $\zeta\theta$. This gives 
us the choice of three free functions and the mass function. In the following sections, 
we shall consider several
choices of these free functions and show that these choices, given the regular initial choice
of collapse, determine the spacetime uniquely.

At the start of the collapse $t_{i}=0$, we implicitly consider only those profiles of the matter cloud
which satisfy the energy conditions and have regular and smooth energy- momentum tensors. At $t_{i}=0$,
we use the gauge freedom of the $R(r,t)$ to fix it, so that $R(r, t_{i})=r$. In general,
this gauge freedom is a scaling, of the form $R=r\,a(r,t)$, where, the function $a(r,t)$
will satisfy certain conditions. Firstly, at $t=t_{i}$, $a(t_{i})=1$, secondly, at the singularity
time $t_{s}$, $a(r,t_{s})=0$, and thirdly to maintain the condition of collapse, $\dot{a}<0$.
It immediately follows from the equation \eqref{1eq1} that, at the initially epoch,
the density $\rho=F^{\prime}/r^{2}$
and hence, the regularity of $F$ at $r=0$ demands that the
$r$ dependence of the mass function take the form $F(r,t)=r^{3}\, m(r,t)$, where 
$m(r,t)$ is a sufficiently smooth and differentiable function inside the gravitating system. 
It's $t$- dependence
is not determined from the equation \eqref{1eq2} and requires the specification
of the $p_{r}$ and other parameters of the energy- momentum tensor. Also, note that, for
physical situations, the function $m(r,t)$ must be a smooth and decreasing function of $r$. 

Let us also note from the first equation \eqref{1eq1}, that the density diverges for $R=0$ as well as 
$R^{\prime}=0$. The $R=0$ implies that the area radius vanishes and hence signifies the collapsing of
shells to form shell focussing singularity at the center of the mater cloud. On the other hand 
$R^{\prime}=0$ indicates shell crossing singularities. 
As is well known, these are gravitationally
weak and point to existence of coordinate singularities. We shall not occupy ourselves with 
shell focussing singularities here.

\section{Pressureless Collapse: OSD and LTB models}\label{sec4}
For the dust collapse scenario, the viscosity coefficients $\eta$ and $\zeta$
may be taken to be zero. To qualify as dust, the fluid must be pressureless.
Furthermore, we impose $p_{r}=p_{t}=0$, which implies that the
radial and tangential pressures are equal and vanishing. This leads to the following equations:
\begin{eqnarray}\label{osd_ltb_eqns}
& F^{\prime}=\rho\, R^2 R^{\prime}\,\, ,& \dot{F}=0\, ,\\
 &\alpha^{\prime}=0\,\, ,~~~~~~~~& \frac{\dot{R}'}{R'}=\dot{\beta}.
\end{eqnarray}
Several conclusions follow from these equations. Firstly, from $\dot{F}=0$, 
we get $F=F(r)$. Since $F(r)$ is the amount of mass enclosed inside the cloud of 
co-moving radius $R$, above condition implies that the mass inside the cloud does not change with time.
This is reasonable since there is no influx or outflux of matter. Secondly, from the
equation $\alpha^{\prime}=0$, it follows that $\alpha=\alpha(t)$.
Note that one may define a new coordinate time $\bar{t}$ such that $d\bar{t}=e^{\alpha}\, dt$, and
hence, the function $\alpha(t)$ may be absorbed by redefinfing the time function. For
our case, we shall consider $t$ to be the comoving time function and hence, $\alpha=0$ effectively.
Thirdly, from the equation $(\dot{R}'/{R'})=\dot{\beta}$, we get that $R^{\prime}=e^{\beta(r,t)+h(r)}$,
where $h(r)$ is any arbitrary function of $r$ only. Let us redefine to call $e^{2h(r)}=1-k(r)$, where $k(r)$
is a function $r$. This implies that the mass function is given by
\begin{equation}
\dot{R}^{2}=\frac{F(r)}{R}-k(r)\, , \label{EOM}
\end{equation}
and the metric becomes:
\begin{equation}\label{metric_interior}
ds^2=-dt^2+\frac{ R^{\prime}(r,t)^{2}}{1-k(r)}\, dr^{2}+R(r,t)^2 \left(d\theta^2+\sin^2{\theta}\, d\phi^2\right).
\end{equation}
Note that the expression for $\dot{R}$ will have two signatures, $+$ve
for the expanding phase and $-$ve for the contracting phase.
The function $k(r)$ is quite significant since it determines the nature of the gravitational collapse.
If $k(r)=0$, we get the marginally bound collapse, where the shells of the matter cloud are 
assumed to have zero initial velocity at infinity or at the beginning of the collapse, $k(r)>0$
signifies bounded collapse, where the matter shells have negative initial velocity, whereas
$k(r)<0$ holds for unbounded collapse where matter at the beginning of collapse is assumed to 
have positive velocity. For later convenience, it is useful to rewrite the function $k(r)$ in a 
scaling form, $k(r)=r^{2}K(r)$.  The general solution of the (\ref{EOM}) is
\begin{eqnarray}\label{shell_solution}
 t&=&t_s-\frac{R^\frac{3}{2}}{\sqrt{F}}\, Y\left[\frac{R\,k(r)}{F}\right]
\end{eqnarray}
where $t_s$ is the time for all the collapsing shells to reach at the central singularity $R=0$ is given
by
\begin{eqnarray}
t_s&=&\frac{r^\frac{3}{2}}{\sqrt{F}}\, Y\left[\frac{rk(r)}{F}\right]
\end{eqnarray}
The function $Y(y)$ is given by the following form \cite{Landau_Lifshitz}:
 \begin{eqnarray}
 Y(y)&=&\frac{\sin^{-1}{\sqrt{y}}}{y^{3/2}}-\frac{\sqrt{1-y}}{y}, ~~~~~~~~~~~1\ge y>0\nonumber\\
 &=&(2/3), ~~~~~~~~~~~~~~~~~~~~~~~~~~~~~~~~~~~y=0 \nonumber\\
 &=&-\frac{\sinh^{-1}{\sqrt{-y}}}{(-y)^{3/2}}-\frac{\sqrt{1-y}}{y} , ~~~~~ 0> y>-\infty .
\end{eqnarray}

One of the most useful quantities in this study is the quantity $C$ as given in 
\eqref{value_of_c}, since it determines the signature of the MTT formed 
during the gravitational collapse. For the case of dust, only the density
appears in $T_{ab}$. Using the Einstein equation \eqref{1eq1}, we get that the 
equation for $C$ simplifies to give 
\begin{equation}
C=\frac{2F(r)^{\,\prime}}{2R(r,t)^{\, \prime}- F(r,t)^{\,\prime}}
\end{equation}
This formula shall be used in the following sections to determine the nature of MTT.
\subsection{Homogeneous collapse}
For homogeneous collapse, the mass function may be written as $F(r)=m\,r^{3}$, where the 
function $m(r)$ is a constant independent of $r$. The scaling variable $a(r,t)$ is also 
reduced to a function of $t$ only and the function $K(r)$, is taken to be a constant. Incidentally, 
the values of $K$ here determines the nature of collapse - $K=0$, indicates marginally 
bound collapse, $K=1$ is for bounded collapse and $K=-1$ signifies unbounded collapse.

\subsubsection{Marginally bound collapse}
For the marginally bound case, the solution of the equation of motion \eqref{EOM} is given by 
eqn. \eqref{shell_solution}:
\begin{eqnarray}
R(r,t)=r\left( 1-\frac{3}{2}\,\frac{\sqrt{F}}{r^{\,3/2}}\,t\right)^{\frac{2}{3}}. \label{tz}
\end{eqnarray}
The abovementioned equation \eqref{tz}, gives the time curve for the collapsing shell. Also 
note that here, $\dot{R}<0$. The time for the shell to reach singularity, denoted by $t_{s}$, 
follows from the equation eqn. \eqref{tz} by putting $R=0$. 
This gives us $t_{s}=(2/3\,\sqrt{m})$. Since $m$ is a constant here, it follows from this relation that 
all shells reach the singularity at the same time. 

\begin{figure}[h]
	\centering
	\includegraphics[width=0.45\linewidth]{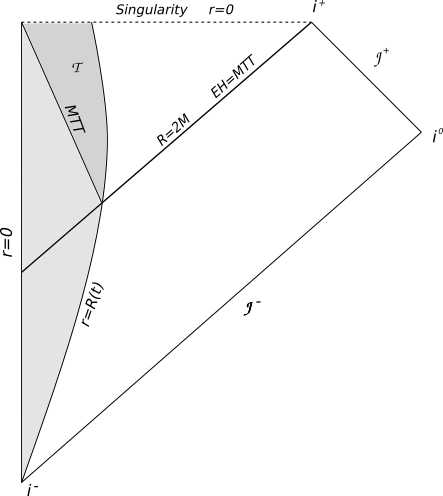}
	\caption{The Penrose diagram for $k=0$ dust collapse model of a black hole. The curve $r=R(t)$ 
	shows the boundary of the dust cloud. The cloud reaches the singularity at $t=0$. All shells
	reach the singularity at the same time. The boundary of the cloud reaches the Schwarzschild radius at
	$t=-4M/3$. at exactly the same time, the null ray of event horizon also reaches that point. Once outside, the event horizon matches with the null marginally trapped tube (MTT). Furthermore, exactly
	at $t=-4M/3$, just as the cloud reaches its Schwarzschild radius, a timelike marginally trapped tube (MTT) forms and begins to shrink at the rate of $\dot{R}_{ah}=-3/2$ to reach the 
	singularity at $t=0$. The region $\tau$ inside the cloud is the trapped region.}
	\label{fig:OSDK0}
\end{figure}

For simplification of solutions to the differential equations, we shall work with the scenario
where the singularity time $t_{s}$ is shifted to $t_{s}=0$. This essentially 
shifts the time coordinates linearly without
changing any physical content. The motion of the collapsing shell after this shift in time
coordinate becomes:
\begin{eqnarray}
R(r,t)=\left[\,(3/2)\,\sqrt{F}\,\,(-t)\,\right]^{\frac{2}{3}}. \label{tz1}
\end{eqnarray}
This equation also gives us the time for the shell to reach the Schwarzschild radius
located at $r=r_{H}$. 
Note that, on the hypersurface, apart from the condition $R=2M$,
one also has the matching conditions at $r=r_{H}$, given by
$F(r_{H})=2M=r_{H}^{3}m$ (see the equation \eqref{matching_boundary_1} in the Appendix). 
To find the time for the shell to reach the Schwarzschild radius, 
denoted by $t_{H}$, we put these conditions in the equation \eqref{tz1} to get:
\begin{eqnarray}
 t_{H}=-\frac{2}{3}F(r_{H})=-\frac{4M}{3}. \label{tzE}
\end{eqnarray}
Let us now find the development of the MTT. 
As is the standard practice now, in the following, we shall sometimes also use 
apparent horizon (AH) to mean a MTT. 
The equation for the MTT/ AH in spherical symmetry is given by the condition 
$g^{ab}\,\nabla_{a}R\,\nabla_{b}R=0$. For the metric being studied here, this implies $R_{ah}=F(r)$.  
The  time curve of the apparent horizon is obtained from the equation \eqref{tz1} by using
$R=F$ and gives:
\begin{equation}\label{ah_trajectory}
R_{ah}(r)=-\frac{3}{2}\, t.
\footnote{From the equation \eqref{tz}, the time curve of apparent horizon is given by
 $t_{ah}(r)=(2/3)\sqrt{r^3/F}-(2/3)R_{ah}$. This clearly shows that 
 the AH starts to form at exactly the same time when shell reaches the Schwarzschild radius, $R=2M$.}
 \end{equation}
Naturally, it follows from this equation \eqref{ah_trajectory} that the AH is formed at $R=2M$, 
and shrinks with time and collapses to $R=0$ at the time of singularity formation $t=t_{s}$. 
More precisely, the apparent horizon starts at $R=2M$ at time $t_{H}$, shrinks 
at a constant rate $\dot{R}_{ah}=-3/2$, and reaches the singularity $r=0$ at $t=0$. The
collapsing spacetime admits two marginally trapped tubes (MTT) formed out of the marginally trapped 
surfaces. Outside the collapsing region, the marginally trapped tube matches with the $R=2M$ null surface,
whereas, inside the collapsing star, the trajectory of the surface follows the equation
\eqref{ah_trajectory}. The trajectory of the AH is shown in the figure (\ref{fig:OSDK0}). Since 
the MTT outside matches with the EH, it is null, whereas, the MTT inside is
timelike.

Let us now find the time development of the event horizon.
Using the metric, we evaluate time evolution of radius along a radial null geodesic.
The radial null geodesic of the outgoing photons give
\begin{eqnarray}
 \left(\frac{dr}{dt}\right)_{Null}= \frac{1}{R'}.
\label{nullkz}
\end{eqnarray}
Again, using $dR/dt=[R' \left(dr/dt\right)_{Null}+\dot{R}]$, the time evolution
of the event horizon reduces to: 
\begin{eqnarray}
\frac{dR}{dt}&=&1-\sqrt{\frac{F}{R}} \label{nullz}
\end{eqnarray}
Since $R(t)^{3/2}=(3/2)\sqrt{m}r^{3/2}(-t)$ and $F(r)=m\, r^{3}$, the previous equation gives:
\begin{eqnarray}
\frac{dR}{dt}&=&1+\frac{2}{3}\frac{\,R}{\,t}. \label{nullz1}
\end{eqnarray}
The solution of this equation gives the event horizon.
The general solution of this equation is obtained by integrating
with the integrating factor $t^{-2/3}$ and gives:
\begin{equation}\label{reh_eqn1}
R_{eh}(t)=3t+C^{\prime}(-t)^{2/3}.
\end{equation}
The constant $C^{\prime}$ is fixed as follows: From equation \eqref{tz1}, the time 
taken by the shell to reach $R=2M$ is $t=-4M/3$. Since the event horizon is the last
null ray reaching the null infinity, we use this condition in equation \eqref{reh_eqn1}
to obtain $C^{\prime}=3(9M/2)^{1/3}$.

So, from equation \eqref{reh_eqn1}, it follows that
the event horizon begins to grow from the non- singular center just as the collapse process 
begins. The time of beginning of event horizon is obtained from equation \eqref{reh_eqn1} as follows:
Let at $t=t_{eh}^{i}$, $R_{eh}=0$. This gives the time of formation of the event horizon
$t_{eh}^{i}=-(9M/2)$. The rate of growth of the event horizon is also obtained from \eqref{reh_eqn1}
and gives
\begin{equation}
\dot{R}_{eh}=3-2(9M/2)^{1/3}(-t)^{-1/3}.
\end{equation}
This clearly shows that initially, 
at $t_{eh}^{i}$, $\dot{R}_{eh}=1$ whereas, at $t=-4M/3$, 
just as the shell reaches the Schwarzschild radius (or the null curve of the event horizon reaches $R=2M$),
the event horizon stops growing, $\dot{R}_{eh}=0$ and remains at the Schwarzschild radius.
To sum up, the event horizon begins to develop just as the matter shells 
begin to fall and then its rate of growth slows down as 
rate of fall of matter begins to slow down, ultimately stopping at time $t=-4M/3$ when matter flow stops 
and accordingly, it matches with the Schwarzschild null horizon (see fig. \ref{fig:OSDK0}). 

\subsubsection*{Example}
In the following we shall consider the examples of a simple mass profile
which shall collapse according to the formalism developed above.
Let us consider the Misner- Sharp mass function to be $F(r)=mr^{3}$, with $m=(1/2)$.
The $t-R(r,t)$ graph for the collapse is given below in 
figure \ref{fig:OSDK0_mass}. We have taken care to exclude
shell- crossing singularities and have ensured that
there are no trapped surfaces on the initial slice. Several points need
to be noticed. First, all the shells collapse to the singularity at the same time at $t=0.94$. 
Second, the shell which begins at $R(r,t)=1$ 
reaches its Schwarzschild radius at time $t=0.61$, and exactly at that
instant, the event horizon (or the last null- ray), 
beginning at the center of the cloud, also reaches that spacetime point. 
Thirdly, the MTT also forms at that point and eventually collapses to singularity,
along with the matter cloud. Further, the graph of $C$ establishes that the MTT is timelike. 
Note that this 
graph is identical to figure \ref{fig:OSDK0}, obtained for a generic mass profile. 
\begin{figure}[htb]
\begin{subfigure}{.62\textwidth}
\centering
\includegraphics[width=\linewidth]{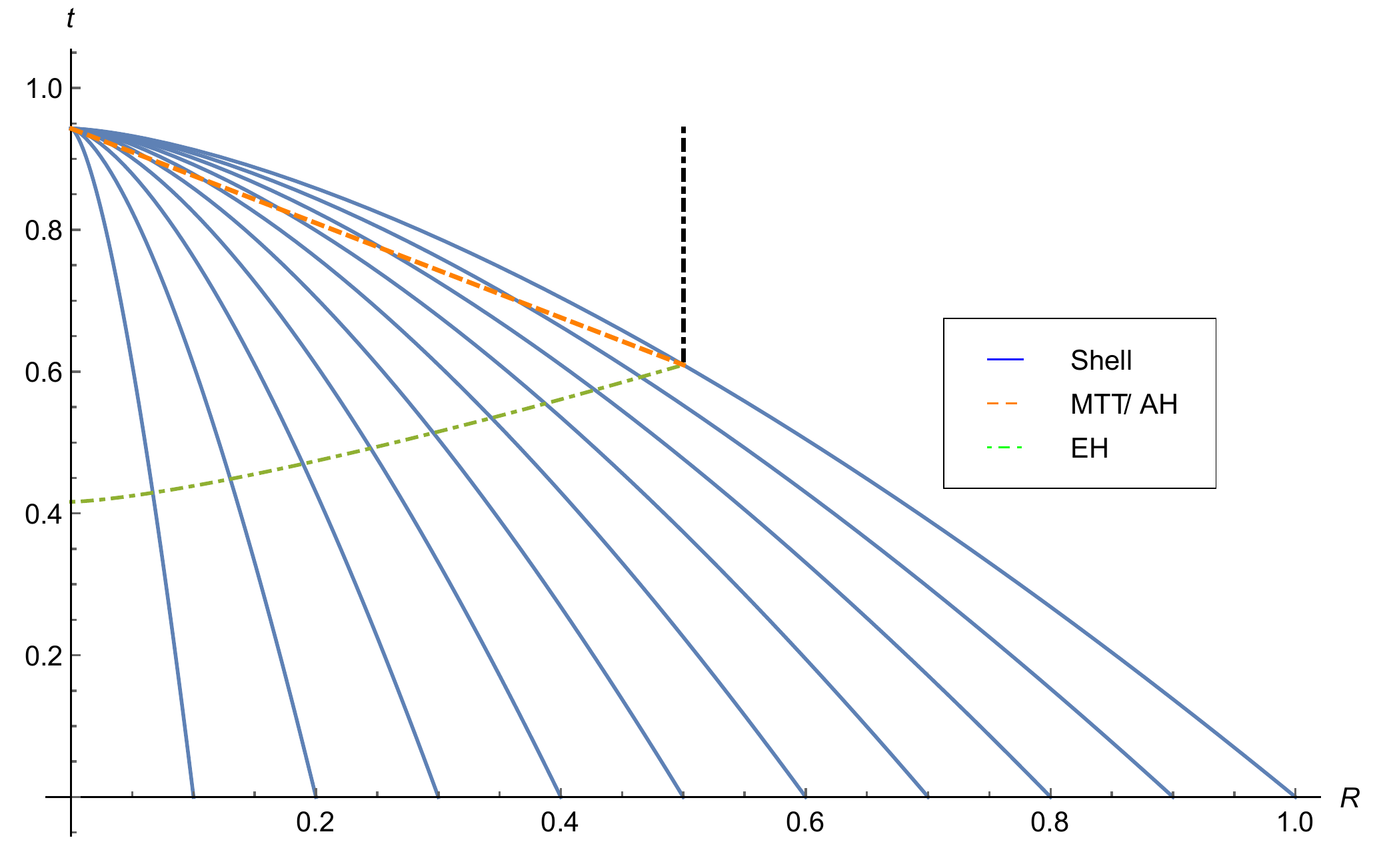}
\caption{}
\end{subfigure}
\begin{subfigure}{.4\textwidth}
\centering
\includegraphics[width=\linewidth]{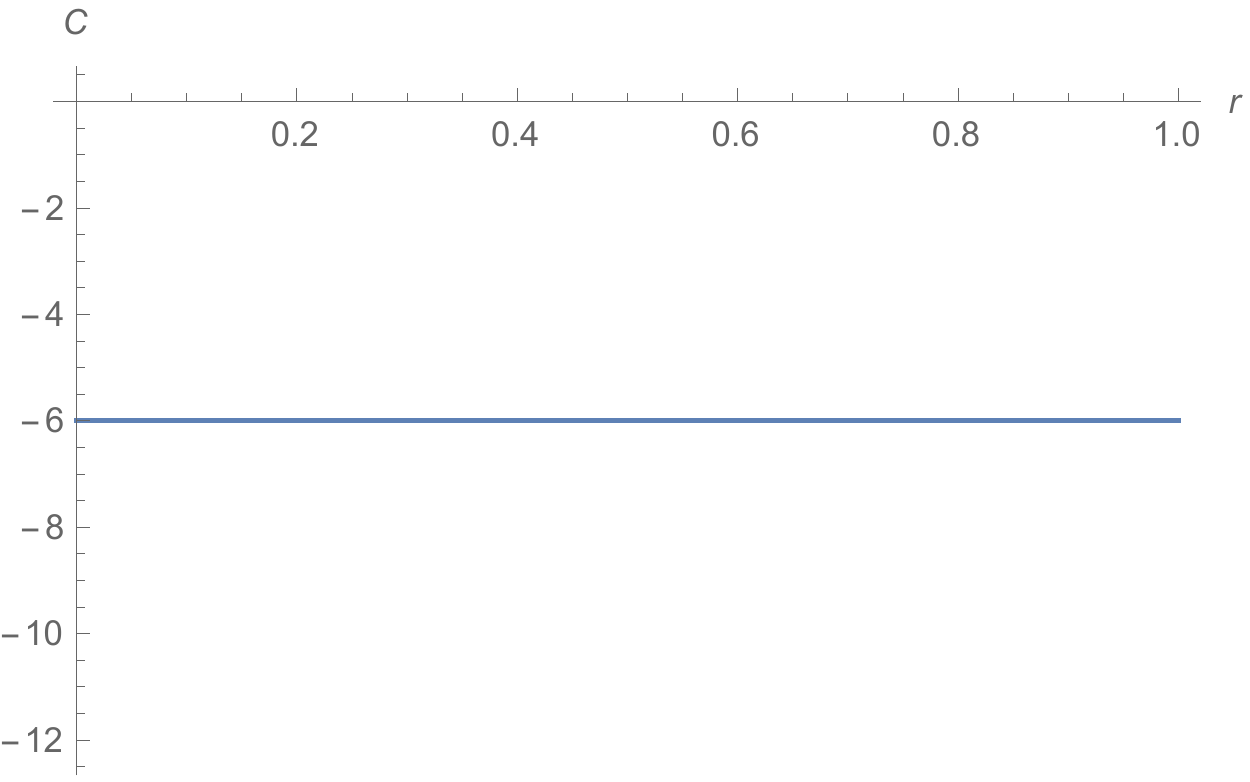}
\caption{}
\end{subfigure}
\caption{The figure (a) gives the plot of $R(r,t)$ vs $t$ for the mass profile discussed above.
The MTT is timelike is also confirmed through the negative value of $C$ in the
graph of $C-r$ in (b). }
\label{fig:OSDK0_mass}
\end{figure}
%

\subsubsection{Bounded collapse}

The line element of the interior metric \eqref{metric_interior} can be rewritten with the parameter
 $r=\sin{\chi}$ as
\begin{eqnarray}
 ds^2=-dt^2+a^{2}(t)\,d\chi^2 +R(r,t)^2 (d\theta^2+\sin^2{\theta}\,d\phi^2).\label{m2-} 
\end{eqnarray}
The parametric solution of the \eqref{EOM} for the $k>0$ is given by the following equations
\begin{eqnarray}
&&R(r,t)=\frac{F(r)}{k(r)}\cos^2\left(\eta/2\right)=r m \cos^2\left(\eta/2\right), \label{Rk0}\\
&&t=\frac{F(r)}{2k(r)^{3/2}}\left( \eta+\sin \eta\right)=\frac{m}{2}\left( \eta+\sin \eta\right), \label{tk=1}
\end{eqnarray}
where $k(r)=r^2$ and $F(r)=m r^3$. The equation \eqref{Rk0} shows that
$a(t)=m\cos^2\left(\eta/2\right)$. Furthermore, the expressions for radius and time coordinates
in \eqref{Rk0} and \eqref{tk=1} respectively show that
collapse begins at $\eta=0$, where $t_{i}=0$ and it reaches the singularity, $R=0$ at
$\eta=\pi$. Thus, the maximum value of the scale factor is $a(t=0)\equiv a_{0}=m$ at 
$\eta=0$ and the minimum value is $a=0$ at $\eta=\pi$, when the matter cloud reaches singularity. 
These relations are the same ones in \cite{Landau_Lifshitz} except that time has been shifted 
by ($\eta \rightarrow \pi-\eta$). A further interesting fact follows directly from 
equation \eqref{Rk0}, which may also be written as:
\begin{equation}
R(r,t)= (R_{0}/2)(1+\cos \eta).
\end{equation}
Consequently, the time taken by the shell initially at $R_{0}=(rm)$ to reach it's Schwarzschild radius at
$R=2M$ is given by:
\begin{equation}\label{timeshell_2m}
\eta_{2M}=\cos^{-1}\left(4M/R_{0}-1\right).
\end{equation}
This is exactly the same moment where the apparent horizon starts forming.

Now, recall that to an outside observer, the collapse process only leads to a Schwarzschild
spacetime. Hence, the metric of the cloud interior (FRW) must be matched with the Schwarzschild spacetime
in the exterior. This matching at the junction, of the metric function
and the extrinsic curvatures (in particular, $K_{\theta\theta}$) give us the following conditions respectively:
\begin{eqnarray}
R(t)=a(t)\sin\chi\\ 
2M=F(r),
\end{eqnarray}
where $M$ is the total mass enclosed by the shell. At the beginning of the shell collapse 
(at $\eta=0$), these conditions imply that 
\begin{eqnarray}
R(t=0)\equiv R_{0}=m\sin\chi_{0}, ~~~~~~~~~~~ 2M=F(r_{0}),
\end{eqnarray}
where $r_{0}=\sin\chi_{0}$ is the radial coordinate of the shell boundary at the beginning of the collapse.
From the above two equations, it follows simply that
\begin{equation}
\chi_{0}= \sin^{-1}\left(2M/R_{0}\right)^{1/2}, ~~~~~~~~~~~~ a_{0}=m=\left[R_{0}^{3}/(2M)\right]^{1/2}.
\end{equation}
We shall label the shell by the radial coordinate it has at $\eta=0$. The time for the 
collapsing shells labeled by '$r$' to reach the central singularity $R=0$ 
is given by
\begin{eqnarray}
t_s=\frac{\pi F(r)}{2k(r)^\frac{3}{2}}=\frac{\pi}{2}m. \label{ts}
\end{eqnarray}
From equation (\ref{ts}), $t_s=(\pi m/2)$ is constant and hence, all
collapsing shells  with different initial 
radius will reach the central singularity at the same time. 

Let us have a look at the trapped surfaces and identify the trapped
region. For spherical symmetry, the outermost trapped surface is identified through 
the equation $R(t,r)=F(r)$. From equation (\ref{Rk0}), we get $\eta=2\cos^{-1}[R(r,t)k/F(r)]^{1/2}$ and 
hence, the AH/MTT is described by  
\begin{equation}\label{eqn_ah_eta}
\eta_{AH}=2\cos^{-1}r=2\cos^{-1}\,(\, \sin \chi\,)=\pi-2\,\chi.
\end{equation}
Thus, for a shell labeled by $r_{0}$ or $\chi_{0}$, the AH/MTT
forms at the time given by the coordinates:
\begin{equation}
\eta_{AH}=\pi-2\chi_{0}=2\cos^{-1}\left(2M/R_{0}\right)^{1/2}=\cos^{-1}\left(4M/R_{0}-1\right),
\end{equation}
where we have used the trigonometric identity $\cos^{-1}x=(1/2)\cos^{-1}(2x^{2}-1)$. 
This equation clearly shows as clouds of larger and larger initial radius 
$R_{0}$ are considered, their AH form at later times.  
Furthermore the time of formation of the horizon is exactly same as the time when the cloud reaches it's 
Schwarzschild radius, given in equation \eqref{timeshell_2m}.
These equations also imply that inside the matter cloud, radius of the AH decreases at a 
constant rate. This follows simply because, for a shell of fixed $r$, the rate 
of decrease of radius of AH is equivalent to
finding  $da/d\eta$. It is easy to see from $a(\eta)=m\cos^2\left(\eta/2\right)$
and the equation \eqref{eqn_ah_eta} that it is indeed a negative constant inside the matter. 
Using the previous equation in (\ref{tk=1}), we may also find the time curve of the apparent horizon:
\begin{equation}
t_{ah}=(m/2)\left(\pi-2\chi +\sin\,2\chi \right).
\end{equation}
Another way to determine the trapped region is to use it's definition, that light rays remain confined in 
this strong gravitational field and the proper area of this bundle
of light rays do not grow with time. To use this formulation, let us 
go to the the proper time coordinates $\eta$ given through equation $\eqref{tk=1}$. 
In the $(\eta- \chi)$ coordinates, the light rays are 
given by the equation $(d\chi/d\eta)=1$. Now, for the metric (\ref{m2-}), apart 
from a constant $4\pi$, the total area is $A=a^2\sin^{2}\chi$ and also, from equation (\ref{Rk0}), 
it follows that $da=-(m/2)\sin\eta\,d\eta$. Thus, we get that the variation of the area
along vector field $(\partial/\partial\eta)$ of proper time coordinate $\eta$ is:
\begin{eqnarray}
\pounds_{\eta}\,A&=& \frac{da}{d\eta}\sin\chi+a\cos\chi \frac{d\chi}{d\eta} \nonumber\\
&=&-m\sin\chi\sin[\eta/2]+m\cos[\eta/2]\cos\chi=m\cos\left[\frac{\eta}{2}+\chi\right].
\end{eqnarray}
Using $\cos(\pi/2)=0$, the expression for non- increase of area, given by $(dA/d\eta)\le 0$ translates to
\cite{Rezzolla_ Zanotti}:
\begin{eqnarray}
\eta\ge \pi-2\chi, \label{tahkg0}
\end{eqnarray}
where the inequality signifies trapped region and the equality 
gives boundary of the trapped region, also called marginally trapped tube or apparent horizon.
Note that this equation \eqref{tahkg0} is exactly same as the one derived in eqn. \eqref{eqn_ah_eta}
using different methods. 
\begin{figure}[htb]
\begin{subfigure}{.62\textwidth}
\centering
\includegraphics[width=1.1\linewidth]{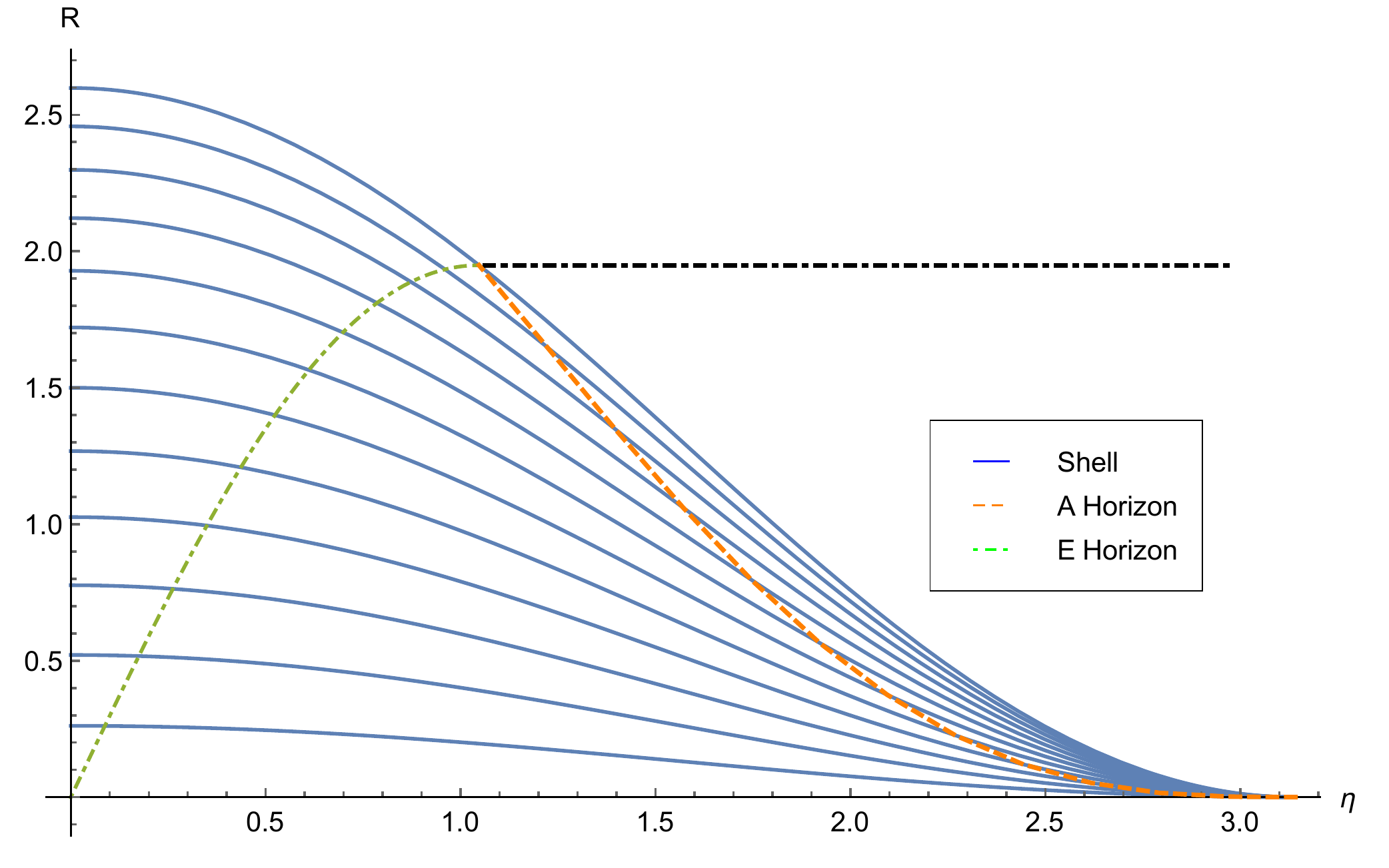}
\caption{}
\end{subfigure}
\begin{subfigure}{.4\textwidth}
\centering
\includegraphics[width=0.9\linewidth]{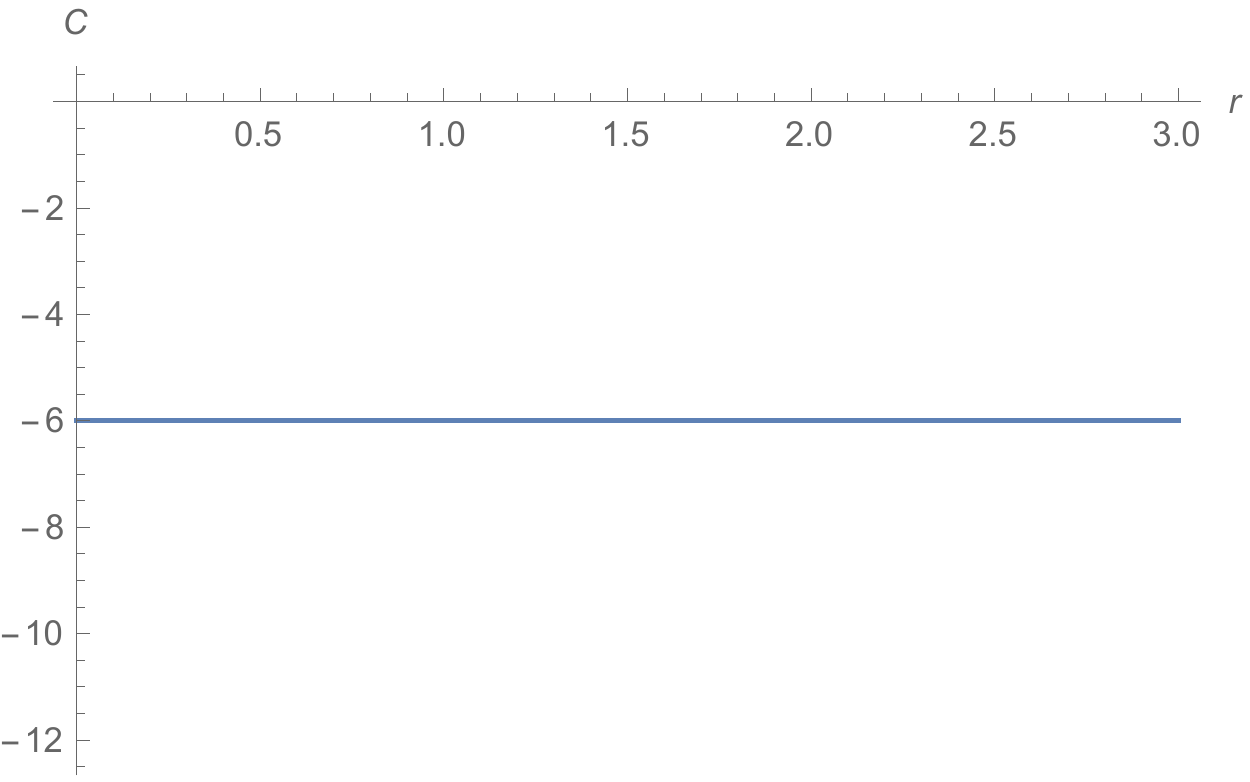}
\caption{}
\end{subfigure}
\caption{Plot of collapsing shell (solid line), MTT (or apparent horizon) (dashed line) and 
	event horizon (dot- dashed) with respect to $\eta$. For OSD model, the equations are 
	given for collapsing shells in (\ref{Rk0}), for apparent horizon in (\ref{eqn_ah_eta}) and 
	for event horizon (\ref{Rehkg0}). We 
	consider the matching of interior to the exterior at the hypersurface $\chi_{0}=\pi/3$ 
	where $r_{0}=0.866$. For simplicity, each collapsing shell has the constant mass $m=3$ for which 
	the Schwarzschild radius is formed at $\left(\eta_{2M},R_{2M}\right)=\left( 1.0472,2\right)$.
	All three curves collapsing shell, apparent horizon and event horizon meet at $R=2$ 
	when $\eta_{_{2M}}=1.0472$. Here too, the MTT is timelike as may be observed from the
	graph of $C$ given above in (b) and hence, it is unstable. }
\label{fig:OSDKg0_mass}
\end{figure}

Now we also want to study the formation and evolution of 
the event horizon, more precisely identify when and where it first starts forming and
how it grows with time. Again, we shall need the null geodesics since
event horizon is also the last ray that reaches the asymptotic null infinity $\mathcal{I}^{+}$.
The outgoing radial null geodesics in the $(r,t)$ coordinates are given by
\begin{equation}
\frac{dr}{dt}=\frac{\sqrt{1-k(r)}}{R'}=\frac{\sqrt{1-r^2}}{m\cos^{2}\left(\eta/2\right)}.
\end{equation}
We shall however use the $(\chi-\eta)$ coordinates in which this equation 
becomes $(d\chi/d\eta)=1$. Now, we use the boundary condition that at the instant the  
the cloud reaches the Schwarzschild radius, the last outgoing null geodesic also reaches that point
at that same instant. For example, just as the shell with initial radius $R_{0}=m\sin\chi_{0}$ (labelled by
$R_{0}$ or $r_{0}$ or $\chi_{0}$) reaches the Schwarzschild radius ($R=2M$), the event horizon also 
reaches there at that instant. So, for this shell, the boundary condition is $\chi=\chi_{0}$ at
$\eta= \eta_{AH}=\eta_{2M}$.  In other words, the coordinates of the event horizon is given by:

\begin{equation}
\chi_{EH}=\chi_{0}+(\eta-\eta_{2M}). \label{chiehkg0}
\end{equation}
Using equation \eqref{Rk0}, the equation of the last outgoing null ray inside the cloud with initial
radius $R_{0}$ is given by
\begin{eqnarray}
R_{EH}&=&m r_{EH}\cos^{2}\left(\eta/2\right)
=m\sin\left(\chi_{0}+\eta-\eta_{2M} \right)\,\cos^{2}\left(\eta/2\right). \label{Rehkg0}
\end{eqnarray}
Note that just as the shell (of radius $R_{0}$)
begins to collapse, the event horizon also begins to form at $\eta=(\eta_{2M}-\chi_{0})$ 
and grows in such a way that at $\eta=\eta_{2M}=(\pi-2\chi_0)$, it's value 
reaches $R_{EH}=m\sin\chi_{0}\cos^{2}\left(\pi/2-\chi_H\right)=m\sin^{3}\chi_{0}=2M$. Also, we can 
find the rate of growth of the event horizon from equation \eqref{Rehkg0} by taking the derivative with
the proper time $\eta$,
\begin{eqnarray}
\frac{dR_{eh}}{d\eta}=m\cos\left[3\left(\chi_{0}+\eta/2\right)-\pi\right] \cos\left(\eta/2\right) \label{raterehkg0}
\end{eqnarray}
 where we have used $\eta_{2M}=\pi-2\chi_{0}$. So, the event horizon grows at a positive rate and 
 at $\eta=\eta_{2M}$, it becomes $dR_{EH}/d\eta=0$ and hence, it stops 
 growing and matches with the outermost trapped surface of the Schwarzschild spacetime.
 These results are summarised in the figure \ref{fig:OSDKg0_mass}.
 
 \subsubsection*{Example} 
\begin{figure}[htb]
\begin{subfigure}{.55\textwidth}
\centering
\includegraphics[width=\linewidth]{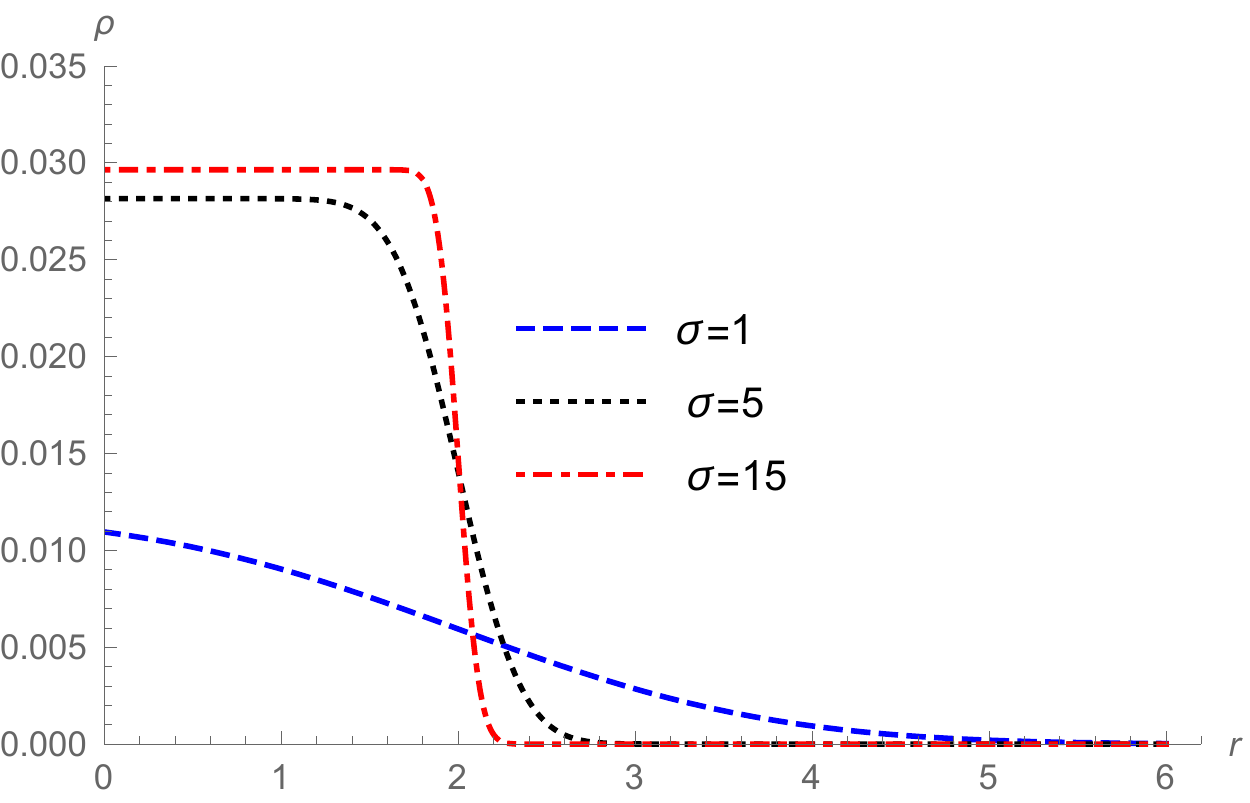}
\caption{}
\end{subfigure}
\begin{subfigure}{.45\textwidth}
\centering
\includegraphics[width=1.1\linewidth]{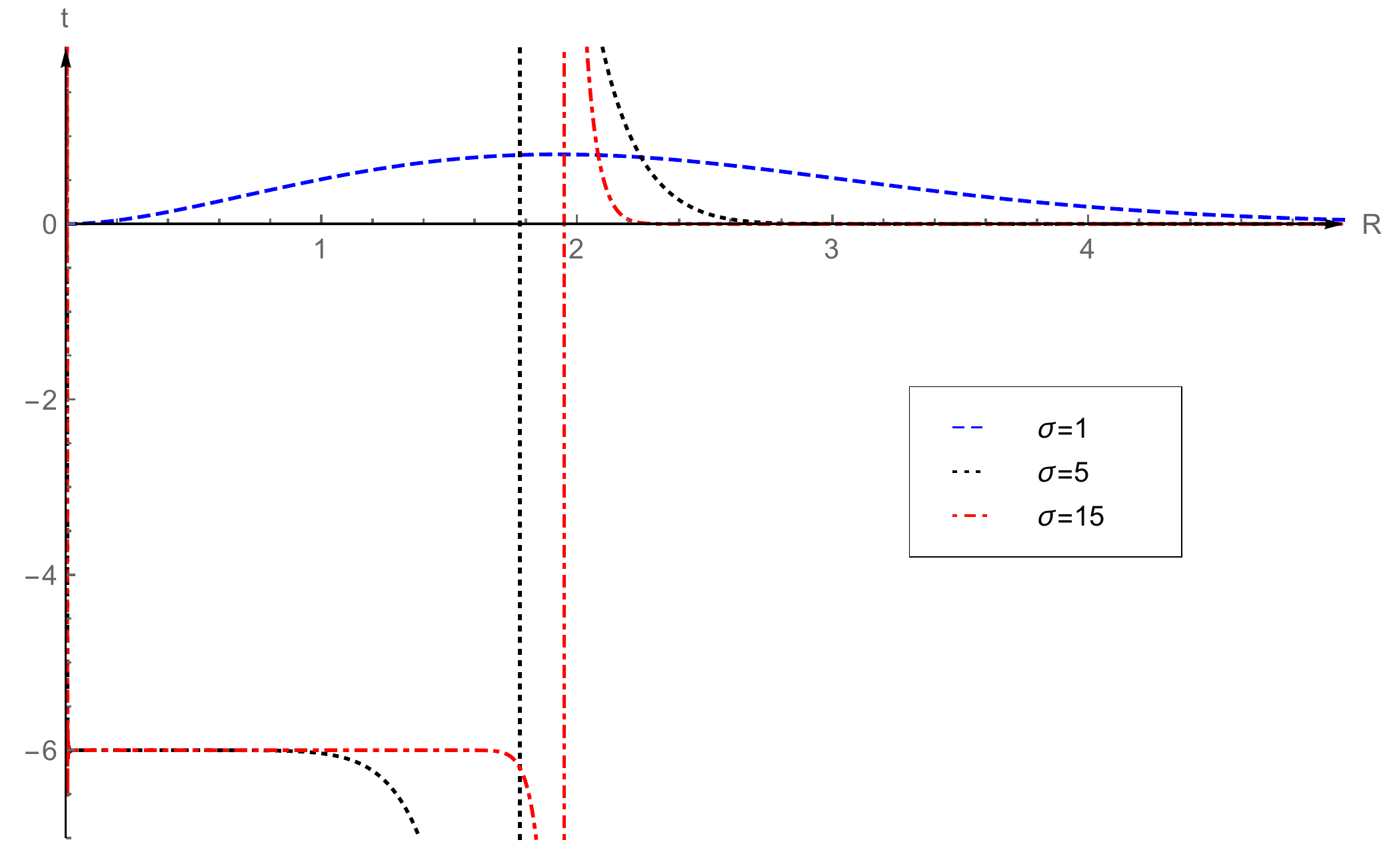}
\caption{}
\end{subfigure}
\begin{subfigure}{.55\textwidth}
\centering
\includegraphics[width=\linewidth]{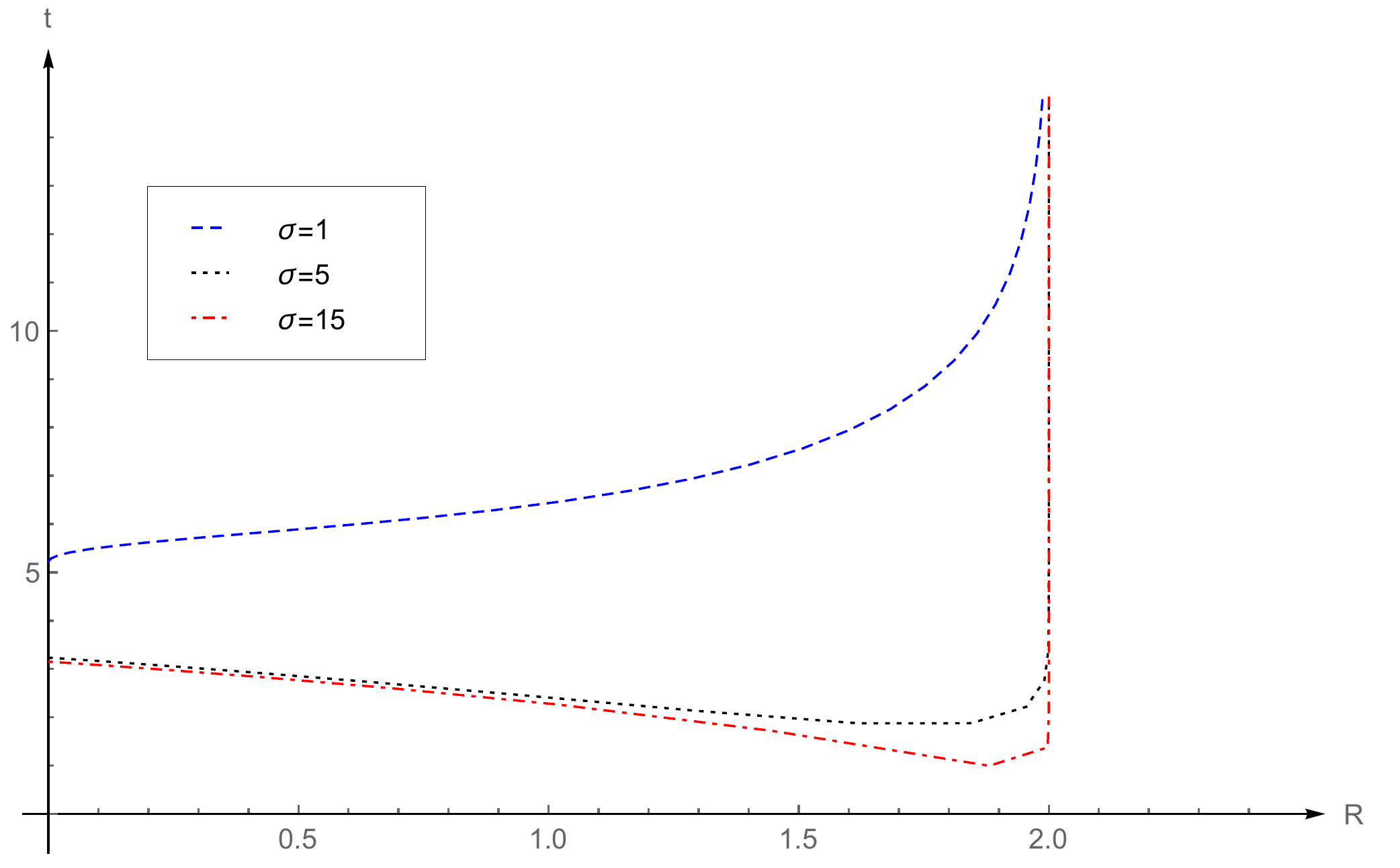}
\caption{}
\end{subfigure}\begin{subfigure}{.62\textwidth}
\centering
\includegraphics[width=0.8\linewidth]{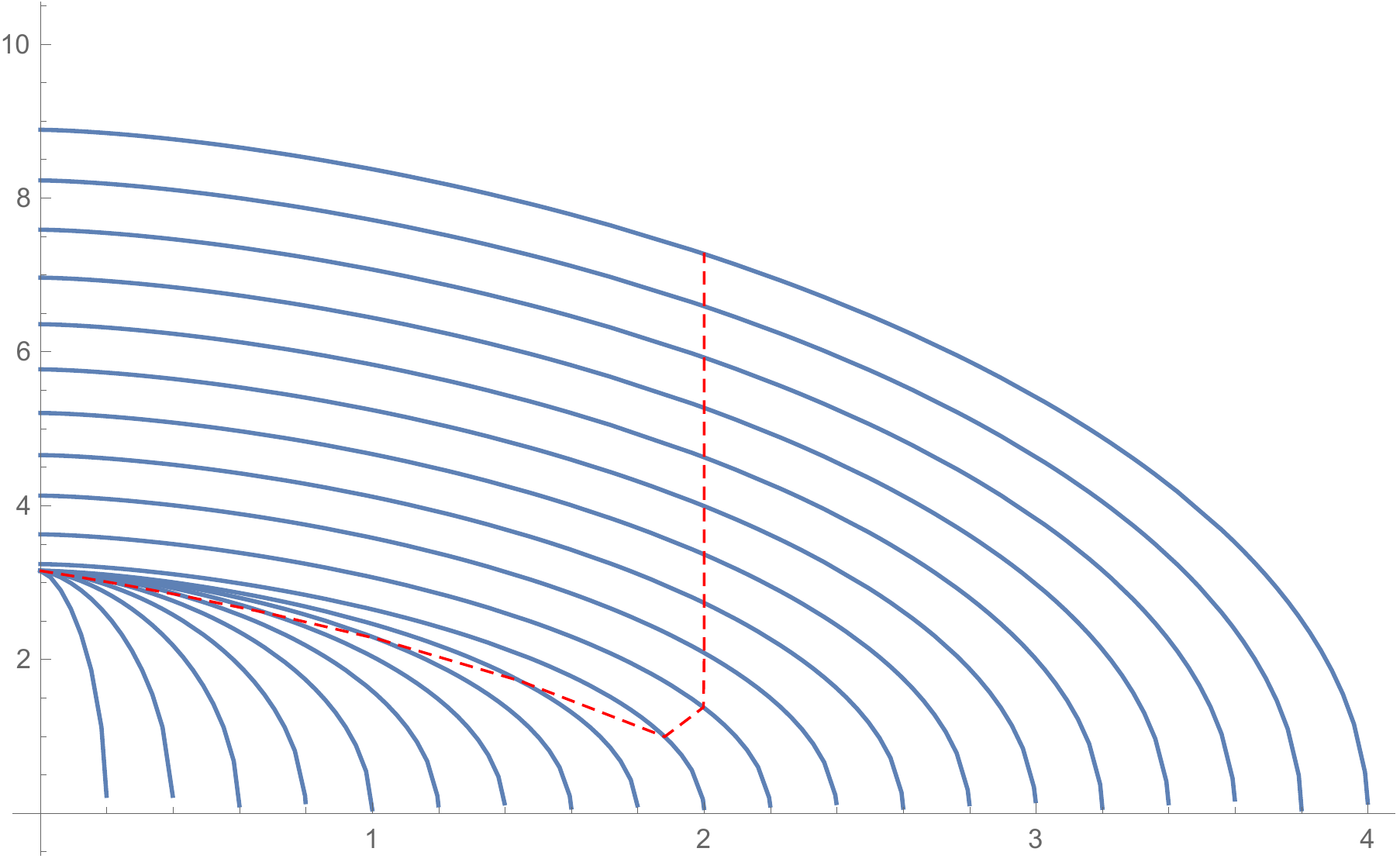}
\caption{}
\end{subfigure}
\caption{This gives the gravitational collapse for the OSD profile discussed above where (a)
gives the density fall-off, (b) gives the values for $C$, (c) gives the formation of MTT 
for $\varsigma=1,5,15$. The graph (d) shows the MTT for $\varsigma=15$ along with the shell coordinates.
The straight lines in (c) and (d) represents the isolated horizon phase.}
\label{fig:OSDKg0_example1}
\end{figure}

 In this example, we consider the density of the cloud to be of the following form \cite{Booth:2005ng}:
 \begin{equation}
 \rho(r)=\frac{m_{0}\,\mathcal{E}(\varsigma)}{r_{0}^{3}}\left[1-\erf\left\{\varsigma\left(\frac{r}{r_{0}}
 -1\right)\right\}\right],
 \end{equation}
 where $m_{0}=m(r\rightarrow \infty)$ is the total mass of the cloud,
 $r_{0}$ is the location on the cloud where it matches the 
 Schwarzschild radius ($r_{0}=2m_{0})$, and the quantity $\varsigma$
  controls the approach to the OSD model- larger the value of $\varsigma$, closer is 
  the density to uniformity. The function $\mathcal{E}(\varsigma)$
  has the following form:
  \begin{equation}
  \mathcal{E}(\varsigma)=3\varsigma^{3}\, \left[2\pi\varsigma(2\varsigma^{2}+3)(1+\erf \varsigma)
  +4\sqrt{\pi}\exp(-\varsigma^{2})(1+\varsigma^{2})\right]^{-1},
  \end{equation}
  and $\erf$ is the usual error function.
  We consider the cases where $\varsigma=1,5$ and $15$. The graphs
  are given in figure \ref{fig:OSDKg0_example1}.

  As seen from the graph of the density profile, the density 
  approaches the step function as $\varsigma$ varies from $1$ to $15$.
  For larger values of $\varsigma$, the density is a step function.  
  The $R(r,t)$ vs $t$  graphs for these various choices of $\varsigma$ are also plotted.
  We have ensured that the parameter space does not have shell- crossing singularities
  and that the initial time slice does not admit any trapped region. From the $R(r,t)-t$
  graph for $\varsigma=15$ (which is closest among these to the OSD model), we note 
  that the MTT forms at approximately $t=0.99$ and $R=1.88$. 
  Wherefrom, the MTT bifurcates into two parts: the timelike membrane collapses to 
  the center of the cloud, while the dynamical horizon asymptotes to the isolated horizon at $R=2$. 
  The timelike nature of the MTT, and it's approach to null through the spacelike phase may also 
  be confirmed through the $C-r$ graph. This graph shows that for smaller $r$
  the signature of MTT is negative, which jumps to $+$ve at $r=1.99$ and eventually
  becoming null at approximately $r=2.3$. Note that these changes are also
  corroborated through the graphs of density $\rho(r)-r$ as well as that of $R(r,t)$ vs $t$.
  For lower values of $\varsigma$, the MTT forms at lower values of $R(r,t)$.
  For example, for $\varsigma=5$, the MTT forms at $t=1.88$ and $R=1.62$. The $C$ value also remains
  $-$ve initally but asymptotes to $0$ as it becomes isolated.  For $\varsigma=1$, the situation is 
  drastically different since, it cannot be called to represent the OSD model. Here, the MTT is
  spacelike, which may also be confirmed from the values of $C$ in the figure below.

\subsubsection{Unbounded collapse}

The forms of the Einstein equations takes simplified form for dust collapse with $k<0$ when we 
use the parametric form of $R$ or $a(t)$. The line element of 
the interior metric \eqref{metric_interior} can be written with $r=\sinh{\chi}$ as
\begin{eqnarray}
 ds^2&=&-dt^2+\frac{R'^2}{1+k(r)}{ dr^2}+R^2 \left(d\theta^2+\sin^2{\theta} d\phi^2\right) \label{m2m-} 
\end{eqnarray}
The parametric solution of the \eqref{EOM} for the $k<0$ is
\begin{eqnarray}
R(r,t)&=&\frac{F(r)}{2|k(r)|}\left(\cosh\eta-1\right), \label{Rk-1}\\
t(\eta)&=&\frac{F(r)}{2|k(r)|^{3/2}}\left(\sinh \eta -\eta\right) \label{tk=-1}
\end{eqnarray}
where $k(r)=-r^2$ and $F=m r^3$. This is the case for unbounded collapse
and hence, the shells begin at some fixed time and follows the collapse process.
The collapse starts at $\eta=\eta_{0}$, where $t=t_{0}=(m/2)(\sinh\eta_{0}-\eta_{0})$
and $R_{0}\equiv R(r,t=t_{0})=(rm/2)(\cosh\eta_{0}-1)\equiv rm\alpha$, and it reaches 
the singularity at $\eta=0$ at $t=0$ where $R=0$. 
Again, these are the same equations as in \cite{Landau_Lifshitz} with the time 
coordinate shifted.

We may rewrite the equation \eqref{Rk-1} to obtain the equation of motion of the shell: 
\begin{equation}
R(t,r)=(R_{0}/2\alpha)(\cosh \eta -1),
\end{equation}
and hence, the shell of initial radius $R_{0}$ reaches its Schwarzschild radius $R=2M$
at the proper time 
\begin{equation}\label{req2mhyper}
\eta_{2M}=-\cosh^{-1}(4M\alpha/R_{0}+1).
\end{equation}
The sign has been kept $-$ve since the $\eta=0$ is the singularity time and the shell must 
reach it's Schwarzschild radius before that time. Also note that at $t_{s}=0$,
the collapsing shells reach the central singularity $R=0$. Thus, all shells with 
different initial radius will reach the central singularity at the same time.

Similar to the junction conditions for the previous two cases, here too the matching of the metric and the 
extrinsic curvatures for FRW spacetime inside the cloud with 
the Schwarzschild spacetime outside the cloud leads respectively to these equations:
\begin{equation}
R(t,r)=a(t)\sinh\chi, ~~~ 2M=F(r).
\end{equation}
So, at the beginning of the collapse of the shell with initial radius $R_{0}$,
these conditions imply that the following two relations hold:
\begin{equation}
R_{0}=m\alpha\sinh\chi_{0}, ~~~~~~ 2M=F(r_{0})=m\sinh^{3}\chi_{0},
\end{equation}
where $r_{0}=\sinh\chi_{0}$ is the radial coordinate of boundary of the cloud at the beginning
of the collapse. Thus, we rewrite these two equations as:
\begin{equation}\label{formula_chi0}
\chi_{0}=\sinh^{-1}(2M\alpha/R_{0})^{1/2}, ~~~~~~ m=[R_{0}^{3}/2M\alpha^{3}]^{1/2}.
\end{equation}

Let us now look at the formation of the trapped surfaces and in particular for MTT/AH.
From equation (\ref{Rk-1}), we have $\eta=\cosh^{-1}[(2R(r,t)k/F)+1]$ and 
hence, the time of formation of AH is obtained by using the condition
$R(r,t)=F(r)$ correspondingly, the time is: $\eta=\eta_{AH}=\cosh^{-1}(2r^{2}+1)
=\cosh^{-1}(2\sinh^{2}\chi +1)=-2\chi$. This just shows that the AH forms 
much before the shells reach the singularity. For a shell labeled by $r_{0}$, $\chi_{0}$ or 
the starting radius $R_{0}$, the apparent horizon forms at the time ($-2\chi_{0}$):
\begin{equation}
\eta_{AH}=-2\sinh^{-1}(2M\alpha/R_{0})^{1/2}=-\cosh^{-1}(4M\alpha/R_{0}+1),
\end{equation}
where we have used equation \eqref{formula_chi0} and the inverse 
hyperbolic identity $2\sinh^{-1}x=\cosh^{-1}(2x^{2}+1)$, for $x>0$.
This shows that the apparent horizon forms at exactly the same time when the matter cloud
reaches it's Schwarzschild radius, given by equation \eqref{req2mhyper}.

Similar to the calculations in the previous subsection, we also may formulate the problem of 
trapped surface by looking at the change of the $2$- sphere areas with proper time.
The outgoing null geodesics will be trapped if their proper area do not 
grow with time. For the metric (\ref{m2m-}), the area 
is also written (apart from some factors of $4\pi$) as $A=a^{2}(\eta)\sinh^{2}\chi$.

From equation (\ref{Rk-1}), $a(\eta)=m\sinh^{2}(\eta/2)$ and hence 
$da=(m/2)\sinh\eta\, d\eta$ and also $(d\chi/d\eta)=1$ thus,
\begin{eqnarray}
\frac{dA}{d\eta}&=& \frac{da}{d\eta}\sinh\chi+a\cosh\chi \frac{d\chi}{d\eta}
=\sinh\left(\eta/2+\chi\right)\le 0 
\end{eqnarray}
This implies that the time of formation of the AH/MTT
is obtained from the equality $\eta_{AH}=-2\chi$ whereas, the points for which 
the inequality is satisfied forms the trapped region. This exactly matches with the expression
of time of formation of apparent horizon derived before.

Let us now check the time of formation of the event horizon. Just as the matter shell starts to fall,
outward directed null rays also begin to proceed towards the asymptotic null infinity. The event horizon
is the last outward directed null ray that reaches the infinity. The outward directed null rays, in the 
$(\eta-\chi)$ coordinates, are given by the equation $(d\chi/d\eta)=1$. Let us use the boundary
conditions that just as the shell with labeled by $R_{0}$ reaches its Schwazschild radius ($R=2M$),
the null ray of the event horizon also reaches there. This is given in the form $(\chi=\chi_{0})$ 
at $\eta=\eta_{2M}$. This gives the equation
\begin{equation}
\chi_{EH}=\chi_{0}+(\eta-\eta_{2M}).
\end{equation}
Just as the cloud of initial radius begins to fall, the event horizon 
also starts to grow inside the cloud. The equation for the radius of the event horizon is
given by:
\begin{eqnarray}
R_{EH}&=&mr_{EH}\sinh^{2}(\eta/2)=m\sinh(\chi_{EH})\sinh^{2}(\eta/2)\nonumber\\
&=& m\sinh(\chi_{0}+\eta-\eta_{2M})\sinh^{2}(\eta/2).
\end{eqnarray}
Notice that at $\eta=\eta_{2M}=-2\chi_{0}$, $R_{EH}=m\sinh^{3}\chi_{0}=2M$, which implies 
that at $\eta=-2\chi_{0}$, the event horizon matches with the Schwarzschild radius of the shell.
Next, we also show that the matching is smooth and the rate of growth of the event horizon becomes zero
at that $\eta=\eta_{2M}=-2\chi_{0}$. This is obtained as follows:
\begin{equation}
\frac{dR_{eh}}{d\eta}=m \sinh(\eta/2)\,\sinh[(3\eta/2) +\chi_{H}-\eta_{2M}].
\end{equation}
So, it follows that at $\eta=\eta_{2M}=-2\chi_{0}$, $(dR_{EH}/d\eta)=0$. The behaviour 
of the graphs of the EH, MTT and the shells may also be studied here. The nature
is similar to the previous two cases, we skip them.

\subsection{Inhomogeneous collapse}
For the inhomogeneous collapse, the $\alpha=\alpha(t)$ in the metric and may again be absorbed through the redefinition of
the time coordinates, see equation (\ref{osd_ltb_eqns}). The mass function still remains a function of $r$ only and is taken
to be of the form $F(r,t)=r^{3}m(r)$ and the metric function is given by $k(r)=K(r)\, r^{2}$. 

\subsubsection{Marginally bound collapse}
The marginally bounded collapse corresponds to $K=0$. The metric is given by
\begin{equation}
ds^{2}=-dt^{2}+R^{\prime\,2}(r,t)\,dr^{2}+R^{2}(r,t)(d\theta^{2}+\sin^{2}\theta\, d\phi^{2}),
\end{equation}
where $R(r,t)$ is the radius of the shell. The equation of motion of the shell is given by
\begin{equation}\label{eom_ltb_1}
\dot{R}^{2}=F(r)/R.
\end{equation}
The solution of the equation of motion is given by the following form:
\begin{equation}
t=(2/3)\left[\sqrt{r^{3}/F} -\sqrt{R(r,t)^{3}/F}\right],
\end{equation}
where the radius of the shell at the beginning of the collapse at $t_{i}=0$ is
$R(r,t_{i})=r$. The shells will be labeled by the value of the radius it assumes
at the initial time $t=0$. For example for the shell being studied above, it shall
be labeled by the coordinate $r$. For this shell, labeled by the coordinate $r$,
the time taken for it to reach the singularity is 
\begin{equation}
t_{s}=(2/3)r^{3/2}/\sqrt{F}.
\end{equation}
Note that since $F(r)$ is inhomogeneous, 
all shells do not reach the singularity at the same time.
The equation of motion given above for the shell may naturally be rewritten as:
\begin{equation}\label{time_curvesol_ltb_1}
t=t_{s}-(2/3)\sqrt{R(r,t)^{3}/F}.
\end{equation}
Let us, for simplification, shift the time coordinate and use the choice $t_{s}=0$. With this simplification,
the time for the shell to reach $R=2M$ is given by $t_{2M}=(-4M/3)$.
The equation of the trapped surface is obtained by using the condition $R(r,t)=F(r)$ giving the 
equation for MTT/AH as:
\begin{equation}
R_{AH}(r,t)=-(3/2)\,t.
\end{equation}
This equation clearly implies that the AH/MTT begins at $R=2M$, shrinks at a 
constant rate of $\dot{R}_{AH}=-(3/2)$, and goes to zero at the 
same time when the singularity forms. The apparent horizon which is outside the shell, matches with
the Schwarzschild null event horizon. Note that the slope of the $R_{AH}(r,t) -t$ graph
is negative. However, this does not mean that the AH/MTT is timelike. In the examples 
that follow, we shall show that even if the MTT behaves as a timelike curve in the
$R(r,t) -t$ graph, it is the constant $C$ which fixes the signature of MTT \cite{Booth:2005ng, Andersson:2005gq}.
  
Let us now look at the formation of the event horizon, for which we need to look at the radial null
geodesic, given by the curve $R[r, t_{n}(r)]$. The event horizon shall be obtained by tracing 
the last radial null geodesic reaching the null infinity. The tangent vector field to this curve
is given by:
\begin{equation}
\frac{dR}{dt}=\dot{R}+R^{\prime}\,\left(\frac{dr}{dt}\right)_{null},
\end{equation}
where the radial null geodesic, as given in the second term on the right side of
the above equation, for $k=0$, is given by $(dr/dt)_{null}=(1/R^{\prime})$. This gives us
$(dR/dt)=(1+\dot{R})$. From the equation \eqref{eom_ltb_1}, this gives us:
\begin{equation}
\frac{dR}{dt}=1-\sqrt\frac{F}{R},
\end{equation}
Using, form equation \eqref{time_curvesol_ltb_1}, the fact that $F(r)=(2/3)R^{3}(-t)^{-2}$, the above 
equation is reduced to the form:
\begin{equation}
\frac{dR}{dt}=1+\frac{2R}{3t}.
\end{equation}
with the solution $R(r,t)=3t+C^{\prime}\,t^{2/3}$, where $C^{\prime}$ is a constant of integration. The constant 
$C^{\prime}$ may be set by the condition that, at the time when the shell reaches its 
Schwarzschild radius ($R=2M$), the null geodesic also reaches that point at exactly the same time. 
This gives the equation of curve of the event horizon:
\begin{equation}
R_{EH}=3t+3(9M/2)^{1/3}\, (-t)^{2/3}.
\end{equation}
Note that just as in the case for homogeneous collapse, the event horizon begins just
as the matter shells begin to fall, growing slowly to smoothly match with the Schwarzschild horizon at 
$R=2M$ at $t=-4M/3$. At that time, the rate of growth of $R_{EH}$ vanishes and for $t\ge -4M/3$,
the event horizon is the null Schwarzschild horizon of radius $2M$.

\subsubsection*{Examples:}
\begin{figure}[htb]
\begin{subfigure}{.55\textwidth}
\centering
\includegraphics[width=\linewidth]{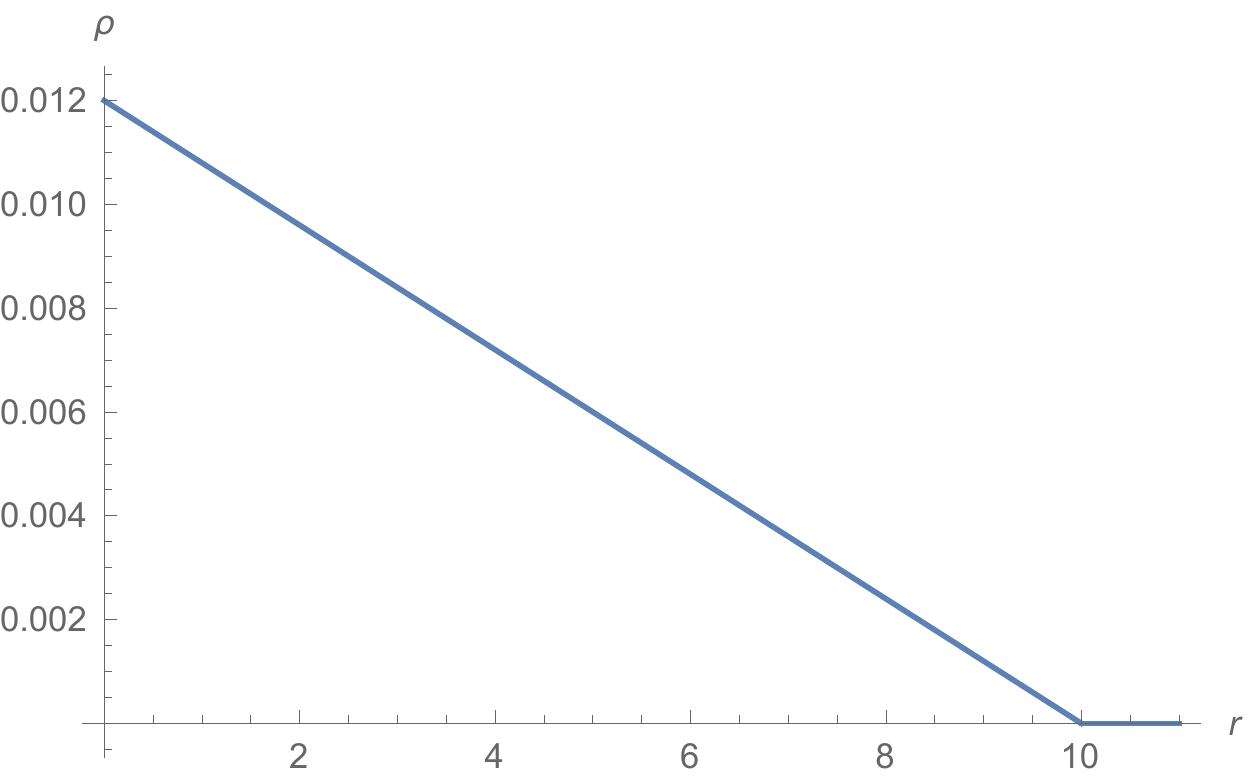}
\caption{}
\end{subfigure}
\begin{subfigure}{.45\textwidth}
\centering
\includegraphics[width=1.1\linewidth]{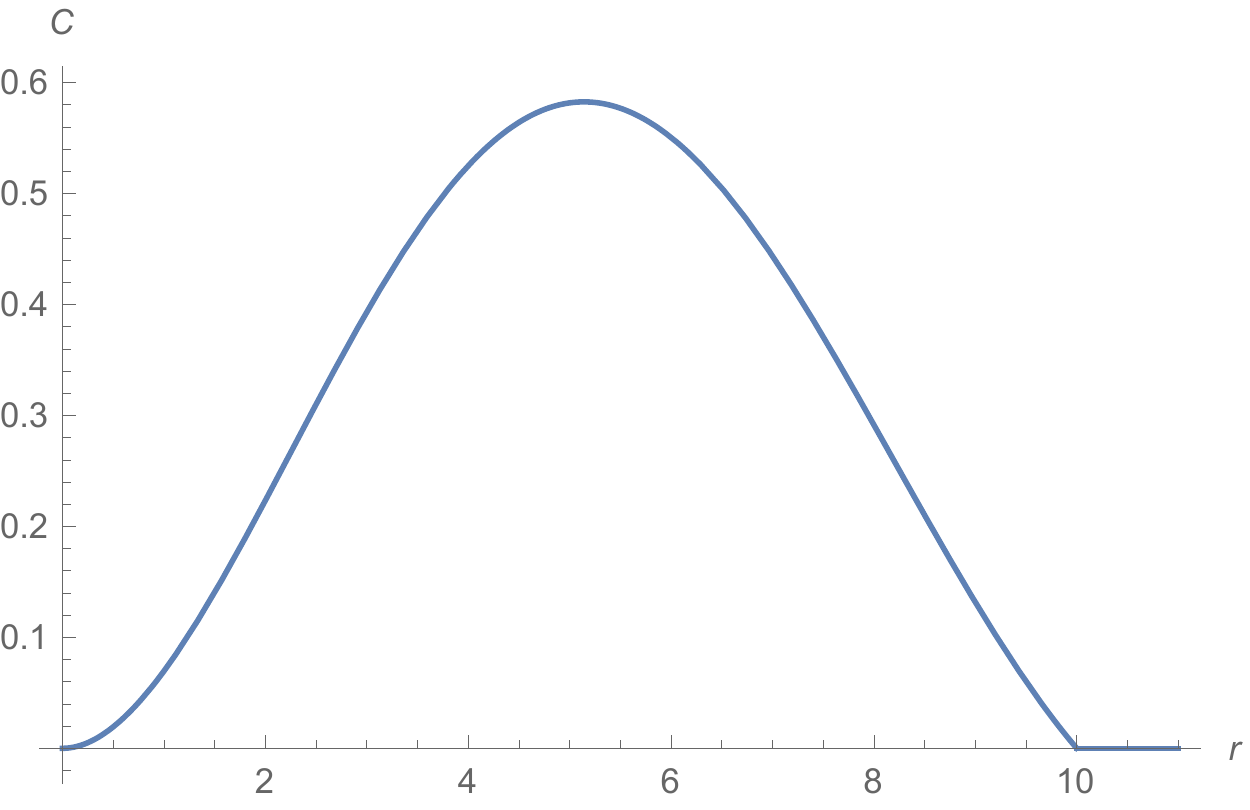}
\caption{}
\end{subfigure}
\begin{subfigure}{.55\textwidth}
\centering
\includegraphics[width=\linewidth]{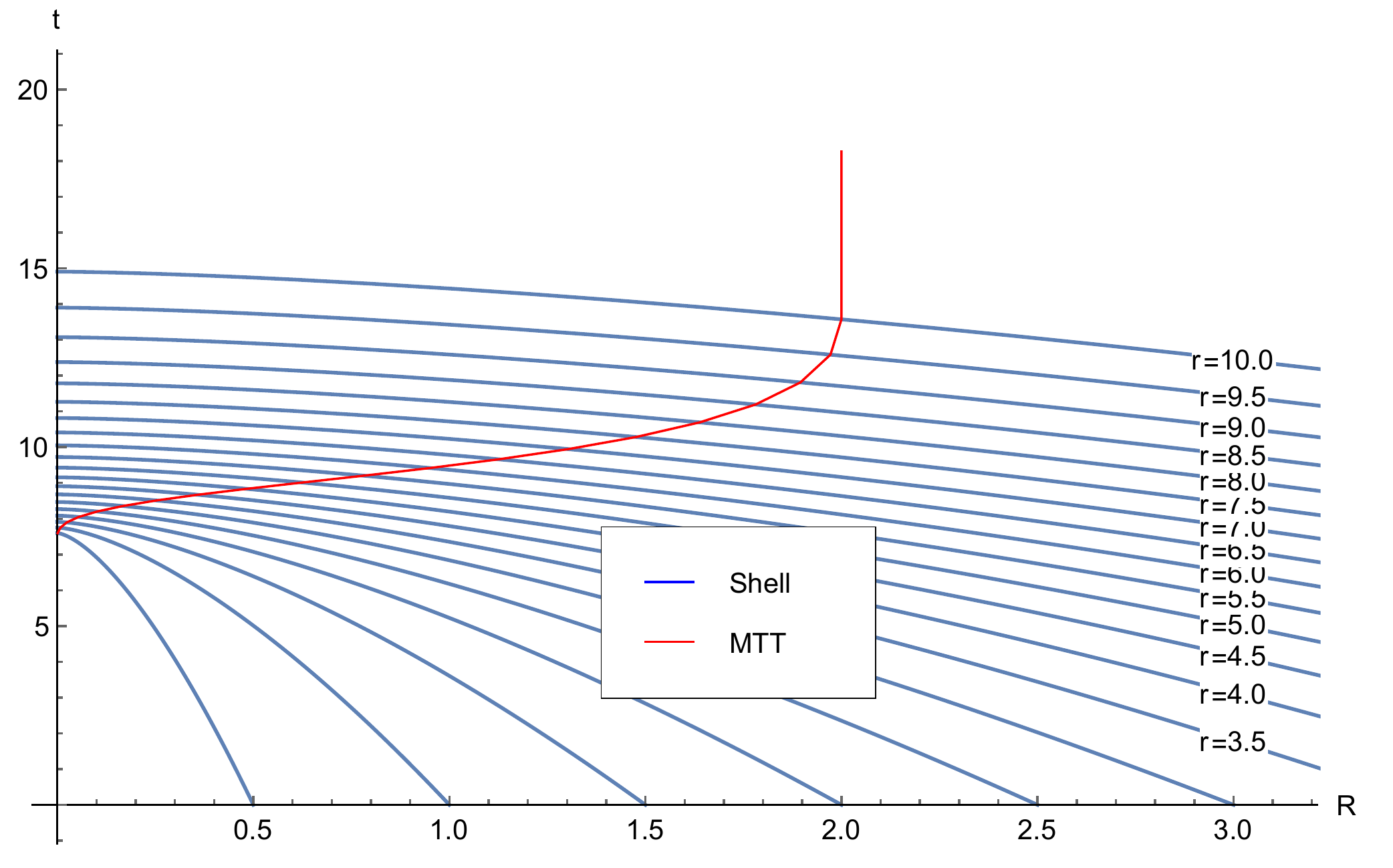}
\caption{}
\end{subfigure}
\caption{The graphs show the (a) density distribution $\rho_{1}$, (b) values of $C$,
and (c) formation of MTT along with the shells. The MTT begins from the center of the cloud.
The straight lines of MTT in (c) after the shell $r=10$ has fallen, represents the isolated horizon phase.}
\label{fig:ltbk0_example1}
\end{figure}

\begin{figure}[htb]
\begin{subfigure}{.55\textwidth}
\centering
\includegraphics[width=\linewidth]{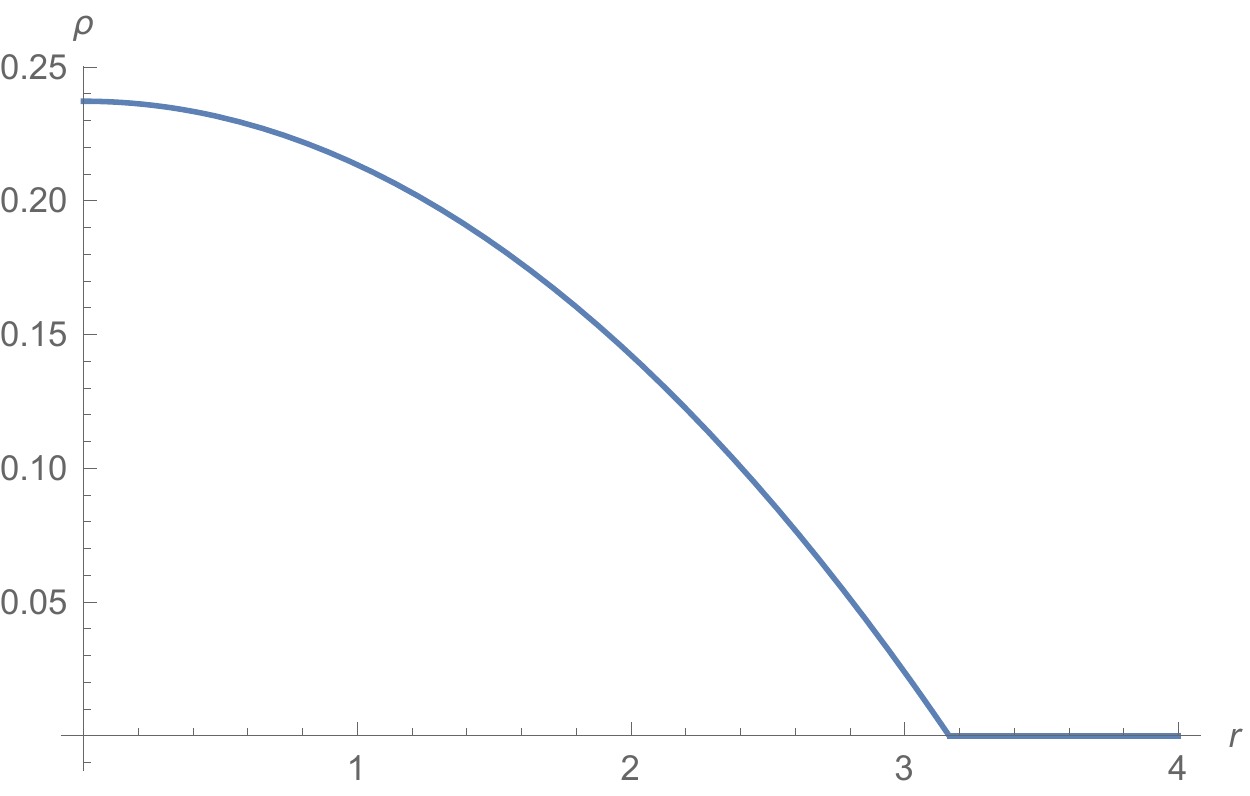}
\caption{}
\end{subfigure}
\begin{subfigure}{.45\textwidth}
\centering
\includegraphics[width=1.1\linewidth]{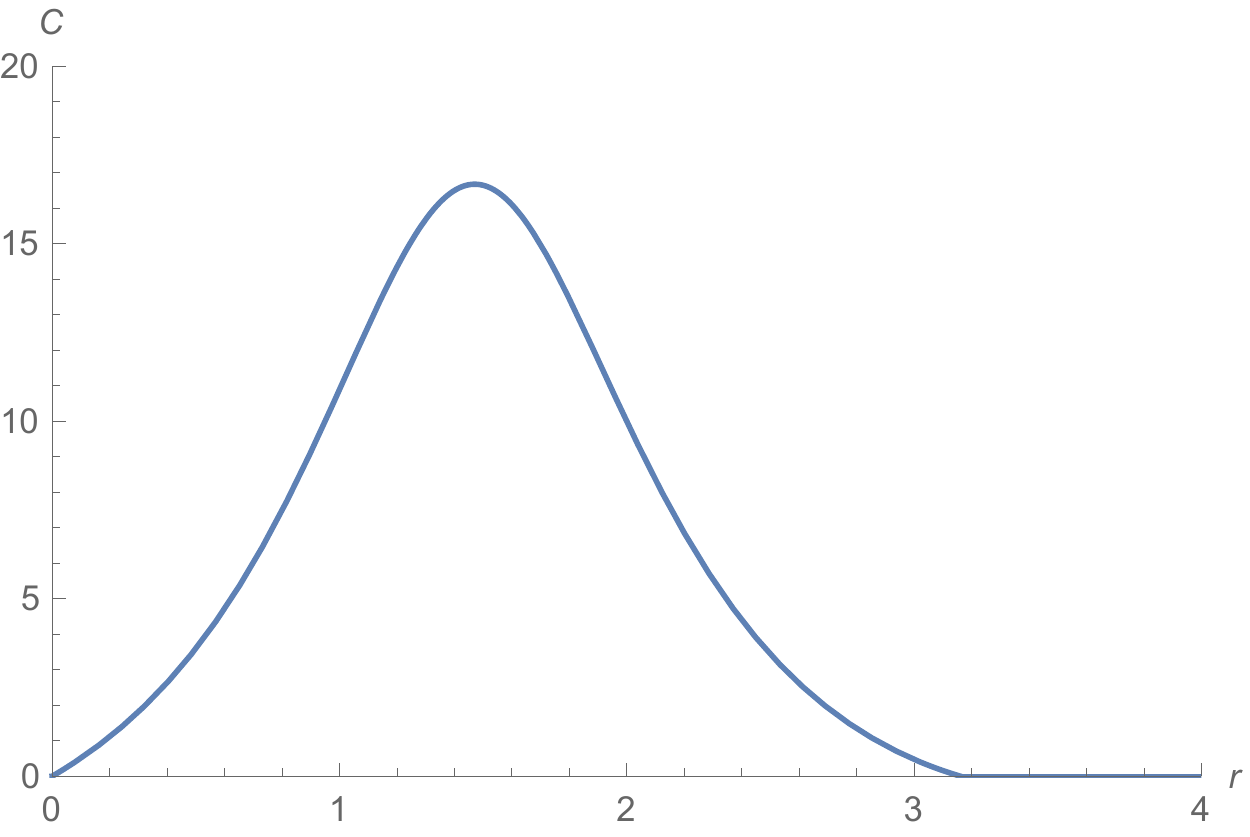}
\caption{}
\end{subfigure}
\begin{subfigure}{.55\textwidth}
\centering
\includegraphics[width=\linewidth]{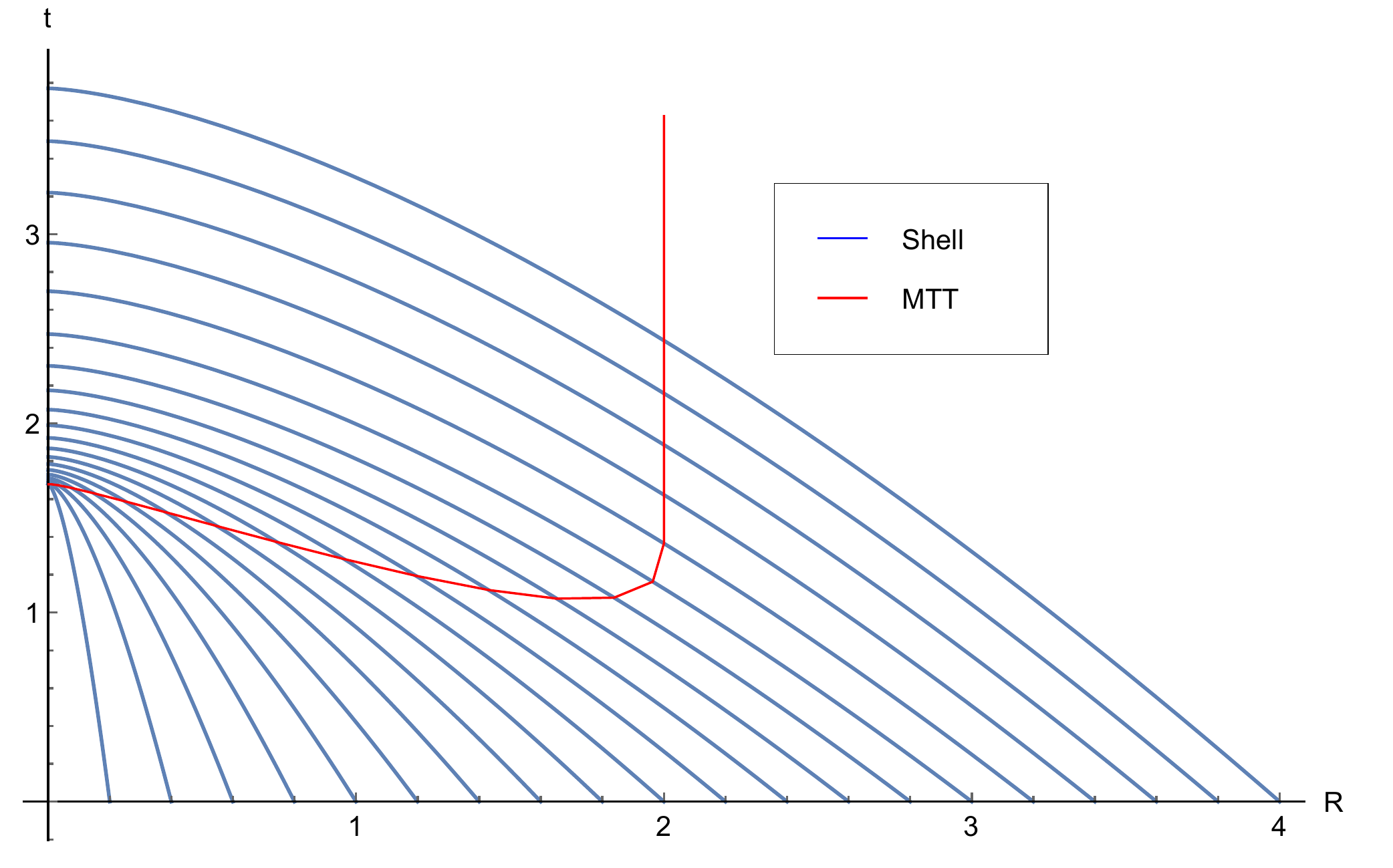}
\caption{}
\end{subfigure}
\caption{The graphs show the (a) density distribution $\rho_{2}$, (b) values of $C$,
and (c) formation of MTT along with the shells. Note that the MTT seems to begin at $r=2.6$
and then bifurcates in a timelike manner to the singularity while another part proceeds towards
the isolated horizon. However, since the signature of $C$ is always positive, the behavior of the MTT
must be like that in figure \ref{fig:ltbk0_example1}. The timelike nature arises due
to non-trivial intersection with the foliation \cite{Booth:2005ng, Andersson:2005gq}.}
\label{fig:ltbk0_example2}
\end{figure}

Here, we shall consider two examples, with the densities having the following forms:
\begin{eqnarray}
&&\rho_{1}(r)=(3M/2500)(10-r)\,\Theta(10-r),\nonumber\\
&&\rho_{2}(r)=(3M/40\sqrt{10})(10-r^{2})\,\Theta(10-r^{2}),
\end{eqnarray}
where $\Theta(x)$ denotes the Heaviside theta function. The factors have been chosen 
to get the isolated horizon at $R(r,t)=2$ and the corresponding 
masses have been normalised with the choice $M=1$. In each of these cases
the MTT are spacelike, as indicated by the values of $C$. Again, there
are no shell- crossing singularities and no trapped surfaces on the initial slice. 
The $R-t$ plots however 
are intricate in these two cases and are markedly different
(see figures \ref{fig:ltbk0_example1} and \ref{fig:ltbk0_example2}). We give each of 
these two cases since
they show the non- trivial ways in which the MTTs cross the foliation. 
For the density profile corresponding to $\rho_{1}$, the MTTs are spacelike. 
%
%
As seen from
the $R(r,t)-t$ graph, the MTT forms out of the central singularity, evolves in a spacelike manner
and approaches the isolated horizon phase at $R=2$. 
Although it may seem from these graphs that timelike membranes arise here,
that it is not so may be verified from the graphs of $C$. This bending of graphs only
indicates that the MTTs cross the foliation is intricate ways \cite{Booth:2005ng, Andersson:2005gq}.
%
%

\subsubsection{Bounded collapse}
For the case of bounded collapse, $k(r)>0$, the parametric solutions are given by:
\begin{eqnarray}
R(r,t)&=&\frac{F(r)}{k(r)}\cos^{2}\left(\eta/2\right)=r \cos^{2}\left(\eta/2\right) \label{Rk0LTBkg0}\\
t&=&\frac{F(r)}{2k(r)^{3/2}}\left( \eta+\sin \eta\right)=\frac{r^{3/2}}{\sqrt{F}}\left( \eta+\sin \eta\right), \label{tk=1LTBkg0}
\end{eqnarray}
where we assume the function $k(r)$ to be of the form
$k(r)=F(r,t)/r$, with $F(r,t)=m(r)r^3$. 

The collapse of the cloud begins at $\eta=0$, where $t=t_{i}=0$ and $R(r,t_{i})=r$ and reaches 
the singularity at $\eta=\pi$ where $R=0$. Note that the time for collapsing shells 
labeled by \emph{r} to reach the central singularity $R=0$ is
\begin{eqnarray}
t_s=\frac{\pi F(r)}{2\,k(r)^{3/2}}=\frac{\pi}{2m(r)^{1/2}} \label{tsLTBkg0}
\end{eqnarray}
It follows clearly from this equation (\ref{tsLTBkg0}), $t_s$ is not a constant (unlike the OSD collapse).
So, shells with different initial radius will reach the central singularity at
different times. From equation (\ref{Rk0LTBkg0}), we have $\eta=2\cos^{-1}(Rk/F)^{1/2}$
and hence, the proper time for the shell to reach the Schwarzschild radius $R=2M$ is
correspondingly given by $\eta_{2M}=2\cos^{-1}(2M/r)^{1/2}$.

Let us now locate the AH/MTT, which for spherical symmetry,
is denoted by the condition $R(t,r)=F(r,t)$. From equations (\ref{Rk0LTBkg0}) and 
(\ref{tk=1LTBkg0}), the equations for formation of trapped surfaces are given by:
\begin{eqnarray}
R_{ah}&=&r_{ah} \cos^{2}\left(\eta_{ah}/2\right) \label{Rk0ahLTBkg0}\\
t_{ah}&=&\frac{1}{2[m(r_{ah})]^{1/2}} \left(\eta_{ah}+\sin{\eta_{ah}} \right) \label{tk=1ahLTBkg0}.
\end{eqnarray}
In order to find $r_{ah}$, we use the trapping equation $R=F$. 
Taking derivative on both sides we have
\begin{eqnarray}
\frac{dr_{ah}}{dt}&=&\frac{\dot{R}}{F'-R'}.\label{rahtLTBkg0}
\end{eqnarray}
To obtain the derivatives appearing in the above equation, we use the equations 
\eqref{Rk0LTBkg0} and \eqref{tk=1LTBkg0} and get, after some straightforward simplifications,
a complicated looking expression, relating the change of the shell radius of the AH with respect to
proper time $\eta$:
%
%
\begin{eqnarray}
\frac{dr_{ah}}{d\eta}&=&-\frac{(\sin\eta)/2+(1-k/k)^{1/2}\cos^{2}(\eta/2)}{D} \label{chiahkg0LTB},
\end{eqnarray}
where the denominator $D$ is given by the following form:
\begin{eqnarray}
D&=&(k F'/F)-\left[(F'/F)-(k'/k)\right]\cos^{2}\eta/2 \nonumber \\
&& ~~~~~~~~~~~~~~~~~~~~ +[(1-k)/4k]^{1/2}\left(F'/F- 3k'/2k\right)\left[\eta+\sin(\eta)\right].
\end{eqnarray}
Using the solutions of (\ref{chiahkg0LTB}) into (\ref{Rk0ahLTBkg0}),(\ref{tk=1ahLTBkg0}) gives 
the equation of the trapped surfaces/apparent horizon.  Also from equation (\ref{tk=1ahLTBkg0}), at
this same time, the apparent horizon should be at $R=2M=F$.

These equations also give the evolution of event horizon. Just as in the previous sections, we 
determine the outgoing radial null geodesics which 
are given by
\begin{eqnarray}
\frac{d r_{eh}}{dt}&=&\frac{[1-k(r)]^{1/2}}{R'}\label{chiehkg0LTB}
\end{eqnarray}
Now using the equations (\ref{Rk0LTBkg0}) into the above equation (\ref{chiehkg0LTB}) we have
\begin{eqnarray}
\frac{d r_{eh}}{d\eta}&=&-\frac{(\sin\eta/2)+(1-k/k)^{1/2}\cos^{2}(\eta/2)}{\bar{D}}\label{chiehkg0LTBeh}
\end{eqnarray}
where the denominator $\bar{D}$ is given by the following form:
\begin{eqnarray}
\bar{D}&=&\left[(F'/F)-(k'/k)\right]\cos^{2}\eta/2 \nonumber \\
&& ~~~~~~~~~~~~~~~~~~~~ -[(1-k)/4k]^{1/2}\left(F'/F- 3k'/2k\right)\left[\eta+\sin(\eta)\right].
\end{eqnarray}
We have considered $(r_H,\eta_{2M})$ to be the point where outer event horizon forms. Thus 
the equation of the interior event horizon is given by
\begin{eqnarray}
R_{eh}&=&r_{eh}\cos\left[ \frac{\eta}{2}\right]^2 \label{Rehkg0LTB}
\end{eqnarray}
We have used numerical integration to solve 
the above equations (\ref{chiahkg0LTB}) and (\ref{chiehkg0LTBeh}).
We have all the required equations to 
study the LTB collapsing shells (\ref{Rk0LTBkg0}), apparent horizon (\ref{Rk0ahLTBkg0}) and event horizon 
(\ref{Rehkg0LTB}). Where we have consider that the matching of interior to the exterior is done at the 
hypersurface when $\chi_{_H}=\pi/3$, $r_H=0.866$ and $m(r)=(n_0+r n_1)$, such that the point where exterior 
even horizon formed is $\left(\eta_{2M},R_{2M}\right)=\left(2.1399,2.0\right)$. Thus all three curves 
collapsing shell, apparent horizon and event horizon should meet at $R=2.0$ when $\eta_{_{2M}}=2.1399$ for 
$n_0=1/11$ and $n_1=1/4$.

\subsubsection*{Examples:}
(i)Let us consider the case where the density
is given by
\begin{equation}
\rho(r)=(3M/5000)(100-r^{3})\,\Theta(100-r^{3}). 
\end{equation}
For this density profile too, the MTT is spacelike. 
%
\begin{figure}[htb]
\begin{subfigure}{.55\textwidth}
\centering
\includegraphics[width=\linewidth]{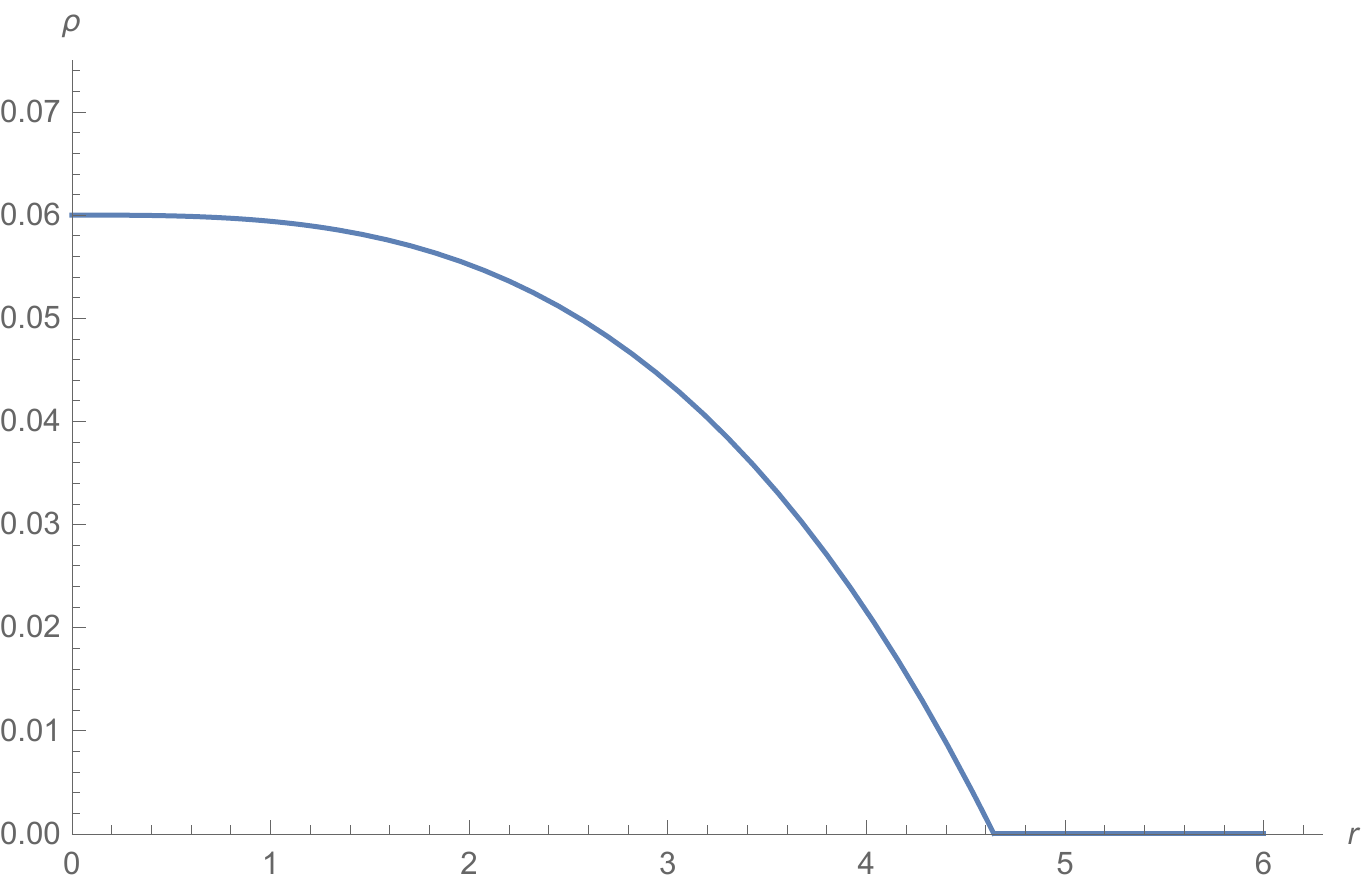}
\caption{}
\end{subfigure}
\begin{subfigure}{.45\textwidth}
\centering
\includegraphics[width=1.1\linewidth]{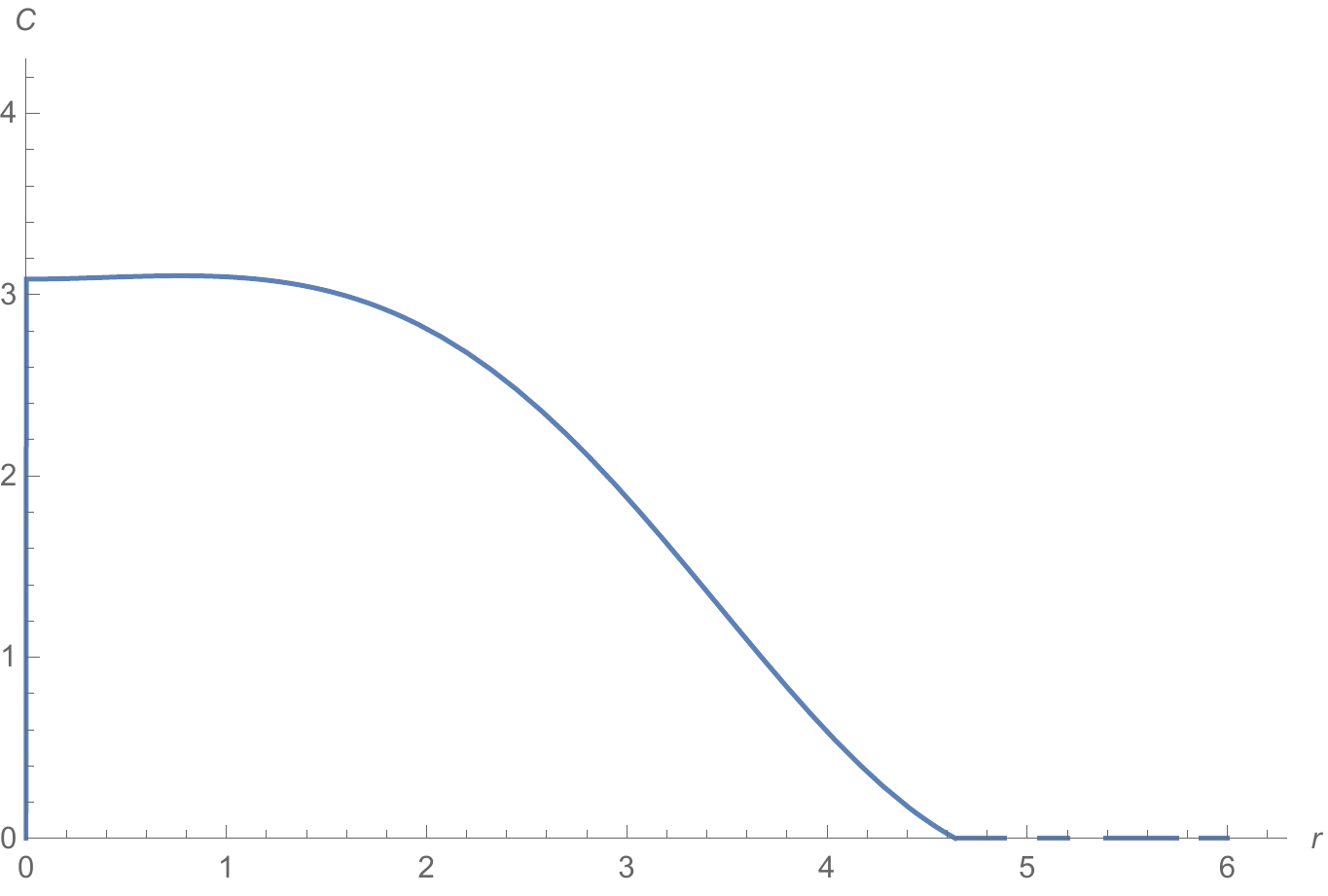}
\caption{}
\end{subfigure}
\begin{subfigure}{.55\textwidth}
\centering
\includegraphics[width=1.2\linewidth]{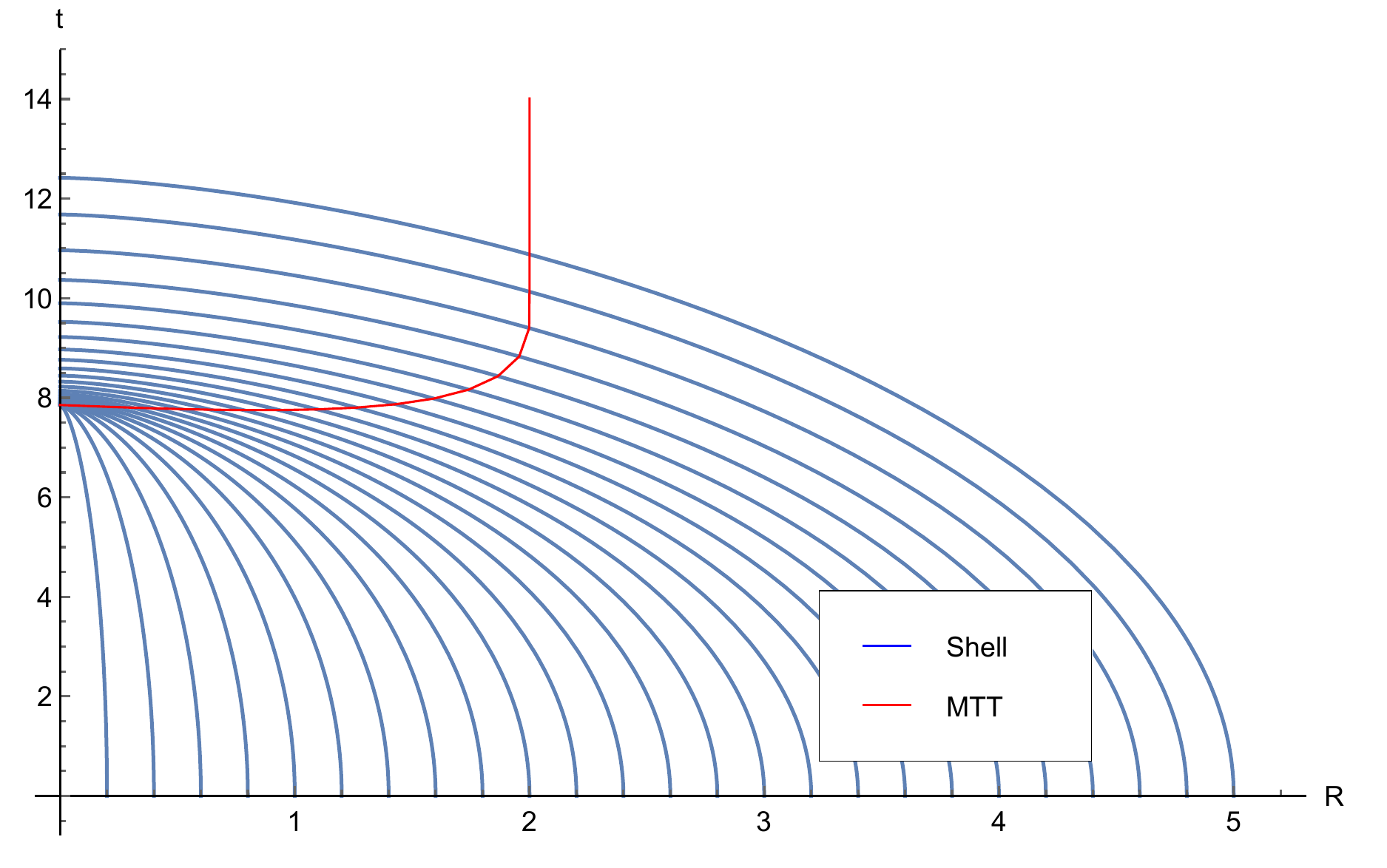}
\caption{}
\end{subfigure}
\caption{The graphs show the (a) density distribution, (b) values of $C$,
and (c) formation of MTT along with the shells. The MTT begins from the center of the cloud.
The straight lines of MTT in (c), after approximately $r=4.6$, represents the isolated horizon phase.}
\label{fig:ltbkg0_example1}
\end{figure}
%
This may be confirmed
though the $C- r$ graph (see the figures in \ref{fig:ltbkg0_example1}). 
The $R-t$ graph shows that it develops from the 
center of the cloud and evolves in a spacelike manner to reach the isolated horizon at 
$R=2$. Although the graph may look to have a timelike evolution in the $R-t$ graph,
it is due to the choice of foliation as explained above.

(ii)Let us consider a Gaussian profile with the density given by the following form \cite{Booth:2005ng}:
\begin{equation}
\rho(r)=\frac{m_{0}}{\pi^{3/2}r_{0}^{3}}\exp (-r^{2}/r_{0}^{2}),
\end{equation}
where $m_{0}$ is the total mass of the matter cloud, $r_{0}$ is a parameter which indicates 
the distance where the density of the cloud decreases to $[\rho\,(0)/e]$. 
%
\begin{figure}[htb]
\begin{subfigure}{.55\textwidth}
\centering
\includegraphics[width=\linewidth]{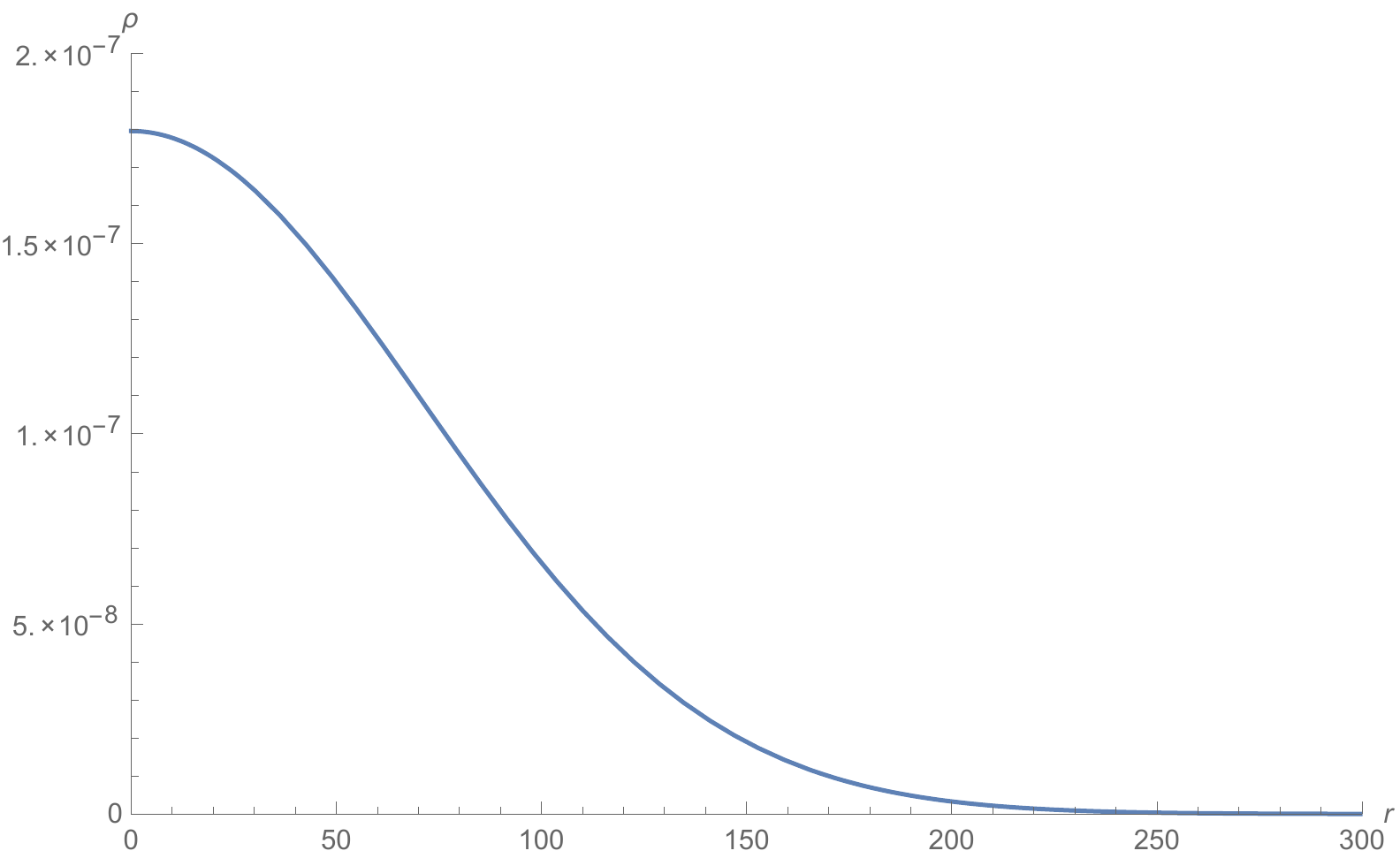}
\caption{}
\end{subfigure}
\begin{subfigure}{.45\textwidth}
\centering
\includegraphics[width=1.1\linewidth]{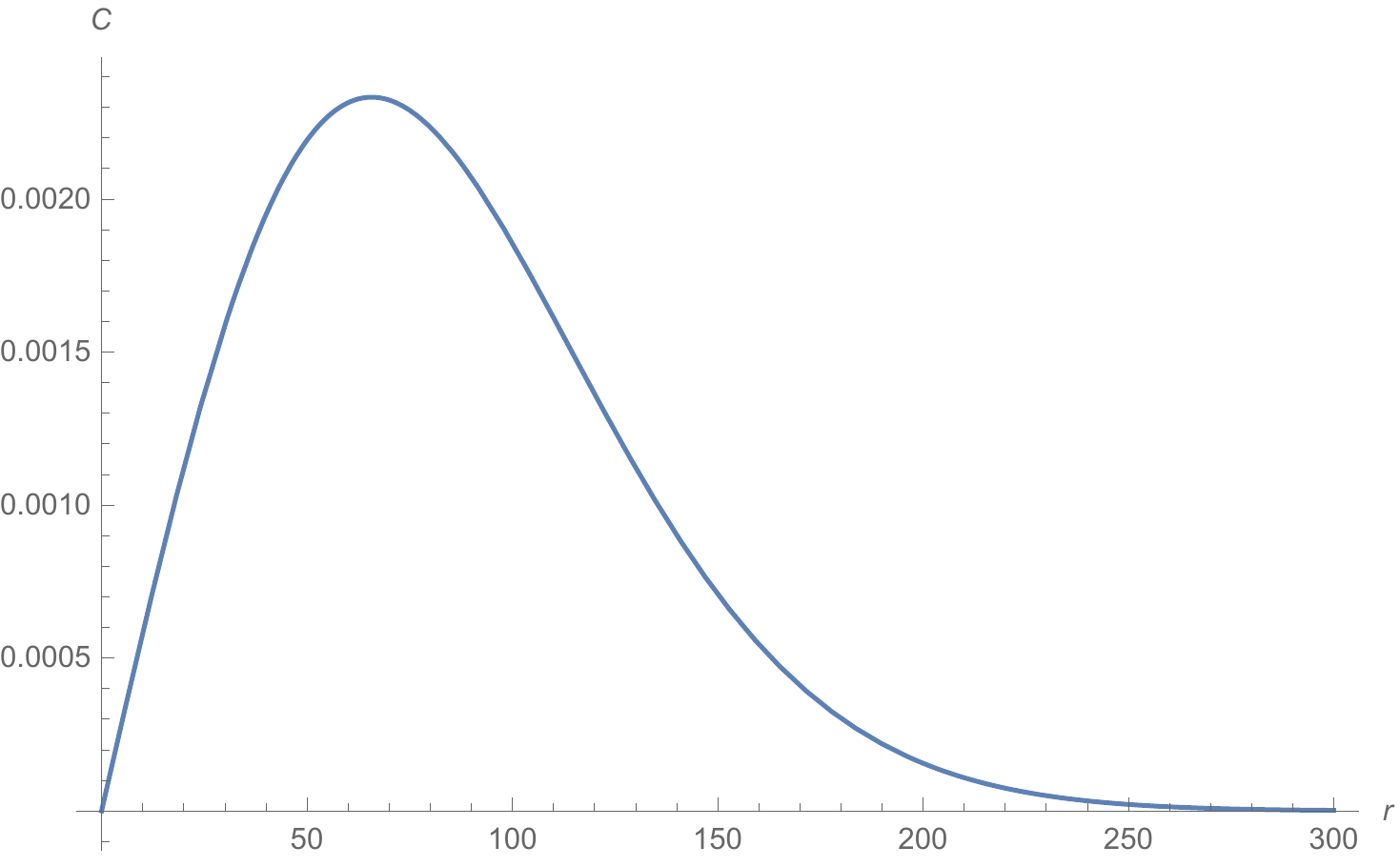}
\caption{}
\end{subfigure}
\begin{subfigure}{.55\textwidth}
\includegraphics[width=1.2\linewidth]{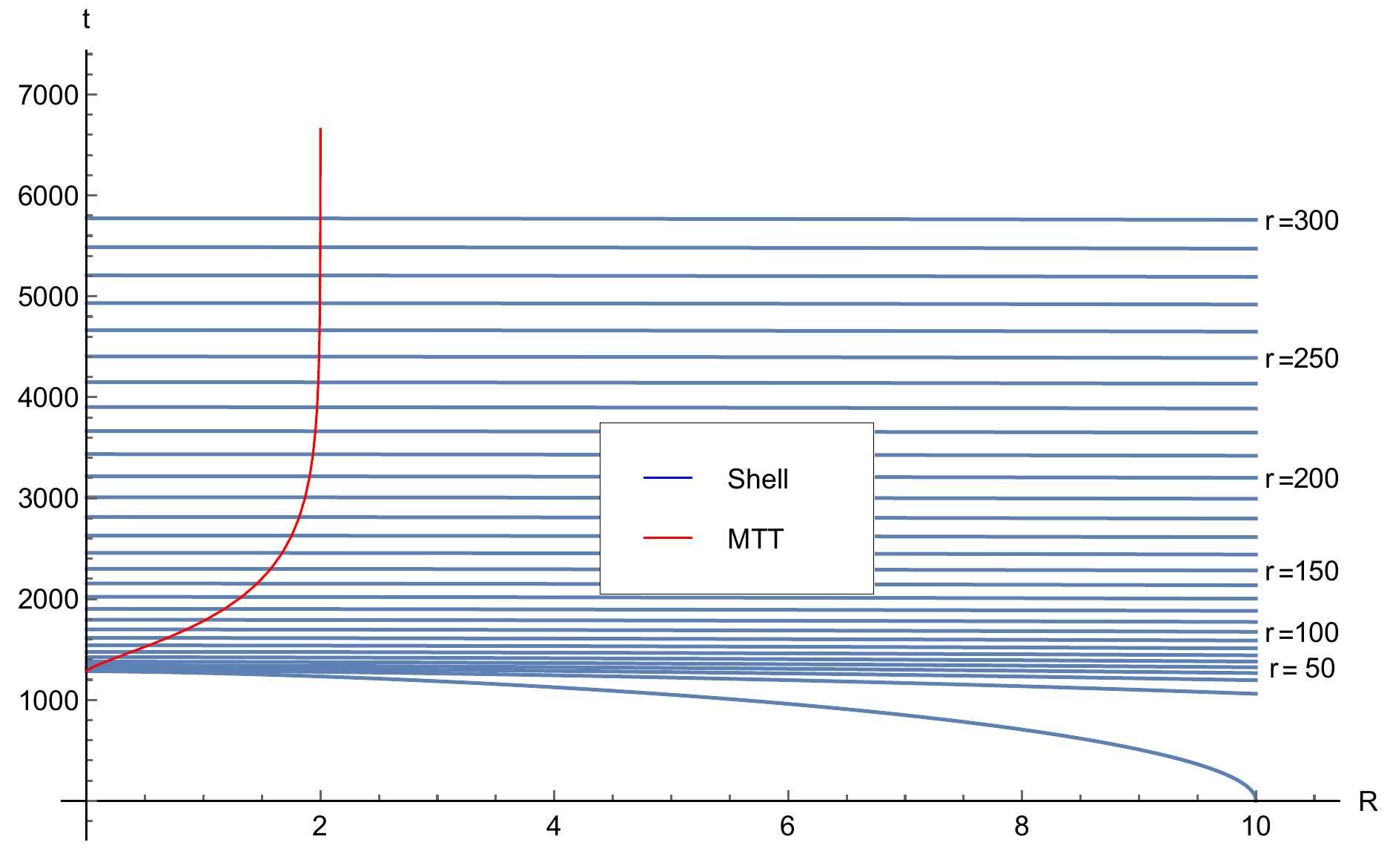}
\caption{}
\end{subfigure}
\caption{The graphs show the (a) density distribution, (b) values of $C$,
and (c) formation of MTT along with the shells. The MTT begins from the center of the cloud.
The straight lines of MTT in (c), after the shell $r=250$, represents the isolated horizon phase.}
\label{fig:ltbkg0_example2}
\end{figure}
%
In our example,
we have chosen $r_{0}=100\, m_{0}$. As usual, the MTT
begins from the central singularity,
and develops as a dynamical horizon until at approximately $r=200$, is begins to resemble
an isolated horizon (see figure \ref{fig:ltbkg0_example2}). 
This may also be confirmed from the fact that the density at $r=200$
is almost negligible. However, since the Gaussian profile almost disappears at $r=380$,
the $R=2$ is also reached at that value of the shell coordinates. 

(iii) Let us consider another density profile with the following form:
\begin{equation}
\rho(r)=\frac{m_{0}}{8\pi r_{0}^{3}}\exp(-r/r_{0}),
\end{equation}
where $m_{0}$ is the total mass of the matter cloud, $r_{0}$ is a parameter which indicates 
the distance where the density of the cloud decreases to $[\rho\,(0)/e]$.
%
\begin{figure}[h!]
\begin{subfigure}{.55\textwidth}
\centering
\includegraphics[width=\linewidth]{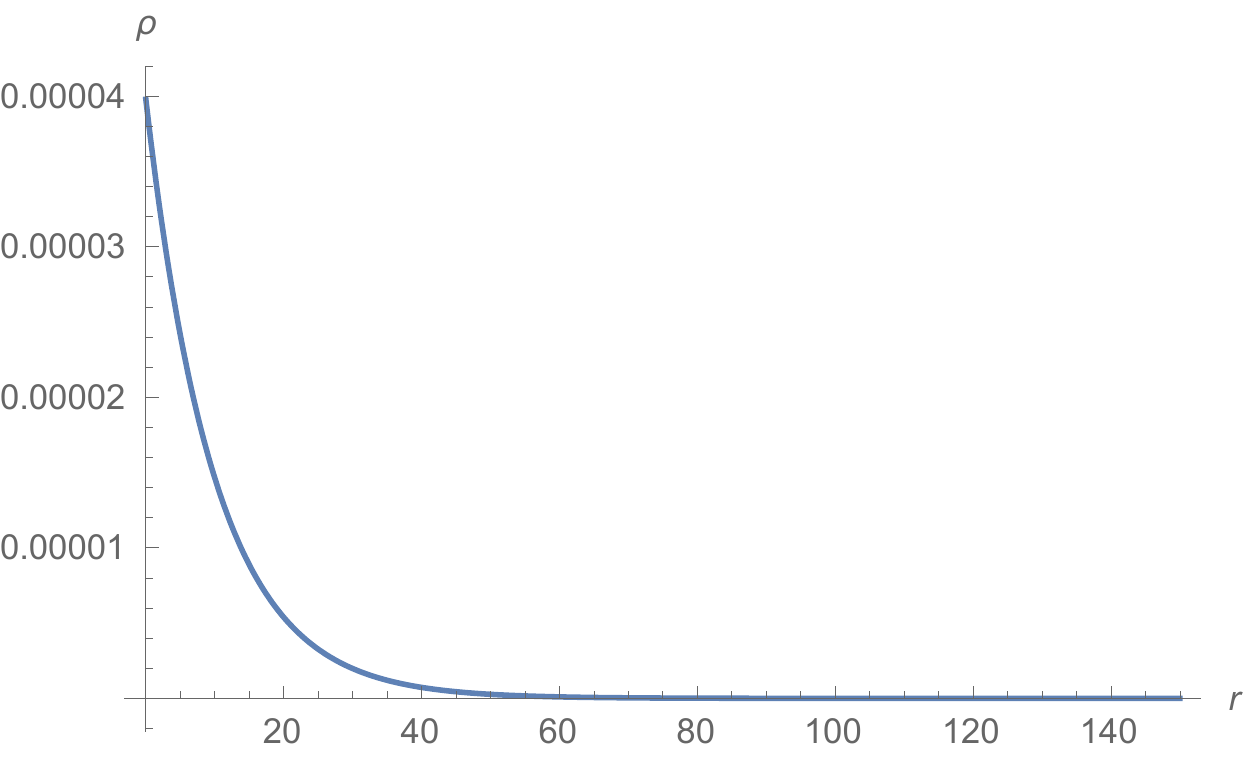}
\caption{}
\end{subfigure}
\begin{subfigure}{.45\textwidth}
\centering
\includegraphics[width=1.1\linewidth]{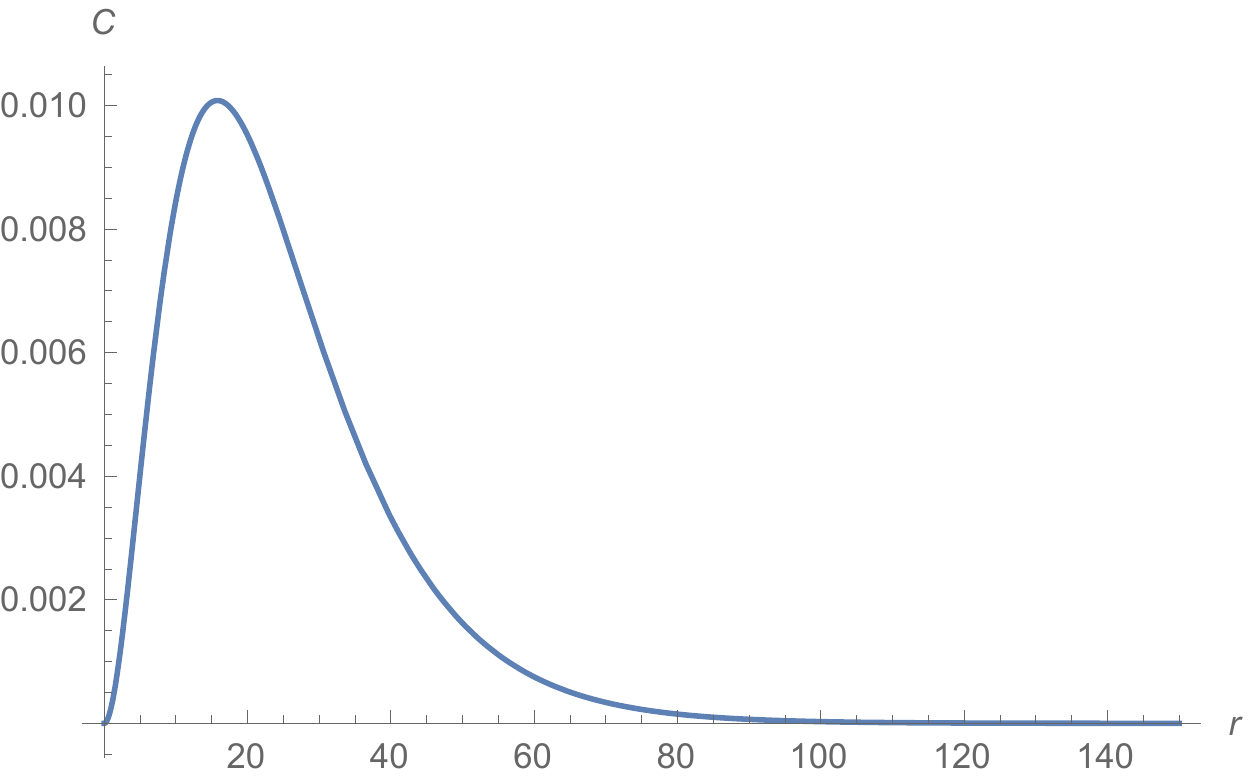}
\caption{}
\end{subfigure}
\begin{subfigure}{.55\textwidth}
\includegraphics[width=1.2\linewidth]{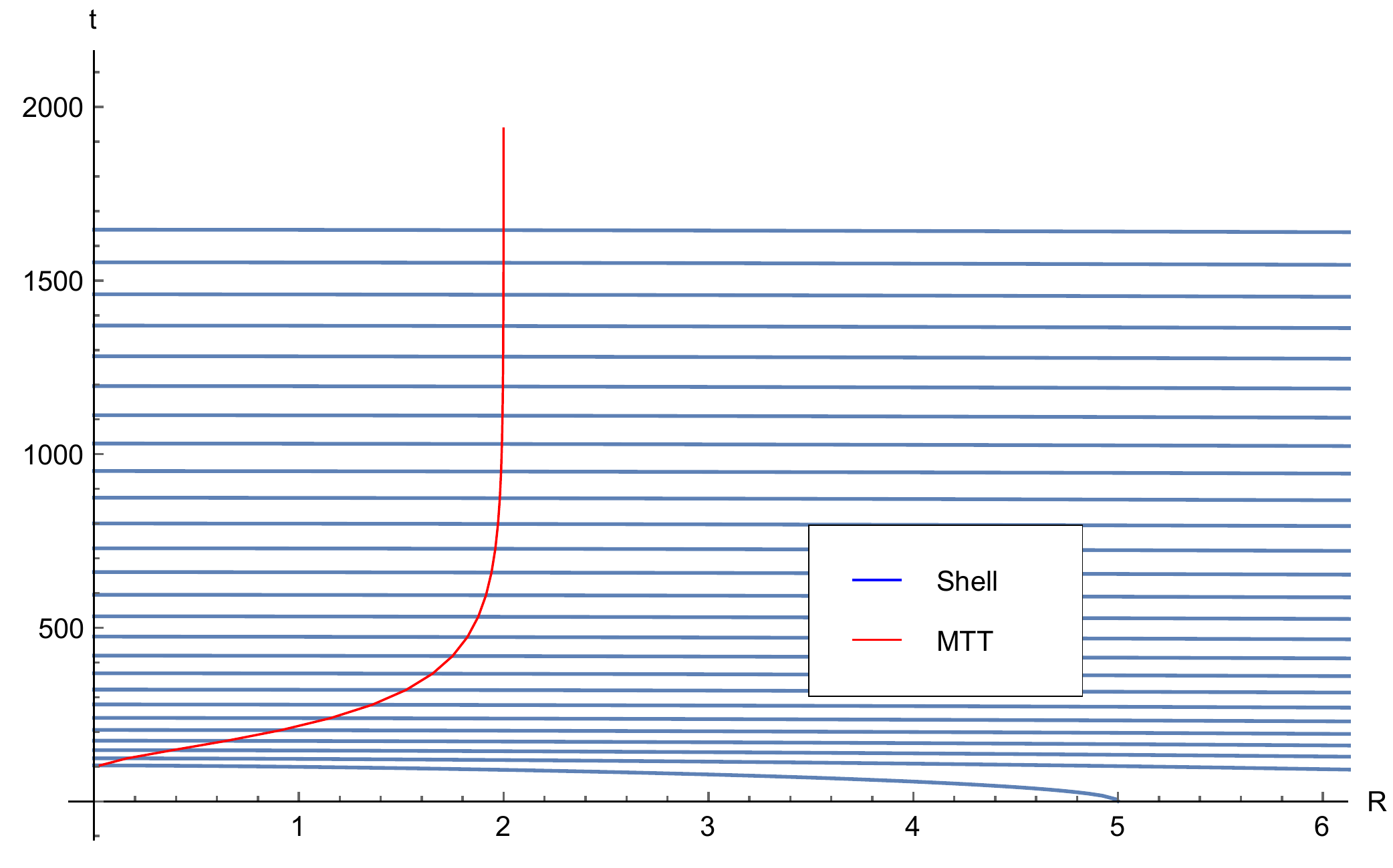}
\caption{}
\end{subfigure}
\caption{The graphs show the (a) density distribution, (b) values of $C$,
and (c) formation of MTT along with the shells. The MTT begins from the center of the cloud.
The straight lines of MTT in (c) represents the isolated horizon phase.}
\label{fig:ltbkg0_example3}
\end{figure}
%
The MTT begins from the central singularity,
and develops as a dynamical horizon until at approximately $r=70$, is begins to resemble
an isolated horizon. This may also be confirmed from the fact that the density at $r=70$
is almost negligible. However, since the Gaussian profile almost disappears at $r=100$,
the $R=2$ is also reached at that value of the shell coordinates. This may 
be seen from figure \ref{fig:ltbkg0_example3}.

(iv) Two shells falling consecutively on a black hole:
Let us assume that a black hole of mass $M$ exists, upon which a density profile of the following form
falls:
\begin{equation}
\rho(r)=\frac{8\,(m_{0}/\,r_{0}^{3}\,)\,[(r/r_{0})-\varsigma]^{\,2}}{[2\varsigma +(3+2\varsigma^{2})\sqrt{\pi}e^{\varsigma^{2}}\{1+\erf(\varsigma)\}]}
\, \exp[(2r/r_{0})\varsigma-(r/r_{0})^{\,2}\,],
\end{equation}
where $m_{0}=M/2$ is the mass of the shell, $2r_{0}$ is the width of each shell. 
%
\begin{figure}[h!]
\begin{subfigure}{.55\textwidth}
\centering
\includegraphics[width=\linewidth]{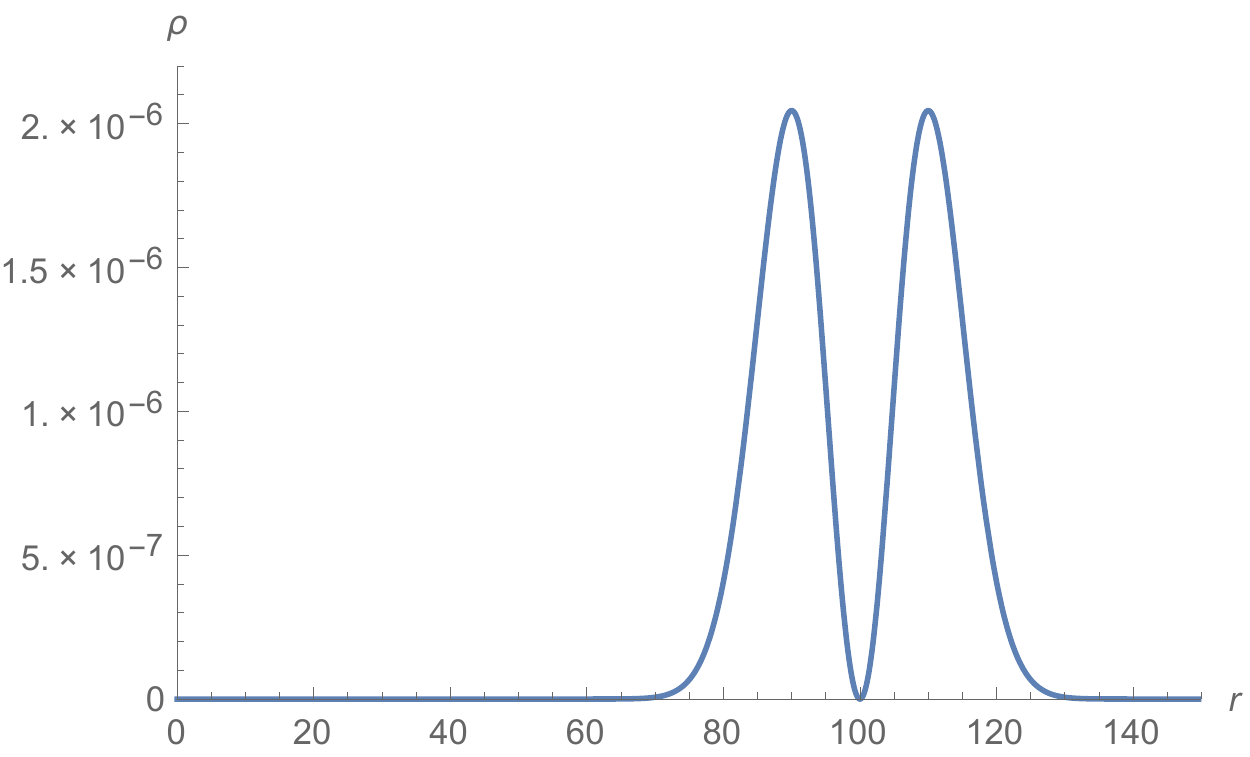}
\caption{}
\end{subfigure}
\begin{subfigure}{.45\textwidth}
\centering
\includegraphics[width=1.1\linewidth]{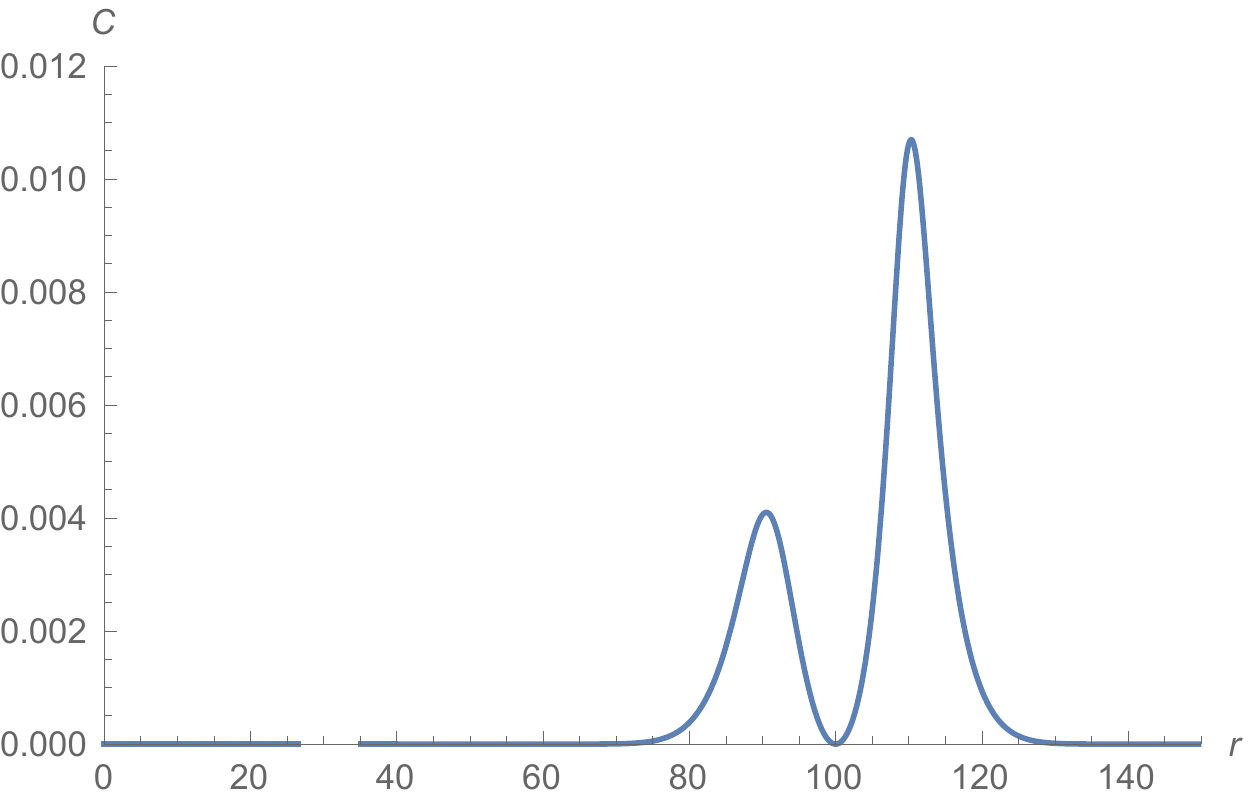}
\caption{}
\end{subfigure}
\begin{subfigure}{.55\textwidth}
\includegraphics[width=1.2\linewidth]{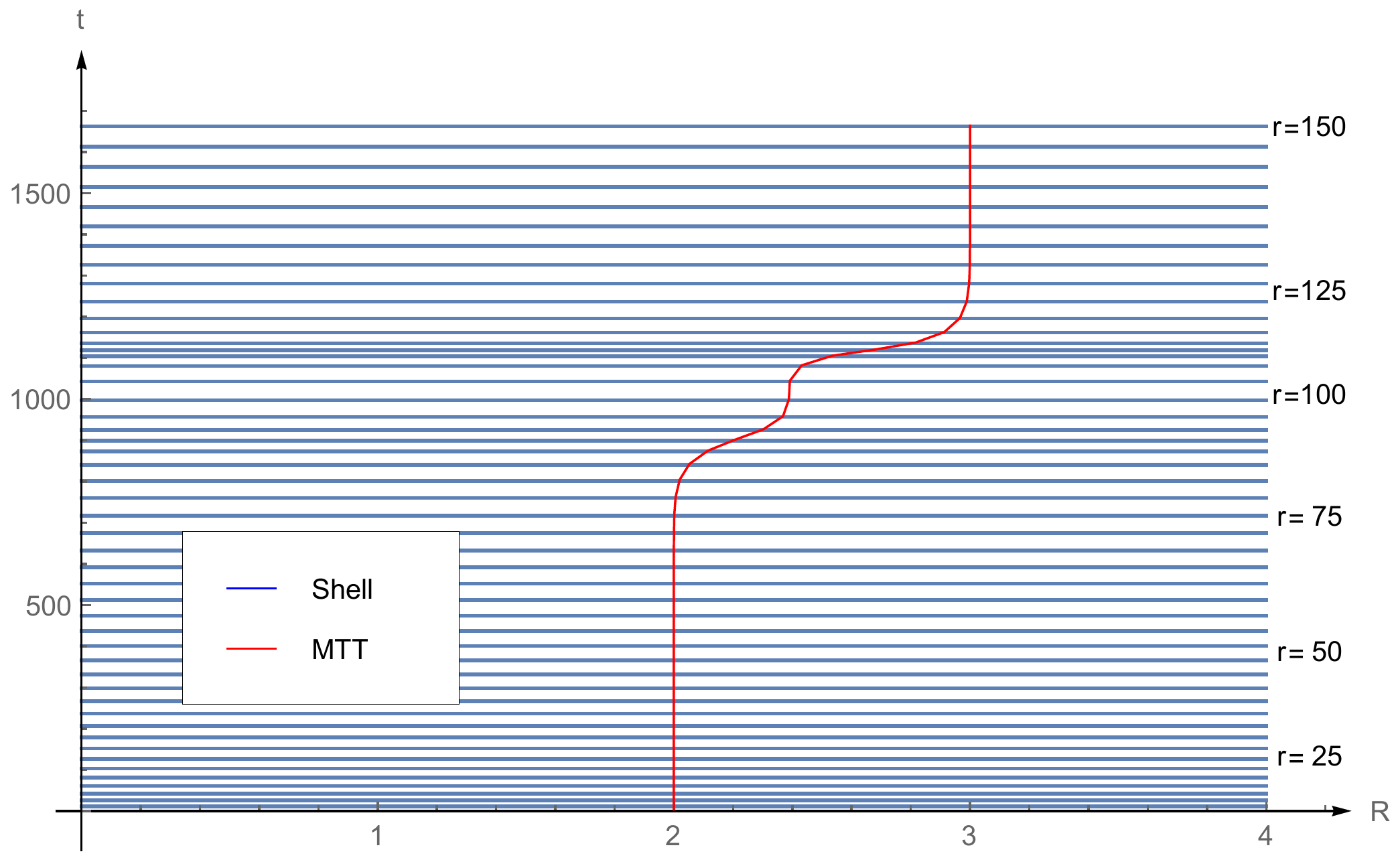}
\caption{}
\end{subfigure}
\caption{The graphs show the (a) density distribution, (b) values of $C$,
and (c) formation of MTT along with the shells which fall consecutively 
on a black hole. The MTT begins from $R=2$, where the previous EH is situated.
The straight lines of MTT in (c) represents the isolated horizon phase.}
\label{fig:ltbkg0_example4}
\end{figure}
%
If we assume that
the initial black hole has the Schwarzschild radius given by $\bar{r}=2M$, then the mass
for each shell of radius $r (r> \bar{r})$ is then $m(r)=M+\int_{\bar{r}}^{r}\rho(\hat{r})
\hat{r}^{2}\, d\hat{r}$. The quantity $\sigma$ is a parameter which denotes the position where
the density vanishes. Here, we have used $M=1$, $r_{0}=10$ and $\varsigma=10$. At around $r=65$,
the MTT starts to grow in a spacelike fashion and reaches approximately at $R=2.4$ at
approximately $t=1014$ when the $r=100$th shell falls. Note that at this time the $C$ vanishes
making the MTT null. This is expected since the density of the shell goes to zero here
(see figure \ref{fig:ltbkg0_example4}). Again, just as 
the next shell starts to fall, the MTT again begins to evolve in a spacelike fashion to reach
$R=3$ at $t=1500$ when the shell denoted by $r=140$ has fallen in.

\section{Spacetimes admitting viscous matter fields}\label{sec5}
In this section, we shall consider spacetimes due to collapse of matter 
whose energy- momentum tensor contains viscous matter fields. The Einstein equations
derived in equation \eqref{1eq1}-\eqref{1eq5} shall be useful in this regard. In the following,
we shall derive some general conditions about the nature of the spacetime from 
these equations. Additionally, it shall also arise that if one assumes some form of
\emph{equation of state} -type relations between some geometric scalar quantities and the density,
then the situation simplifies. We shall show below that such relations are indeed possible.

First, note that one may envisage some exact relations involving the 
the Newman- Penrose scalar $\psi_{2}$ and the Misner- Sharp mass function \cite {Glass, banerjee}.
The quantity $\psi_{2}$ for this spacetime is given by:
\begin{eqnarray}
\psi_{2} &=& \frac{e^{-2\beta}}{6}\left[\alpha''+{{\alpha}'}^{\,2}-\alpha'\beta'+R^{\prime 2}/R^2
-R^{\prime\prime}/R+(R'\beta')/R-(R'\alpha')/R \right] -\frac{1}{6R^2}\nonumber\\
&-&\frac{e^{-2\alpha}}{6}\left[\ddot{\beta}+{\dot{\beta}}^2-\dot{\alpha}\dot{\beta}+{\dot{R}}^2/R^2
-\ddot{R}/{R}-(\dot{R}\dot{\psi})/R+(\dot{R}\dot{\alpha})/R \right].\label{Psi1}
\end{eqnarray} 
Using the Einstein equations, above equation \eqref{Psi1} can be written in terms of the mass function 
$F(r,t)$:
\begin{equation}
F(r,t)= \left( \rho+\bar{p}_{t}-\bar{p}_r+2\eta\sigma\right)(R^3/3)-(\psi_{2}/2)\,R^3,\label{F}
\end{equation}
where $\bar{p}_{r}=(p_{r}-\zeta\theta)$ and $\bar{p}_{t}=(p_{t}-\zeta\theta)$. 
The quantity $\mathcal{F}(r,t)=-\psi_{2}\, R^3$, has a similar stature as the
mass function \cite{Glass}. The equations similar to those for the mass function $F(r,t)$ given in
\eqref{F_equations} shall play an important role here and are given by: 
\begin{eqnarray}
\dot{\mathcal{F}}&=&-(1/6)\left[R^3\left\{ \rho+\bar{p}_{t}+(2/3)\eta\sigma\right\}\right]_{,\,t}
-(R^{3}/6)\left[\bar{p}_r-(4/3)\eta\sigma \right]_{,\,t}\label{dotE}\\
\mathcal{F}^{\prime}&=&-(1/6)\,R^{3}\rho^{\prime}-(1/6)\left[R^3\left(\bar{p}_{t}-\bar{p}_r
+2\eta\sigma \right) \right]^{\prime}\label{E'}
\end{eqnarray}
These two equations may be combined to extract an expression for the time derivative of the 
density $\dot{\rho}$:
\begin{equation}\label{rt}
\dot{\rho}e^{-\alpha}+\left[\rho+\bar{p}_r-(4/3)\eta\sigma\right]\left(\Theta-\sigma \right)=0.
\end{equation} 
On the other hand, the expression for $\dot{\rho}$ may also be derived from the  
Bianchi identities, given in the equation \eqref{t-Bianchi} and \eqref{r-Bianchi}, and rewritten as:
\begin{eqnarray}
\dot{\rho}&=&-\dot{\beta}\left[\rho+\bar{p}_r-(4/3)\eta\sigma\right]-
(2\dot{R}/R)\left[\rho+\bar{p}_{t}+(2/3)\eta\sigma\right],\label{B1}\\
p_{r}^{\prime}&=&\left\{(4/3)\eta\sigma \right\}^{\prime}+
(2R^{\prime}/R)\left(\bar{p}_{t}-\bar{p}_r+2\eta\sigma\right)
-\alpha'\left\{\rho+\bar{p}_r-(4/3)\eta\sigma\right\}.\label{B2}
\end{eqnarray}
Using the Bianchi identity \eqref{B1} into the equation \eqref{rt}, we have a relation involving the
matter variables and the geometric variables given by:
\begin{equation}\label{eqn_rel_shear}
\left(\rho+\bar{p}_r\right)=\frac{8}{3}\eta\sigma-\frac{4}{3}\eta\theta+
e^{-\alpha}\,\frac{2\dot{R}}{R\sigma}\left(\bar{p}_{t}-\bar{p}_r\right).
\end{equation}
This equation gives some crucial input regarding the pressure anisotropy 
$\left(\bar{p}_{t}-\bar{p}_r\right)$ and its relation to the shear scalar $\sigma$.
The pressure anisotropy must be interpreted as the generator of the shear scalar and 
hence must be proportional to it.
Indeed, using the equation \eqref{eqn_rel_shear} in \eqref{B1}, we have:
\begin{equation}\label{rho_dot_2}
\dot{\rho} =\frac{6\dot{R}^2}{R^2}e^{-\alpha}\left[2\eta+\frac{\bar{p}_{t}-\bar{p}_r}{\sigma} \right],
\end{equation}
which makes our claim, that pressure anisotropy must lead to shear, explicit.
However, we shall show below that the claim still
holds even if we assume the matter density to be 
spatially uniform throughout the collapsing cloud. We must point out that
such an assumption is not contradictory to the presence
of shear or pressure anisotropy. We shall elaborate on this issue below
as well as in the following sections when we take specific examples.
To show this, we first derive another expression for the time change of density which
involves the radial pressure only. If the density 
is uniform, simple integration of equations \eqref{E'} implies that 
$\mathcal{F}(r,t)=-(R^{3}/6)\left(\bar{p}_{t}-\bar{p}_r
+2\eta\sigma \right)$. Using this in \eqref{dotE} we have
\begin{equation}
\dot{\rho}=-(3\dot{R}/R)\left[\rho+\bar{p}_r-(4/3)\eta\sigma\right].\label{rdot}
\end{equation}
Now, let us rewrite the Binachi identity \eqref{B1}, using the equations \eqref{theta_shear_def}
which gives us:
\begin{equation}\label{B1_rewrite}
\left(\rho+\bar{p}_r \right)\theta =-\dot{\rho}e^{-\alpha}-\frac{2\dot{R}}{R}\left(\bar{p}_{t}-\bar{p}_r\right)e^{-\alpha}+\frac{4}{3}\eta\sigma^{2},
\end{equation}
which may also be written in the following form, equivalent to \eqref{rho_dot_2}: 
\begin{equation}
\dot{\rho}e^{-\alpha}=\left[(4/3)\eta\sigma^2-\left(\rho+\bar{p}_r\right)\Theta\right]
\left[1-\frac{(2/3)\left(\bar{p}_{t}-\bar{p}_r\right)}{\rho+\bar{p}_{r}-
(4/3)\eta\sigma}\right]^{-1}. \label{b11}
\end{equation}

The radial derivative of the equation \eqref{B1_rewrite} along with \eqref{b11}, gives 
the following equation:
\begin{eqnarray}
\left(\rho+\bar{p}_r \right)\theta^{\, \prime}+\left(\rho+\bar{p}_r \right)^{\,\prime}\theta  
&=&\left(\dot{\rho}e^{-\alpha}\right)^{\prime}\left[1+\frac{(2/3)\left(\bar{p}_{t}-\bar{p}_r\right)}
{\rho+\bar{p}_r-(4/3)\eta\sigma}\right]+\left[(4/3)\eta\sigma^2\right]^{\,\prime}\nonumber\\
 && \,\,~~~~~~~~-\dot{\rho}e^{-\alpha}\left[\frac{(2/3)\left(\bar{p}_{t}-\bar{p}_r\right)}
{\left(\rho+\bar{p}_r-(4/3)\eta\sigma\right)}\right]^{\prime},\label{rpp}
\end{eqnarray}
which along with the radial part of Bianchi identity \eqref{B2} give the following elaborate form:
\begin{eqnarray}\label{rpp_1}
&\left(\rho+\bar{p}_r \right)\theta^{\,\prime}&=-\left[(4/3)\eta\sigma^{\prime}
-(4/3)\eta\sigma\,\alpha^{'}+(2R^{'}/R)\left(\bar{p}_{t}-\bar{p}_r +2\eta\sigma\right)\right]{\theta}
\nonumber\\
&&+\left[\frac{(2/3)}{\left(\rho+\bar{p}_r-(4/3)\eta\sigma\right)
-\frac{2}{3}\left(\bar{p}_{t}-\bar{p}_r\right)}\right]\left[\alpha^{'}\left(\rho+\bar{p}_r\right)
\{2\eta\sigma^{2}-\bar{p}_{t}\theta\left(\bar{p}_{t}-\bar{p}_r\right)\}\right.\nonumber\\
&&~~~~~~~\left. +\left\{( \rho+\bar{p}_{r})\Theta+(4/3)\eta\sigma^{2}\right\}\left\{
\left(\bar{p}_{t}-\bar{p}_r\right)^{'}-\frac{2\left(\bar{p}_{t}
-\bar{p}_r\right)^2}{\left(\rho+\bar{p}_r-4/3\cdot\eta\sigma\right)} \right.\right. \nonumber\\
&&~~~~~~~~~~~~~~
\left.\left.- \frac{4\eta\sigma\left(\bar{p}_{t}-\bar{p}_r\right)}{\left(\rho+\bar{p}_{r}
-4/3\cdot\eta\sigma\right)}\right\}-8/3 \cdot\alpha^{\prime}\eta^{2}\sigma^{3}\right]
+\left\{(4/3)\eta\sigma^{2}\right\}^{\prime}.
\end{eqnarray}
Several results follow directly from equation \eqref{rpp_1}.
Let the pressure anisotropy $(p_{t}- p_{r})$, and 
the viscosity parameters $\eta$ and $\zeta$ vanish. If we write
$p_{r}=p_{t}\equiv p$, then the above equation reduces to:
\begin{equation}
\left(\rho+p \right)\theta^{\,\prime}=0. \label{rpp_2}
\end{equation}
This implies that for the gravitational collapse of uniform density perfect fluid
with irrotational motion, the expansion scalar must be spatially uniform. Furthermore,
the spacetime must be isotropic as well as conformally flat \cite{banerjee, Raychaudhuri, misra}.
These results hold true even if the spacetime has bulk viscosity but negligible
shear viscosity \cite{banerjee}. However, if the fluid is dissipative, with non- vanishing 
shear viscosity, these results donot hold and expansion becomes a scalar function.
The situation however alters significantly if the pressure anisotropy arising through 
$(p_{t}-p_{r})$ is also taken into account. As may be seen from the
equation \eqref{rpp_1}, these anisotropies are in the same footing
as the shear terms. Indeed, then one may envisage situations
where the quantities arising from the dissipative forces like the shear and bulk viscosity 
cancel those due to anisotropy, leading to spatially uniform expansion scalar, just like
for perfect fluids. Although that situation would  be highly fine tuned, 
it is not unlikely. To summarise, we have shown that if 
the fluid has shear and bulk viscosity, as well as pressure anisotropy, then \emph{generically}
the spacetime will not admit isotropy, conformal flatness or spatially uniform expansion scalar.

This brings into question the possibility if the local anisotropy
of fluids may be identified as the source of viscous effects. Given the form of these quantities
in \eqref{eqn_rel_shear}, \eqref{rho_dot_2} and in \eqref{rpp_1}, this expectation holds ground. 
In the following, we shall assume that relation of this kind do exist, and to give
form to this expectation, we assume simple linear relation among these quantities, like 
$(p_{t}-p_{r})\propto \sigma $. To put it on a firmer perspective,
they are to be related to the density function through the following constraints:
$p_{t}=k_{t}\rho $, $\sigma=k_{\sigma}\rho $, and $\theta=k_{\theta}\rho$.
The values of the constants $k_{t}, k_{\sigma}$ and $k_{\theta}$ 
shall be chosen in such a way that the spacetime preserves the spherical symmetry and that 
any deviation due to shear will be negligible.

\subsection{Time independent mass function}

To study a realistic collapse phenomena,
pressure and viscosity contributions to energy momentum tensor of
the collapsing cloud must be included. To begin with, let us assume that
the collapsing cloud has a certain fixed radial pressure, given by
$p_{r}=(4/3)\eta \sigma+\zeta \theta$. This particular combination
is chosen so that the viscosity terms in the equation of motion cancel
the effects of radial pressure during the collapse.
This choice also keeps continuity with the study of 
pressureless collapse carried out in the previous sections.
However, to retain the physical importance of our model,
we continue to retain the combination $[p_{t}+(2/3)\eta\sigma-\zeta\theta]$
to be non- zero. This particular term includes tangential part of 
the pressure along with certain viscosity terms.
The reasons for these choices is only 
mathematical simplicity. Also, as we shall see, 
this choice gives us a time independent Misner-Sharp mass function. 

The set of the Einstein equations for gravitational collapse of matter cloud
which satisfies these conditions  are given by:
\begin{eqnarray}
 F^{\prime}&=&\rho\, R^{\prime}\,R^{2}\\
 \dot{F}&=&-(R^{2} \dot{R})\,({p}_{r}+\frac{4}{3}\eta \sigma-\zeta \theta )=0,
 \label{2eqt}\\
 \alpha^{\prime}&=&(2R^{\prime}/\rho R)\left[p_{t} +(2/3)\eta \sigma-\zeta \theta\right],
 \label{Etpv}  \\ 
(\dot{G}/{G})&=&2\alpha^{\prime} (\dot{R}/R^{\prime}),  \label{Et1pv} \\
 F(r,t)&=&R(r,t)(1-G+H), \label{EMppv}
 \end{eqnarray}
where $H(r,t)=e^{-2\alpha(r, t)}\dot{R}^2$ and  $G(r,t)=e^{-2\beta(r, t)}R'^2$. 
The number of unknowns to be determined here are more than the independent 
Einstein equations \eqref{2eqt}-\eqref{EMppv}, we close 
the system with the constraints $p_{t}=k_{t}\rho $, $\sigma=k_{\sigma}\rho $, 
and $\theta=k_{\theta}\rho$ given above.

Using these equations of state, solutions of the Einstein equations 
\eqref{Etpv} and \eqref{Et1pv} become
\begin{equation}
\exp(\,2\alpha)=R^{\,4a_{1}}, \hspace{1cm} \exp(\,2\beta)=\frac{R'^2}{b(r)R^{\,4a_{1}}},
\end{equation}
where we have introduced the constants $a_1=k_{t}+(2/3)\eta k_{\sigma}-\zeta k_{\theta}$. Using 
these redefinitions, the line element \eqref{1eq1} may be rewritten as: 
\begin{equation}
ds^{2}=-R^{\,4a_1}dt^2 + \frac{R'^2}{b(r)R^{\,4a_1}}dr^2 + R(r,t)^2 d \theta^2+ R(r,t)^2 \sin^2{\theta}\,d\phi^2 \label{metppv}
\end{equation}
The equation of motion \eqref{EMppv} is also simplified to have the following form:
\begin{equation}
\dot{R}=-R^{2a}\left[\frac{F(r)}{R}-1+b(r)R^{\,4a_1}\right]^{1/2}.
\label{empppv}
\end{equation}
To study evolution of the horizon and the outgoing null geodesics (and the event horizon), and
to simplify the solutions of the equation of motion \eqref{EMppv}, we choose 
the parameter to be $a_{1}=-(1/4)$. This choice simplifies the solution of the
equation of motion \eqref{EMppv}, and the time curve of the collapsing shell is given by
\begin{eqnarray}
dt&=&-\frac{R\, dR}{[F(r)+b(r)-R(r,t)]^{1/2}}.\label{tViscollapse}
\end{eqnarray}
To solve the integral, we choose a parametric form to 
relate the functions $R(r,t)$, $F(r)$ and $b(r)$. A
particular simple choice is given as
\begin{equation}
R=(F/b)\cos^2\left(\eta/2\right). \label{Rpara}
\end{equation}
Using this form, the equation of collapse simplifies and the time curve is obtained from
the equation: 
\begin{equation}
dt=\left(\frac{F^2}{2b^2}\right)\frac{\sin\eta\,\cos^2(\eta/2)}
{[F+b-(F/b)\cos^2 (\eta/2)]^{1/2}}\,\,  d\eta.
\end{equation}
The solution of this equation is the time curve of the collapsing shell and is given by
\begin{eqnarray} 
t&=&\frac{4}{3}[F+b-(F/b)\cos^2 (\eta/2)]^{1/2}\left[F+b+(F/2b)\cos^2 (\eta/2) \right]\nonumber \\
&&~~~~~~~~~~~~~~~~~~~~~~-(4/3)\{F+b-(F/b)\}^{1/2}\left[F+b+ (F/2b)\right].
\end{eqnarray}
%
The boundary conditions are chosen such that
the collapse begins at $\eta=0$ and reaches the central singularity 
at $\eta=\pi$. At the beginning of the collapse, $\eta=0$ we have 
$R(t_{i},r)=[F(r)/b(r)]$ with $t_{i}=0$. At the end state of the collapse process
when $\eta=\pi$, we naturally have $R=0$. Note that the time of formation 
of central singularity, or the time the shell reaches singularity, is also obtained from 
the above equation:
\begin{equation}
t_{s}=\frac{4}{3}\left[\left(F+b\right)^{\frac{3}{2}}-
\{F+b-(F/b)\}^{1/2}\left\{F+b+(F/2b)\right\} \right].
\end{equation}
From these equations, it is also possible to track the formation of apparent horizon and determine the
exact time the shell reaches it's Schwarzschild radius. However, for the present purposes, we shall
utilize the numerical techniques from the previous sections to track the formation of the MTTs,
determine its signature and eventually note how the cloud settles down to the 
null isolated horizon.

The dynamics of the marginally trapped surfaces (whether they are timelike, spacelike or null) 
depends upon the sign of the expansion parameter $C$ defined in equation \eqref{value_of_c}. 
We take the timelike vector field to be $u^\mu=\chi l^\mu+(2\chi)^{-1} n^\nu$ and
the spacelike vector field to be $x^\mu=\chi l^\mu-(2\chi)^{-1} n^\nu$.  
\begin{eqnarray}
T_{\mu\nu}l^{\mu}l^{\nu}&=&(1/4\chi)\left[\rho+p_{t}-(4/3)\eta\sigma-\zeta\theta
+\left( p_{t}-p_{r}\right) \right], \label{Tll}\\
T_{\mu\nu}l^{\mu}n^{\nu}&=&(1/2)\left[\rho-\left(p_{t}-(4/3)\eta\sigma-\zeta\theta+\left( p_{t}-p_{r}\right)
\right) \right].\label{Tln}
\end{eqnarray}
Using $p_{r}=(4/3)\eta\sigma+\zeta\theta$, the equations lead to the following form of $C$:
\begin{eqnarray}
C&=&(1/2\chi)\left[\frac{\rho+2\left\{p_{t}-(4/3)\eta\sigma-\zeta\theta \right\}}{4\pi/\mathcal{A}-
(1/2)\left[\rho-2\left\{p_{t}-(4/3)\eta\sigma-\zeta\theta \right\} \right]} \right].\label{ptC}
\end{eqnarray}

\subsubsection*{Examples}
(i) Let us consider a Gaussian profile. The density profile is same as before and
is given by:
\begin{equation}
\rho(r)=\frac{m_{0}}{\pi^{3/2}r_{0}^{3}}\exp (-r^{2}/r_{0}^{2}),
\end{equation}
where $m_{0}$ is the total mass of the matter cloud, $r_{0}$ is a parameter which indicates 
the distance where the density of the cloud decreases to $[\rho\,(0)/e]$. 
%
\begin{figure}[h!]
\begin{subfigure}{.5\textwidth}
\centering
\includegraphics[width=\linewidth]{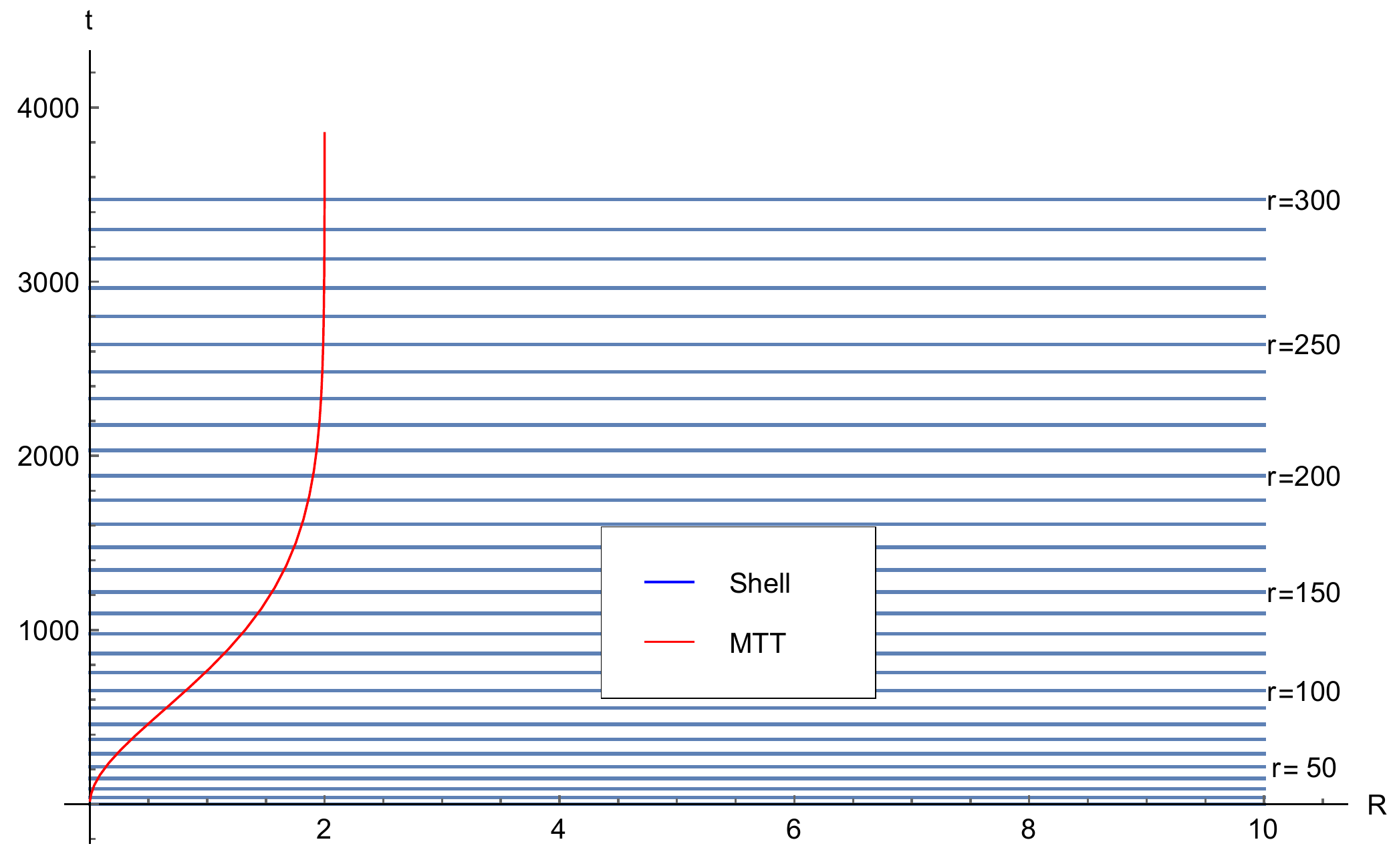}
\caption{}
\end{subfigure}
\begin{subfigure}{.5\textwidth}
\centering
\includegraphics[width=\linewidth]{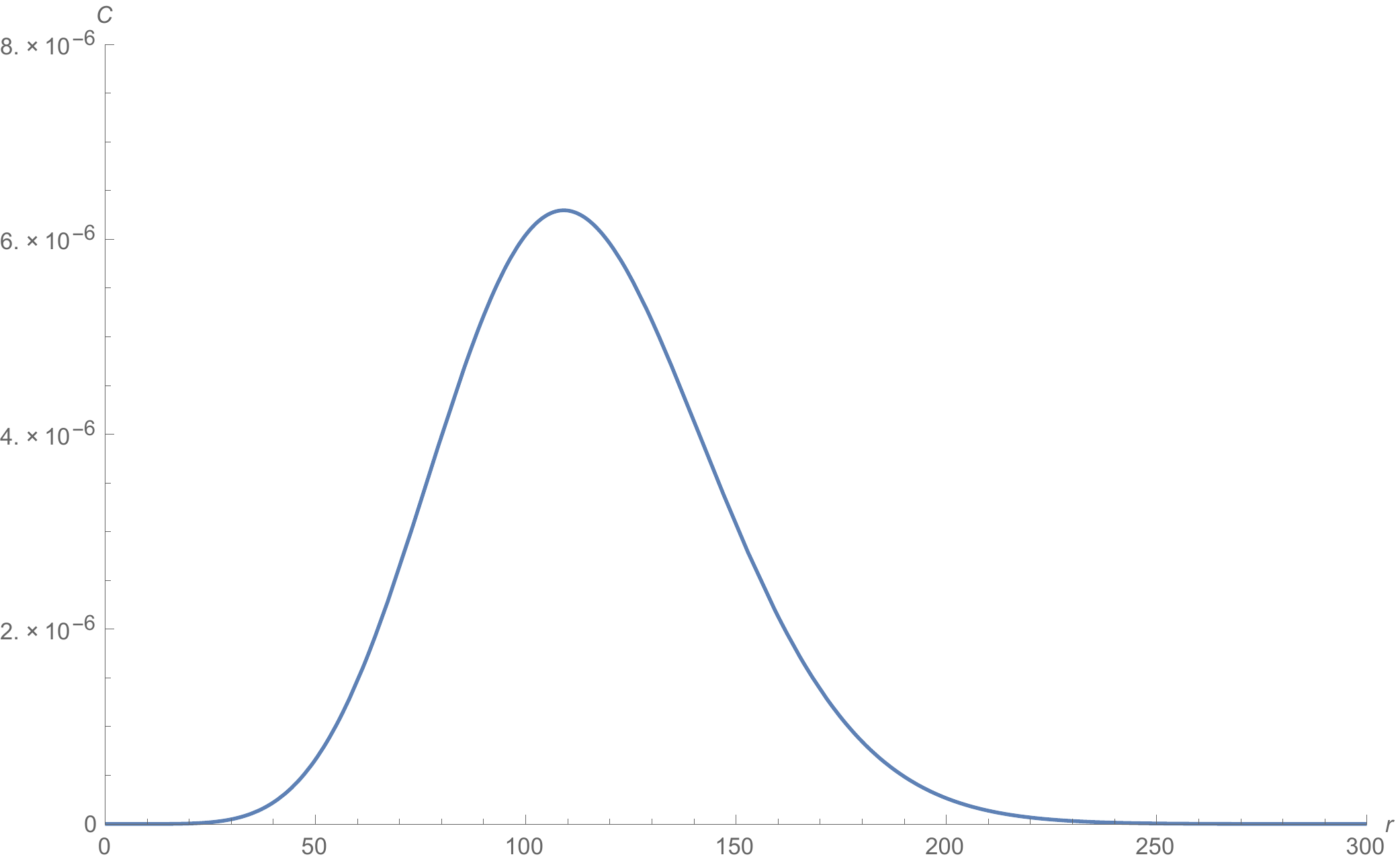}
\caption{}
\end{subfigure}
\caption{The graphs show the (a) formation of MTT along with the shells,
(b) values of $C$. 
The MTT begins from the center of the cloud and remains spacelike. 
The straight lines of MTT in (a), after the shell at $r=250$, represents the isolated horizon phase.}
\label{fig:viscosity1_example1}
\end{figure}
%
Just as before,
we choose $r_{0}=100\, m_{0}$. Here also, the MTT begins from the central singularity,
and develops as a dynamical horizon until it approaches
the isolated horizon at approximately $r=200$. 
This may also be confirmed from the fact that the density at $r=200$
is almost negligible. 
The density profile almost disappears at $r=380$,
and beginning at that value of shell coordinate, the MTT remains at $R=2$. 
The nature of the formation of singularity is identical to
the LTB case discussed in the previous sections. However the difference
is now with respect to the time at which the MTT forms. 
For the LTB, the shell at $r=200$ forms the MTT at $t=3215$, whereas
for the same choice of the density parameters, 
but with the choice of the parameter $a_{1}=-(1/4)$,
the same shell forms the MTT at $t=1912$. The reason is that for 
these choices, the $p_{r}$ is now non- zero and hence contributes to faster
formation of the MTT (see figure \ref{fig:viscosity1_example1}). 
There also exists contributions from the $p_{t}$
terms to the proper time of the observer falling along the shell.

(ii) For the density profile given by the following form,
\begin{equation}
\rho(r)=\frac{m_{0}}{8\pi r_{0}^{3}}\exp(-r/r_{0}),
\end{equation}
the situation is identical to the above case for the Gaussian profile. 
%
\begin{figure}[h!]
\begin{subfigure}{.55\textwidth}
\centering
\includegraphics[width=\linewidth]{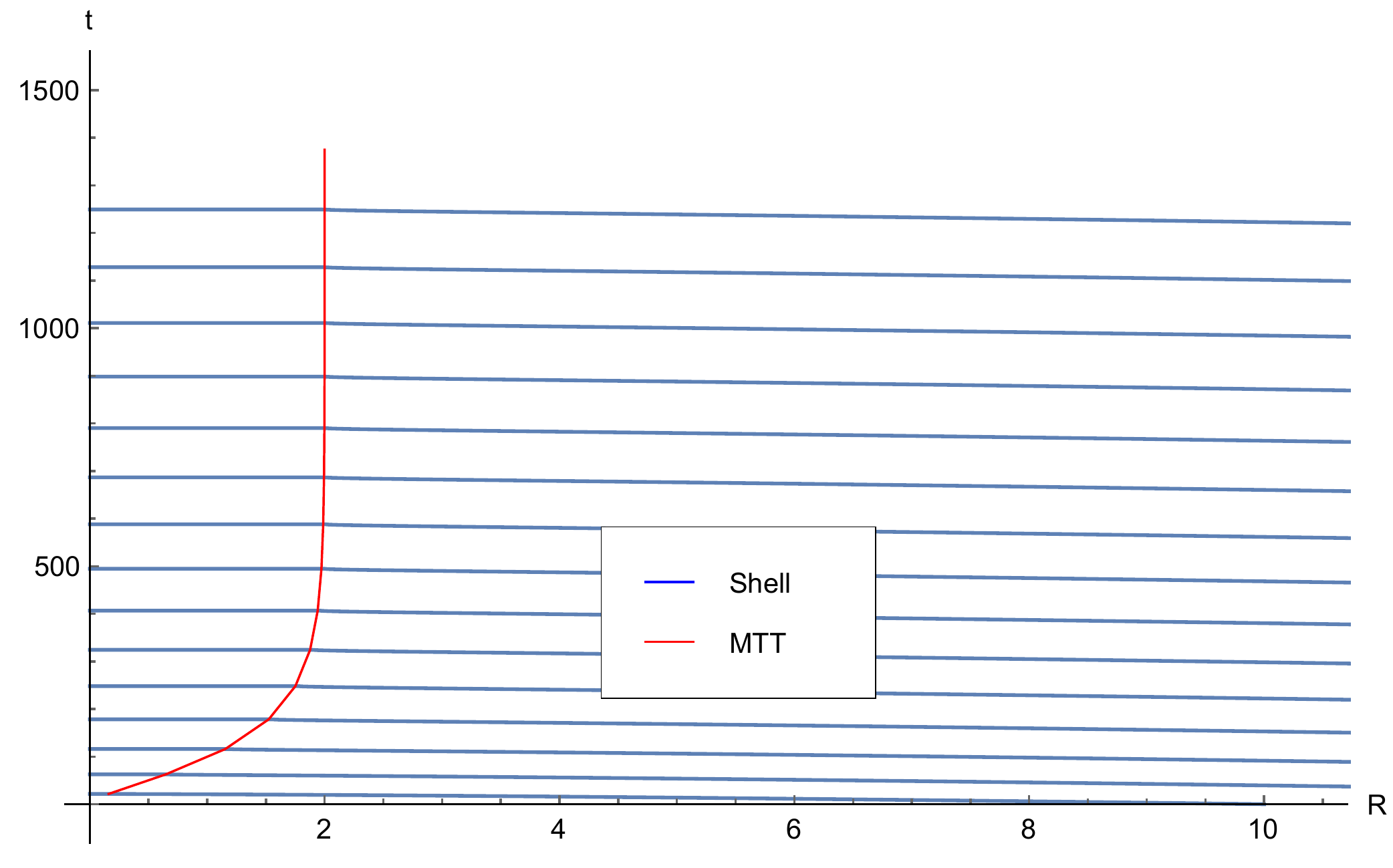}
\caption{}
\end{subfigure}
\begin{subfigure}{.45\textwidth}
\centering
\includegraphics[width=1.1\linewidth]{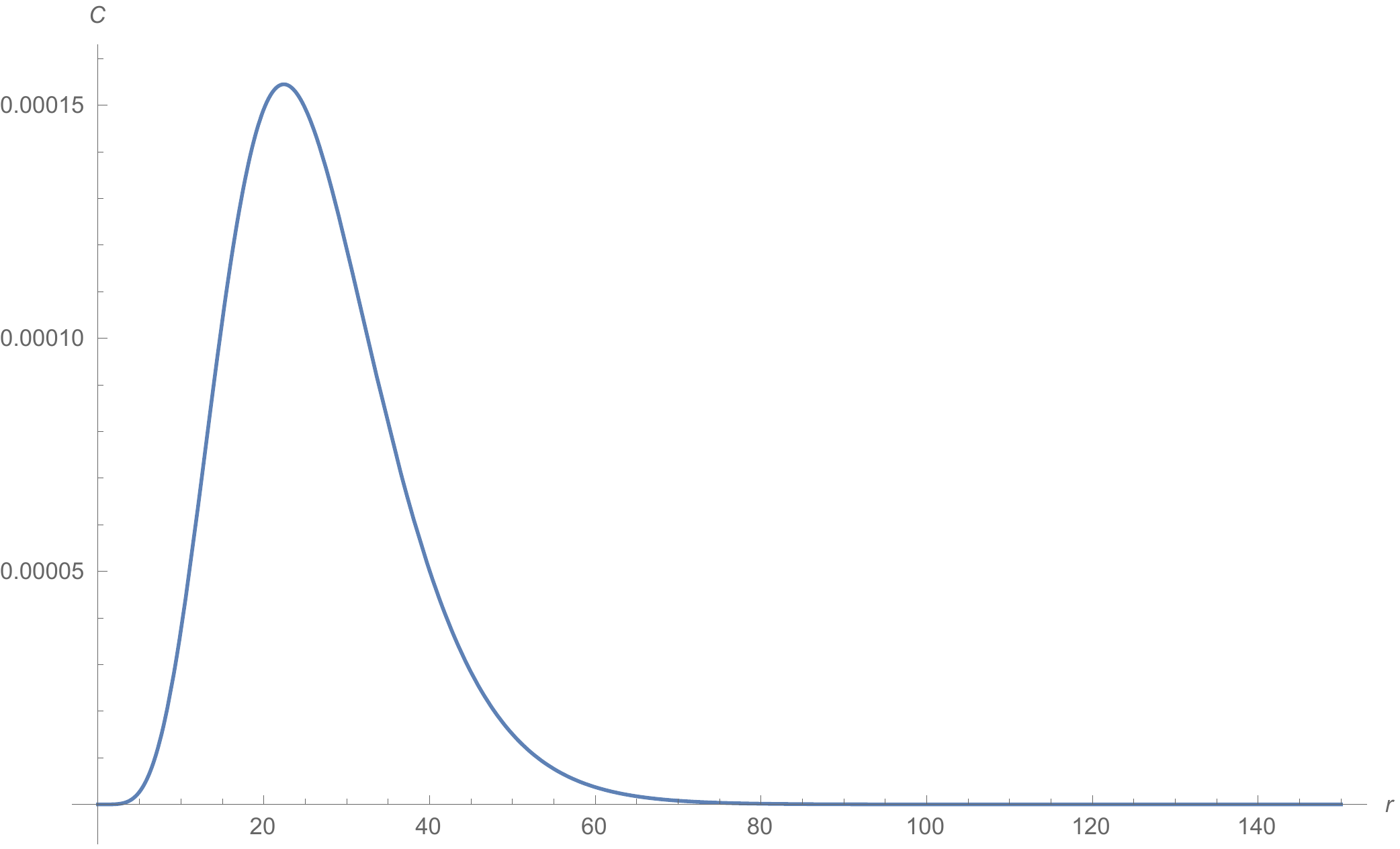}
\caption{}
\end{subfigure}
\caption{The formation of MTT (a),
and (b) values of $C$. }
\label{fig:viscosity1_example2}
\end{figure}
%
The time of formation of the MTT is lower than that obtained for the LTB case.
For the LTB collapse, the MTT begins from the central singularity,
and develops as a dynamical horizon until at approximately $r=70$, is begins to resemble
an isolated horizon. Here, the MTT formation and it's spacelike nature
is retained although the time of formation of the isolated horizon is lowered at 
$t=406$ from that in the LTB case which happens at $t=660$ (see figure \ref{fig:viscosity1_example2}).

\subsection{Time dependent mass function}
Let us consider the system in it's full generality. The Einstein equations shall have all following terms:
\begin{eqnarray}
 \rho&=&\frac{F'}{R^2 R'};\hspace{.5cm}{p}_{r}=-\frac{\dot{F}}{R^2 \dot{R}}+(4/3)\eta \sigma+\zeta \theta
 \label{1eq22}\\
 \alpha'&=&\frac{2R'}{R}\frac{p_{t}-p_{r}+2\eta \sigma}{\rho+p_r-(4/3)\eta\sigma-\zeta\theta}-
 \frac{{p}_{r}'-(4/3)\eta\sigma'-\zeta\theta^{\prime}}{\rho+p_r-\frac{4}{3}\eta\sigma-\zeta\theta} 
 \label{1eq33} \\
(\dot{G}/G)&=& (2\alpha^{\prime})(\dot{R}/R^{\prime})\, ;\, \hspace{.5cm} F(r,t)=R(1-G+H).\label{FGH}
\end{eqnarray}
Note that due to our generality, the Misner- Sharp mass function shall
acquire time dependence. To solve this set of highly nonlinear coupled equations, we assume
a set of constraints on the dynamical quantities: 
$p_{r}=k_r\rho$, $p_{t}=k_{t}\rho$, $\sigma=k_{\sigma}\rho$
and $\theta=k_{\theta}\rho$. By using these conditions, the solutions of metric functions 
are
\begin{equation}
\exp(2\alpha)=\frac{R^{\,4a_{1}}}{\rho^{\,2a_{2}}}, \hspace{1cm} \exp(2\beta)=\frac{R^{\prime}{}^{2}}
{1+r^2B(r,t)},
\end{equation}
where the parameters $a_{1}$ and $a_{2}$ are defined as 
$a_{1}=[k_{t}-k_{r}+2\eta k_{\sigma}]/[1+k_{r}-(4/3)\eta k_{\sigma}-\zeta k_{\theta}]$ 
and $a_{2}=[k_{r}-(4/3)\eta k_{\sigma}-\zeta k_{\theta}]/[1+k_{r}-(4/3) \eta k_{\sigma}-\zeta k_{\theta}]$. 
The line element for this spacetime may thus be written as:
\begin{equation}
ds^{2}=-\frac{R(r,t)^{4a_1}}{\rho(r,t)^{2a_2}}dt^2 + \frac{R(r,t)'^2}{1+r^2B(r,t)}dr^2 
+ R(r,t)^{2}\,[\,d \theta^2+  \sin^2{\theta}\, d\phi^2] \label{metprt}
\end{equation}
The equation of motion obtained from the equation \eqref{FGH} is reduced to the form:
\begin{equation}
\dot{R}=-R^{\,2a_{1}}\rho^{\,-a{_2}}\left[\frac{F(r,t)}{R}+r^2B(r,t)\right]^{1/2}.\label{emppt}
\end{equation}

For exact analytical solution, we introduce simplifications. Let us assume that the mass function $F(r,t)$, 
the metric function $B(r,t)$
and the density $\rho(r,t)$ are of the separable type:
\begin{equation}
F(r,t)=F_{1}(r)F_{2}(t),\,\, ~~B(r,t)=B_{1}(r)B_{2}(t),\,\,~~ \rho(r,t)=\rho_{1}(r)\rho_{2}(t),
\end{equation}
where some of these functions are related, with the following conditions: 
$B_{1}(r)=k(r)/r^2$, $B_2(t)=-F_{2}(t)=-\rho_2(t)^{2a_2}$.
Now, with the choice of the parametric form of $R(r,t)$, given by
\begin{equation}
R=[F_1(r)/k(r)]\,\cos^{2}\,(\eta/2), \label{Rpara}
\end{equation}
the equation of motion of the collapsing cloud \eqref{emppt}, gives the following time curve:
\begin{equation}
dt= \frac{[F_{1}(r)\,\cos^{2}(\eta/2)]^{(1-2a_{1})}\,\rho_{1}^{\,a_2}}{k(r)^{(3/2-2a_{1})}}\,d\eta.
\end{equation}
The solution of this equation which determines motion of the collapsing cloud is 
given by complicated relations involving the Hypergeometric functions
\begin{eqnarray}
t_{{shell}}&=&\frac{2F_1(r)^{1-2a_1}\rho_1(r)^{a_2}\cos(\eta/2)^{3-4a_1}}{(4a_1-3)k(r)^{3/2-2a_1}}
\,\,{}_{2}F_{1}\left[\frac{1}{2},\frac{3}{2}-2a_1;\frac{5}{2}-2a_1;\cos^2(\eta/2)\right]\nonumber\\
&& ~~~~~~~~~~~~~~~~~~~~~~~~~~~~~ -\frac{2\sqrt{\pi}F_1(r)^{1-2a_1}\rho_1(r)^{a_2}}{(4a_1-3)k(r)^{3/2-2a_1}}
\frac{\Gamma[5/2-2a_1]}{\Gamma[2-2a_1]},\label{tGencollapse}
\end{eqnarray}
where ${}_{2}F_{1}(a,b;c;z)$ is the Gauss Hypergeometric function, and $\Gamma(x)$ is the Gamma function.
The boundary conditions are chosen such that collapse starts at $\eta=0$,
where $R(t_{i},r)=[F(r)/k(r)]$ and $t=0$. The cloud reaches
the central singularity at $\eta=\pi$ where $R=0$.
In the $t$- coordinates, the time of formation of central 
singularity is 
\begin{equation}
t_s=\frac{2\sqrt{\pi}F_1(r)^{1-2a_1}\rho_1(r)^{a_2}}{(3-4a_1)k(r)^{3/2-2a_1}}\frac{\Gamma[5/2-2a_1]}
{\Gamma[2-2a_1]}.
\end{equation}

The dynamics of the marginally trapped surfaces (whether they are timelike, spacelike or null) depends upon 
the sign of the expansion parameter $C$, and is given by:
\begin{eqnarray}
C&=&\frac{1}{2\chi}\left[\frac{\rho+p_{t}-(4/3)\eta\sigma-\zeta\theta+\left( p_{t}-p_{r}\right)}
{(4\pi/\mathcal{A})-(1/2)\{\rho-\{p_{t}-(4/3)\eta\sigma-\zeta\theta+\left( p_{t}-p_{r}\right)\}\}} \right].
\label{GC}
\end{eqnarray}

\subsubsection*{Examples}
\begin{figure}[h!]
\begin{subfigure}{.55\textwidth}
\centering
\includegraphics[width=\linewidth]{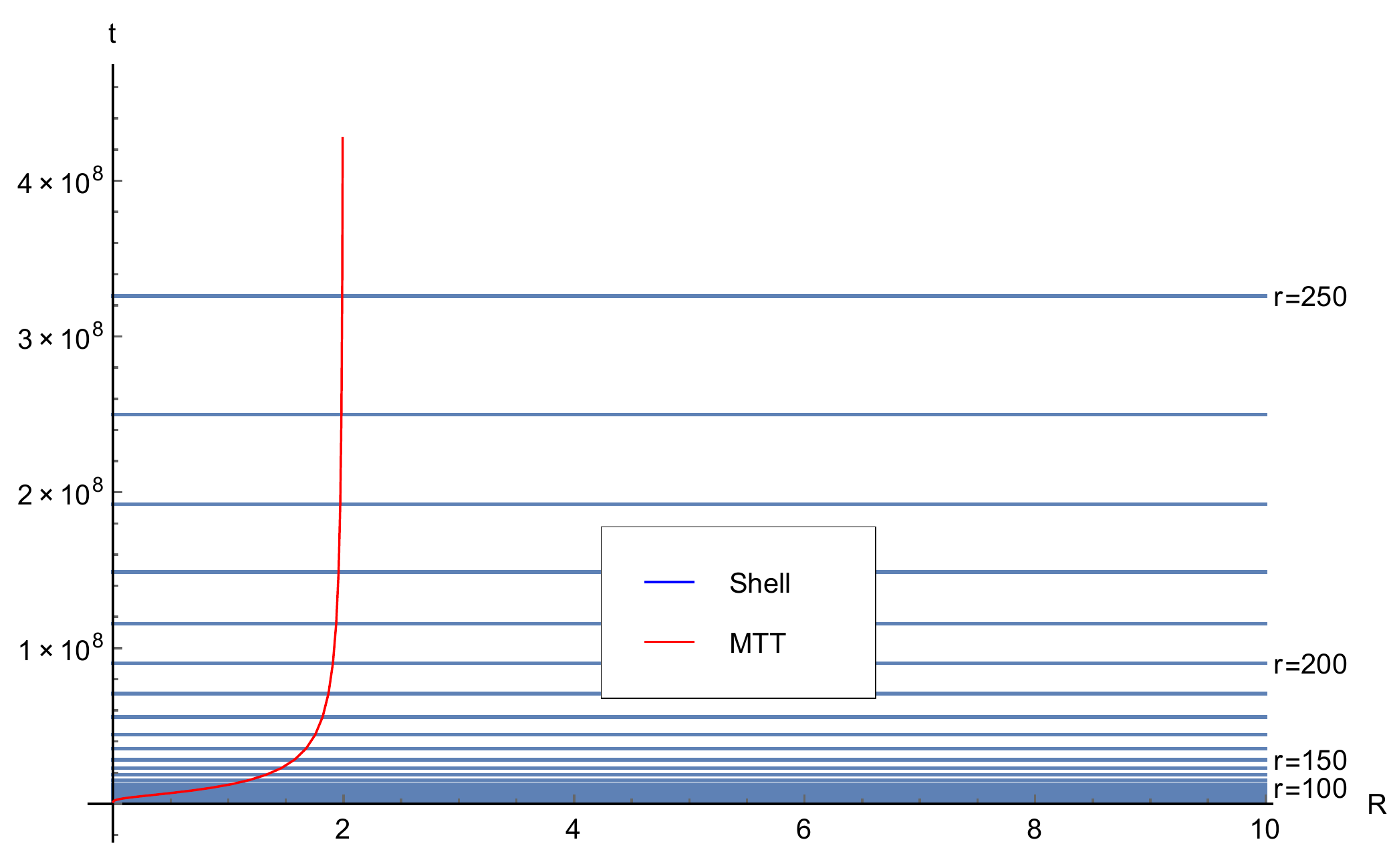}
\caption{}
\end{subfigure}
\begin{subfigure}{.45\textwidth}
\centering
\includegraphics[width=\linewidth]{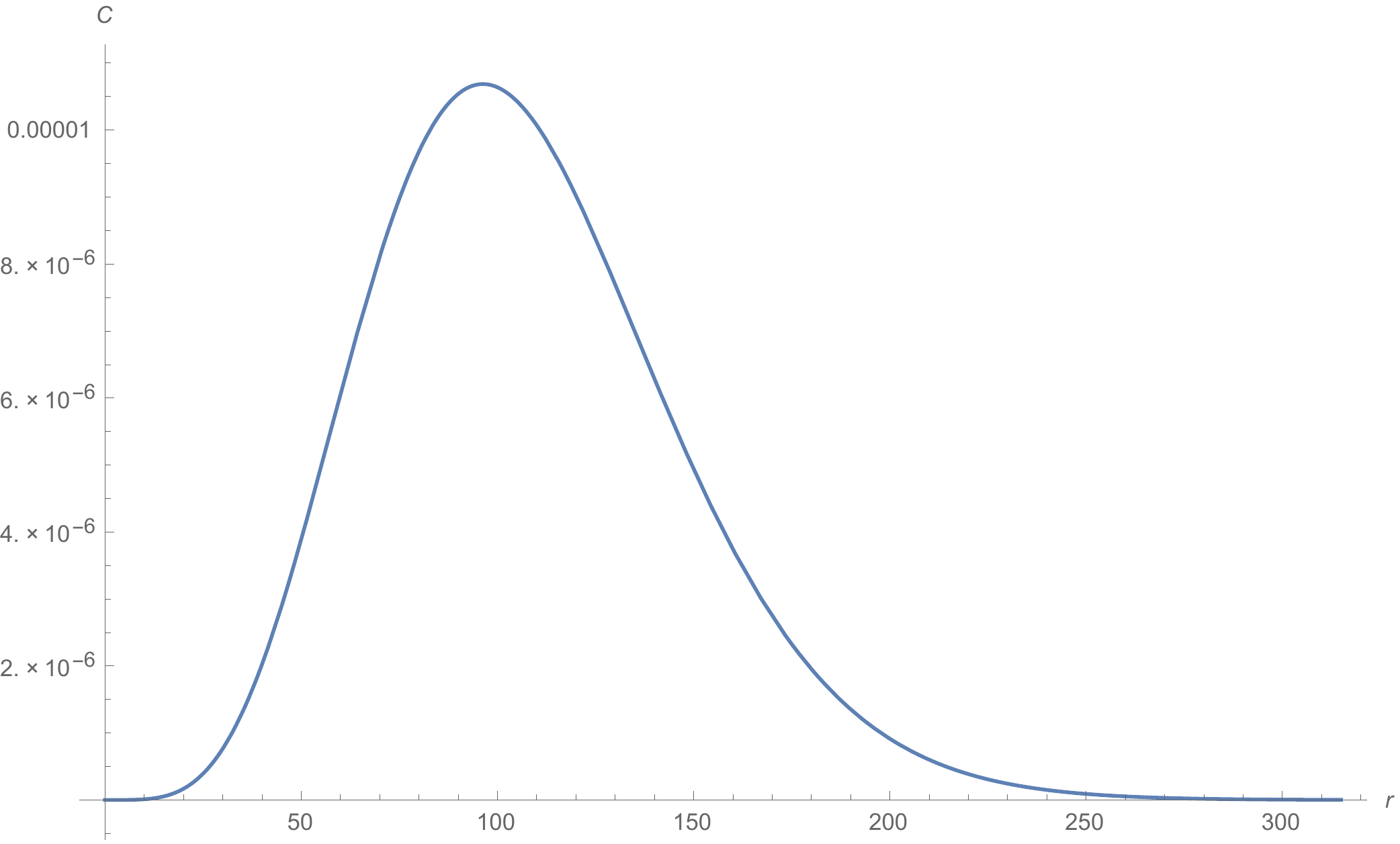}
\caption{}
\end{subfigure}
\caption{The graphs show the (a) formation of MTT along with the shells,
(b) values of $C$. 
Again, note that the MTT begins from the center of the cloud and remains spacelike until
it reaches the isolated horizon phase.}
\label{fig:viscosity2_example1}
\end{figure}

(i) Gaussian: The radial pressure has decreased, and hence the time of formation of singularity or the MTT
is at a larger time (see figure \ref{fig:viscosity2_example1}). 
%
%
The MTT is still spacelike. $k_{r}=(1/2)$, $k_{t}=1/4$, $\eta=1/16$, 
$k_{\sigma}=1/4$, $\zeta=(1/2)$, $k_{\theta}=(3/2)$, giving $a_{1}=-0.3$ and $a_{2}=-0.37$.
Notice that for $r=200$, which for the LTB case reached the isolated horizon
at $t=3215$, here it happens at $t=9\times 10^{7}$, which is approximately $10^{3}$ factor higher.
The reason is that with the choice of a time dependent $F(r,t)$, the radial pressure has decreased
considerably and hence, the time of formation of the MTT for each shell also goes up. 
The nature of formation of MTT however remains identical.

(ii) Large shell: The nature of formation of MTT here
is drastically different in nature from that described in the previous
examples. Here, we observe formation of timelike MTTs. 
%
\begin{figure}[h!]
\begin{subfigure}{.55\textwidth}
\centering
\includegraphics[width=\linewidth]{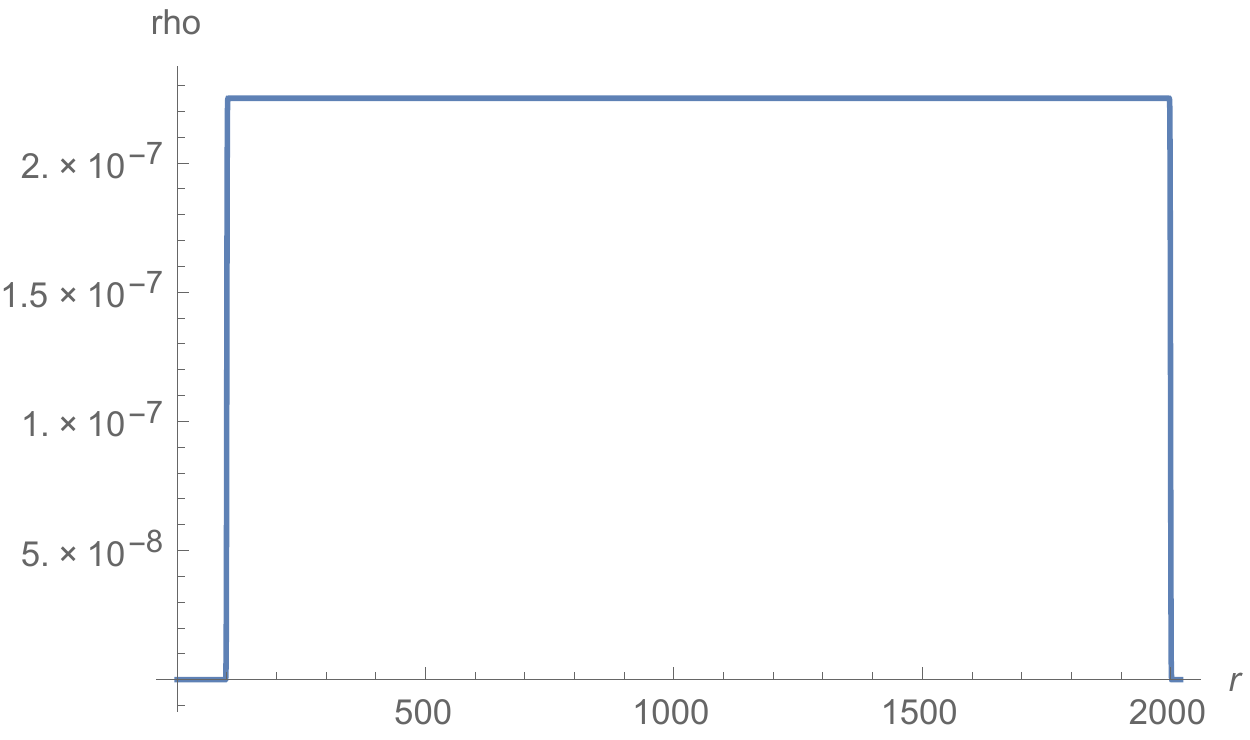}
\caption{}
\end{subfigure}
\begin{subfigure}{.45\textwidth}
\centering
\includegraphics[width=\linewidth]{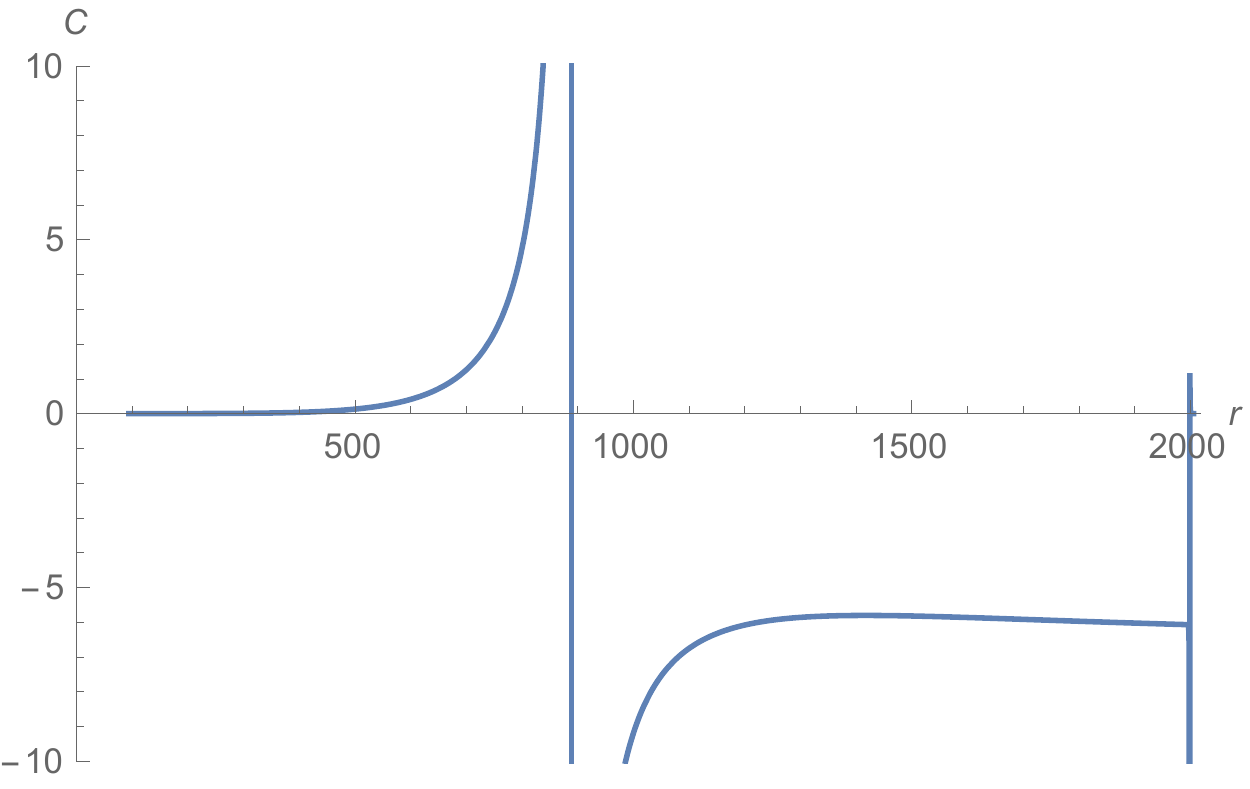}
\caption{}
\end{subfigure}
\begin{subfigure}{.55\textwidth}
\centering
\includegraphics[width=\linewidth]{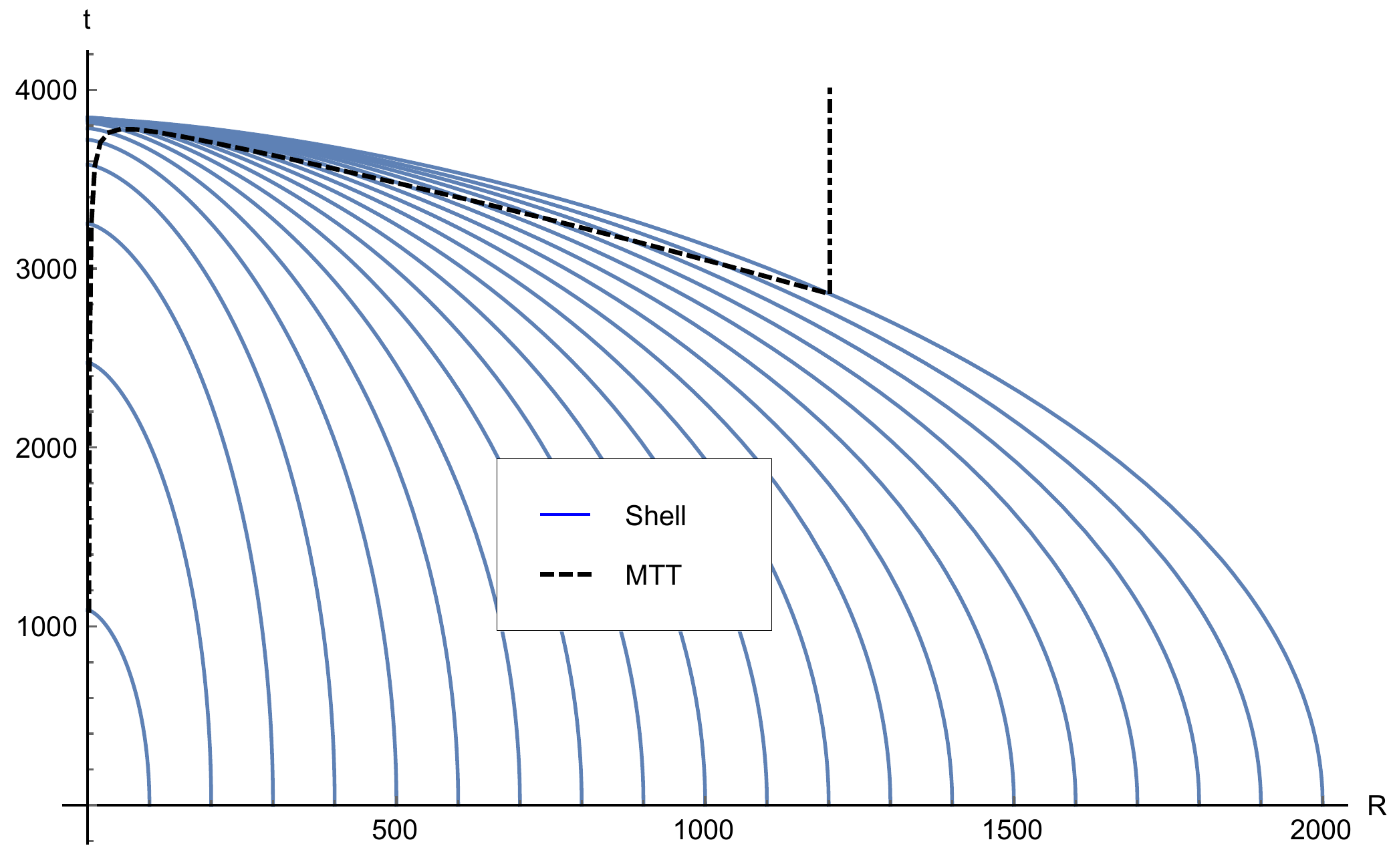}
\caption{}
\end{subfigure}
\caption{The graphs show the (a) formation of MTT along with the shells,
(b) values of $C$. The MTT is timelike. It first forms at $t=3000$ approximately,
and bifurcates in one direction to reach the IH and in another direction to match with the
DH evolving from the previous black hole. }
\label{fig:viscosity2_example2}
\end{figure}
%
Here,
for a more complicated collapse dynamics, the density profile
generate a timelike tube. The MTT forms at about $r=1800$
and bifurcates to reach the initial black holes. The timelike nature of the MTT
is also confirmed from the values of $C$ (figure \ref{fig:viscosity2_example2}). 
Here $k_{r}=(1/80)$, $k_{t}=1/81$, $\eta=1/16$, 
$k_{\sigma}=1/20$, $\zeta=(1/12)$, $k_{\theta}=(1/8)$, giving $a_{1}=0.006$ and $a_{2}=-0.002$. 
 
%
\begin{figure}[h!]
\begin{subfigure}{.45\textwidth}
\centering
\includegraphics[width=\linewidth]{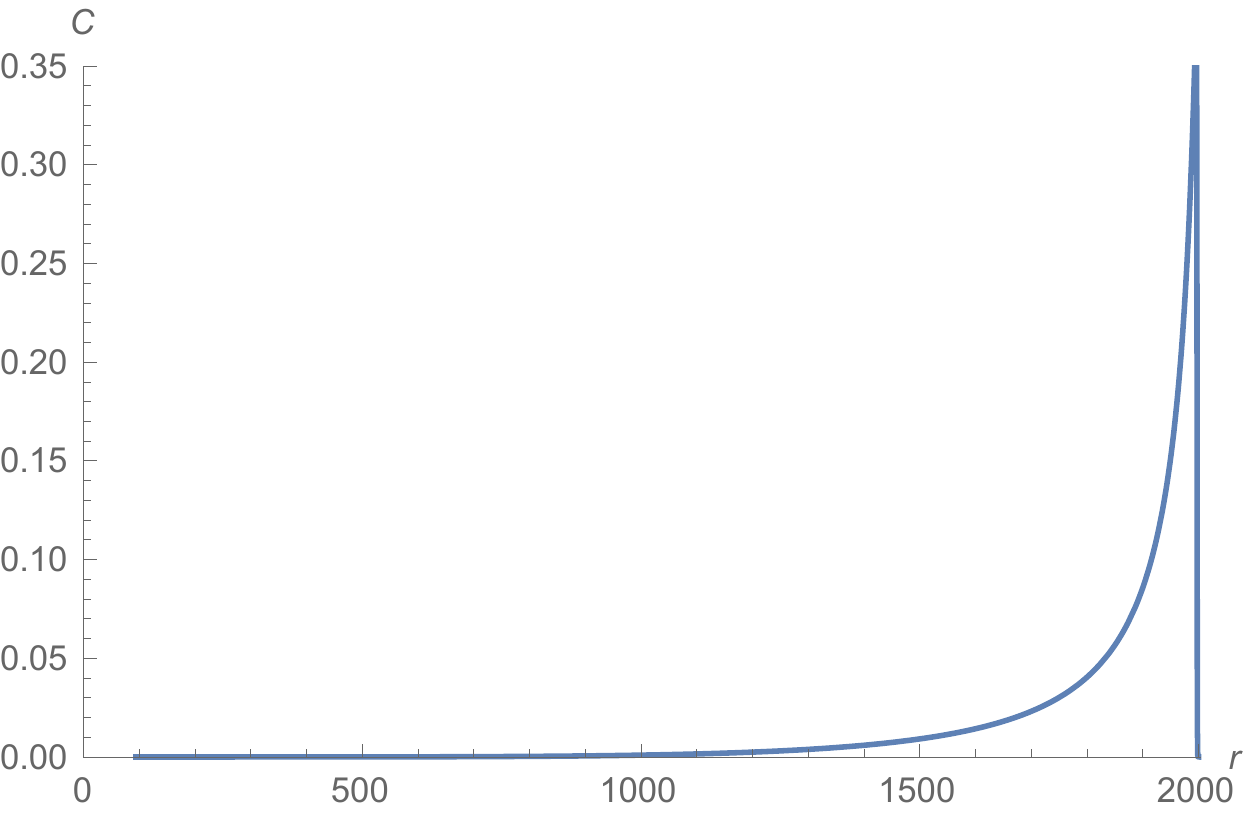}
\caption{}
\end{subfigure}
\begin{subfigure}{.55\textwidth}
\centering
\includegraphics[width=\linewidth]{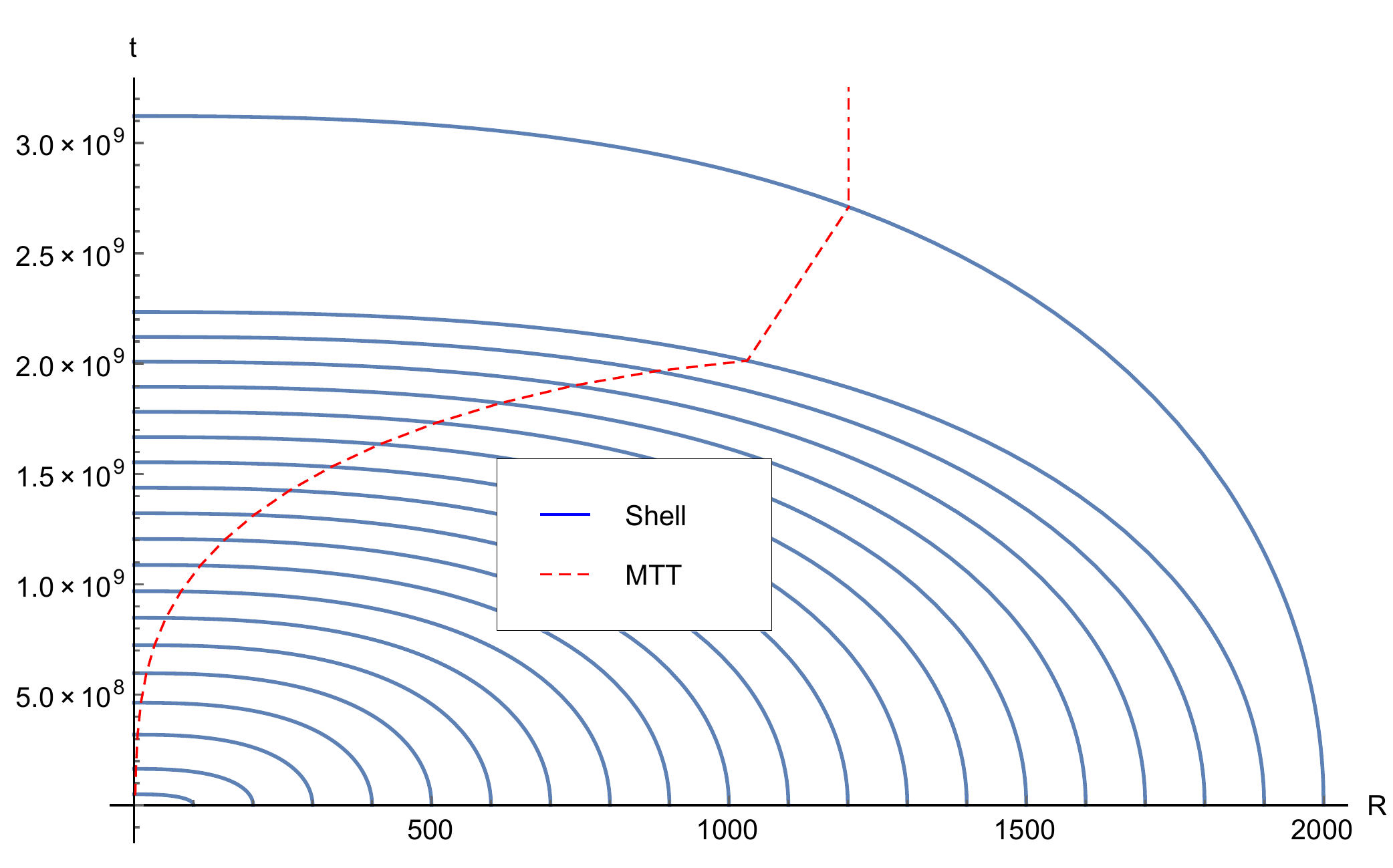}
\caption{}
\end{subfigure}
\caption{The graphs are for the density profile of figure \ref{fig:viscosity2_example2},
but with different parameters. Here, the MTT is spacelike.}
\label{fig:viscosity2_example3}
\end{figure}
%
However, if the following choice of parameters is made, the MTT is spacelike: 
$k_{r} = 1/2$, $k_{t} = 1/10$,  $\eta = 1/8$,  $k_{\sigma} = 1/4$, 
$\zeta = 1/2$, $k_{\theta} =  3/2$, with $a_{1}=-0.48$ and $a_{2}=-0.41$ (see figure 
\ref{fig:viscosity2_example3}).

\begin{figure}[h!]
\begin{subfigure}{.55\textwidth}
\centering
\includegraphics[width=\linewidth]{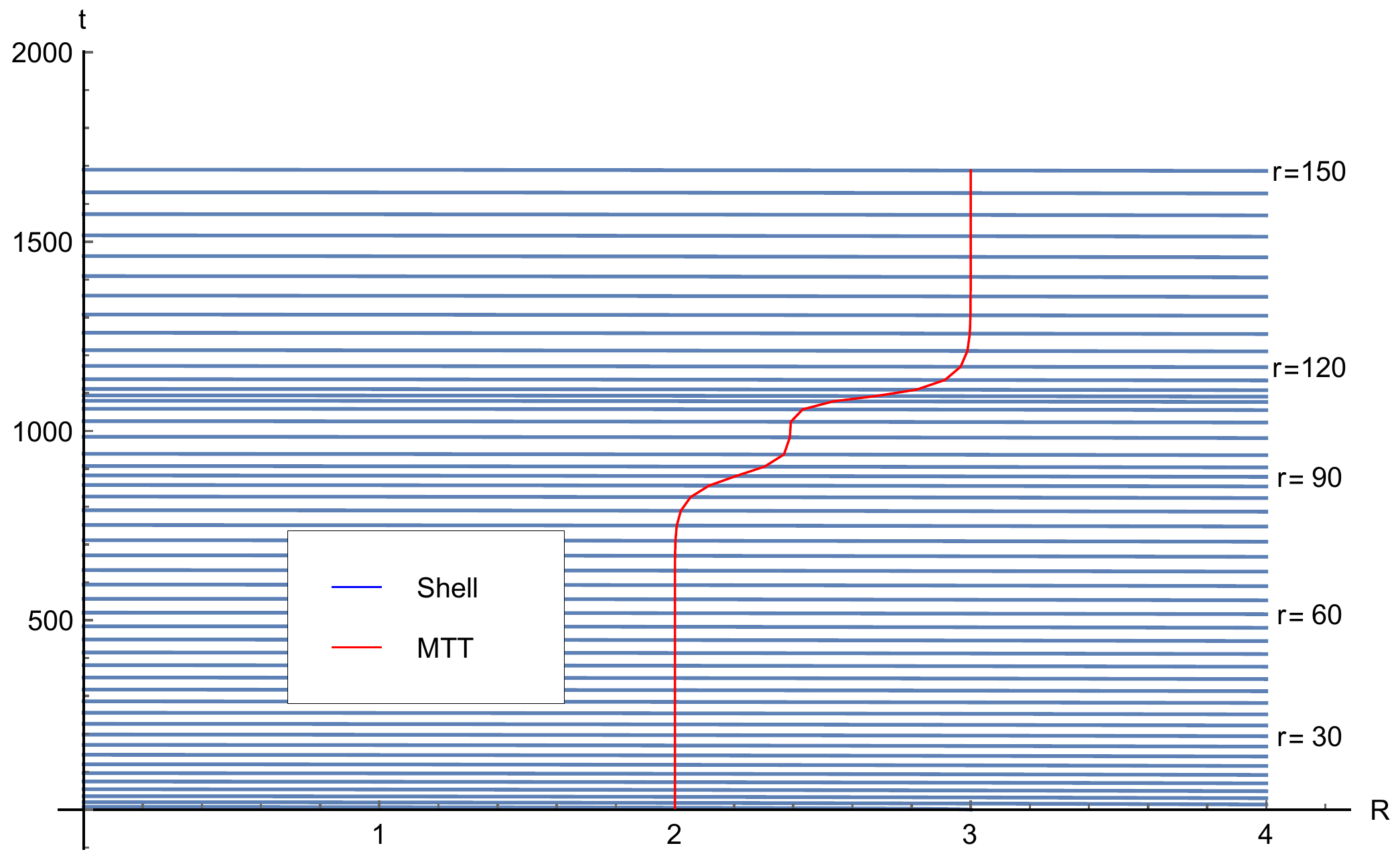}
\caption{}
\end{subfigure}
\begin{subfigure}{.45\textwidth}
\centering
\includegraphics[width=\linewidth]{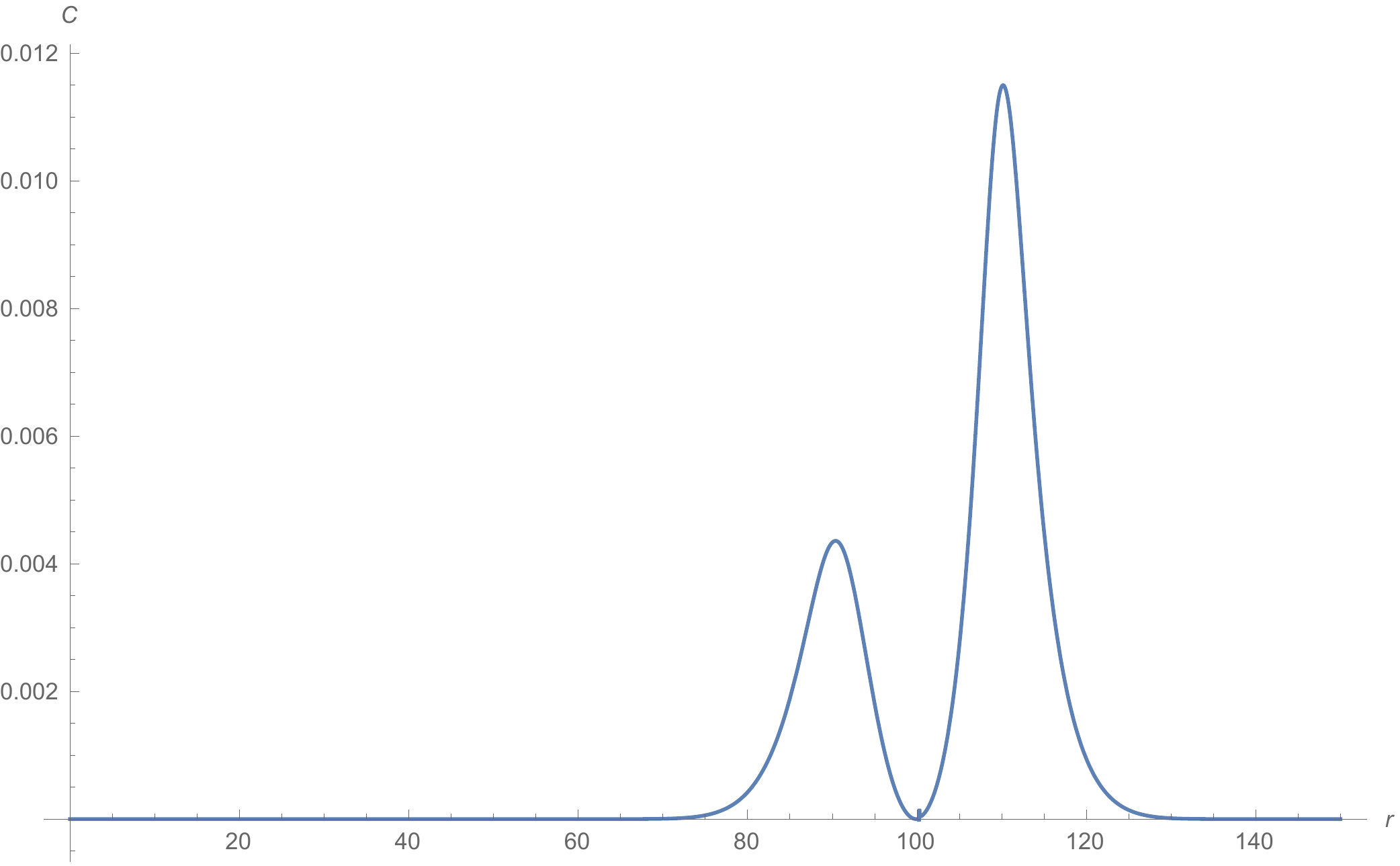}
\caption{}
\end{subfigure}
\caption{The graphs show the (a) formation of MTT along with the shells,
(b) values of $C$. 
The MTT begins from the center of the cloud and remains spacelike. 
The straight lines of MTT in (a) represents the isolated horizon phase.}
\label{fig:viscosity2_example4}
\end{figure}

(iii) Two consecutive shells falling on a black hole: Here $k_{r}=(1/80)$, $k_{t}=1/81$, $\eta=1/16$, 
$k_{\sigma}=1/20$, $\zeta=(1/12)$, $k_{\theta}=(1/8)$, giving $a_{1}=0.006$ and $a_{2}=-0.002$
(see figure \ref{fig:viscosity2_example4}).
%
%
The nature of formation of MTT is identical to that described for the LTB model although 
the time of formation of the MTT is not delayed for these choice of parameter fields.

\section{Discussions}\label{sec6}
In this paper, we have developed analytical and 
numerical techniques to study gravitational collapse of a large class of matter
fields in Einstein's theory. The main focus was to obtain the
trapped regions and locate the marginally trapped surfaces for some 
general class of energy- momentum tensors, including 
fluids admitting bulk and shear viscosity. 
For the purpose of generality, we have included the homogeneous
as well as inhomogeneous dust models. While the dust models have been studied earlier 
\cite{Landau_Lifshitz}, 
the detail study of the formation and time- development of the EH and the MTTs
for a generic class of energy- momentum tensors, through analytical 
as well as numerical means, to our knowledge, have not been carried out in the literature.   
Our analytical methods focus on two specific aspects. (see however,
\cite{Booth:2005ng, Kanai:2010ae,krasinski_hellaby}). 
The first aspect is to use the equations 
of gravitational collapse in the $R(r,t)-t$ coordinate system, to trace the
formation of the EH and the MTTs, simultaneously with the collapse of the matter cloud.
The use of the $R(r,t)$ is specially advantageous, since it is possible to
track the horizons as well as the collapsing sphere at each moment.
The second aspect is the development of numerical codes
to locate trapped regions and marginally trapped surfaces for each of these
matter fields. Through these numerical techniques, we have
ascertained the validity of the analytical calculations as well as 
obtained a faithful representation of the general expectations 
during gravitational collapse. In particular, we have obtained the
signature of the MTTs during each of the collapse scenarios and a general conclusion may be reached:
The MTTs in the OSD models are timelike. The situation for LTB- like collapse
is more complicated. Here, for generically for $\dot{m}(r)>0$, 
the MTTs are spacelike and hence, are all MTTs are dynamical horizons
and reach the isolated horizon phase in equilibrium. Some variations
are however observed, although the effects are most likely unobservable to the observer
outside the black hole. Thus, although the results are
valid for spherical symmetric spacetimes, 
some general conclusions may possibly be drawn about the behavior of MTTs during 
the collapse.
      
While dealing with 
the viscous fluid, we have however kept the 
shear coefficient \emph{low} so that the spacetime does not
deviate drastically from spherically
symmetry.  These parameter ranges, of the coefficients arising
in the energy- momentum tensor, have been utilised to numerically
study the evolution of the MTTs in these cases. Out of this parameter ranges,
we have further restricted to a smaller set of initial data so that we do not
encounter shell- crossing during gravitational collapse or have trapped surfaces at the beginning of the process.
We observe that, within a particular set of assumptions used here,
it is possible to exploit the freedom of choice of equation of state and the viscosity parameters 
to manipulate the nature as well as time of formation of the MTT. Indeed, in the
previous section, we have shown through examples, that alternate choices
of initial data may lead to MTTs which are either timelike or spacelike. Furthermore,
these choices also alter the time of formation of MTTs compared to the dust models,
in the sense that MTT formation may be delayed or accelerated, compared to the dust models,
by suitable choices in the fluid parameters.
We believe that the results obtained in this paper may help in forming a general outlook about the time development
of Marginally Trapped Surfaces during gravitational collapse.

An important aspect of study of the MTTs or trapped surfaces involve identifying boundary of a black 
hole region. The boundary of a trapped region is not known, although the Eardley conjecture
claims that the event horizon of a black hole spacetime may be thought of as the boundary of
(marginally outer) trapped surfaces \cite{Eardley:1997hk}. Indeed, it has 
been shown that for Vaidya- type null
collapse scenarios with mass $m(v)$ having upper bound, and accreting mass 
such that $\dot{m}(v)\ge 0$, the conjecture
holds \cite{BenDov:2006vw}. However interestingly, it has also been found that given a trapping horizon, 
trapped surfaces (or parts of it)
may extend outside of the horizon, and into 
the initial flat region of the Vaidya- spacetime, and  
furthermore non-spherically symmetric trapped surfaces may also
extend outside the standard spherically symmetric trapping horizon
\cite{Bengtsson:2008jr,Bengtsson:2010tj,Schnetter:2005ea}.
So, the exact boundary of a trapped region is not clearly specifiable as of now. A related question
is then the following: if the Eardley conjecture holds, does the event horizon allow a 
local description? One should expect from the global nature of the EH that this should not be so. 
Again for Vaidya- type collapse processes, it has been shown that trapped surfaces
may be constructed which extend into the future and hence acquires non- local nature. Thus, the 
location and nature of the boundary of a strictly trapped region remains unknown. It seems
that the process of further development needs numerical study of MTTs in GR as well
as in other alternate gravity theories to gain insight into the properties of MTTs. These issues will
be addressed in future studies. 


\section*{Acknowledgements}
The authors AC and AG are also supported through the DAE-BRNS project $58/14/25/2019$-BRNS. AC is also
supported by the DST-MATRICS scheme of government of India through MTR$/2019/000916$.


\section*{Appendix: Junction Conditions}
The Israel- Darmois junction conditions provide a set of rules
and boundary conditions which has been used in the previous sections. 
These boundary conditions are summarized below for
some simple cases. In the following, we provide the junction conditions
for a simple model: a $K=0$ OSD model as the interior spacetime 
joined to an exterior Schwarzschild spacetime (denoted by $\mathcal{M^{-}}$)
of mass $M$ (denoted by $\mathcal{M^{+}}$), along a spacelike hypersurface $\Sigma$. 
Let us denote the coordinates on this surface to be $(\tau, \theta, \phi)$.  
From $\mathcal{M^{-}}$, we can write down the surface $\Sigma$ as
$f_{-}(r,t)=r-r_{b}=0$, and hence, the induced metric on $\Sigma$ is
\begin{eqnarray}
ds^{2}_{-}=a(\tau)^2\left(-d\tau^2+ r_{b}^{2}\, d \Omega^2\,\right).
\label{M_1}
\end{eqnarray}
Note that the coordinates $t$ and $\tau$ are related through the relation $dt/d\tau =a(\tau)$.
From the point of view of the exterior spacetime, the hypersurface
may be described by $r=R(\tau)$ and $t=T(\tau)$, with no change in the angular variables. 
The line element of the exterior manifold is then given by
\begin{equation}
ds^{2}_{+}=-\left(Z\dot{T}^2-Z^{-1}\dot{R}^2\right)d\tau^2 
+ R(\tau)^2\left( d \theta^2+\sin^2{\theta}d\phi\right)
\label{m11+}
\end{equation}
where $Z=\left(1-2M/R\right)$. The matching of the metric immediately implies that the 
following two conditions hold:
\begin{equation}
R(\tau)= r_{b}\, a(\tau)\, ; ~~~~a(\tau)^{\,2}=\left[Z\left(dT/d\tau\right)^2-
Z^{-1}\left(dR/d\tau\right)^2\right].  
\end{equation}
The normal vector field $n^{a}$ for $\mathcal {M^{-}}$ and the external spacetime $\mathcal{M^{+}}$ 
are given respectively by:
\begin{equation}
n_{a}=a(\tau)\,(dr)_{a}, ~~~~~ n_{a}=-(dR/d\tau)\,(d\tau)_{a} +(dT/d\tau)\, (dr)_{a}.
\end{equation}

The velocity of observer on the cloud is also determined for these two patches of 
spacetime separately
and are given by:
\begin{equation}
\frac{dR(r_b,\tau)}{d\tau}=-a(\tau)\left[F(r_{b})/R\right]^{\,1/2}, \label{Rrs}
\end{equation}
where the $-$ve sign is chosen to signify collapse. The velocity as observed from the 
external coordinates is then:
\begin{equation}
\frac{dR}{dT}=-\left(2M/R\right)\left(1-2M/R \right). \label{RrT}
\end{equation}
Note also that the second of the metric matching condition, along with the equation (\ref{Rrs}) 
implies that
\begin{equation}
\left(\frac{dT}{d\tau}\right)= a/(1-2M/R)\label{Vt1}
\end{equation}

Using the normal vector fields, the extrinsic curvatures of the interior and the exterior spacetime 
may also be determined, and they give:
\begin{eqnarray}
K^{-}_{\tau\tau}&=&0, \label{td1}\\
K^{+}_{\tau\tau}&=&-\sqrt{\frac{2M}{R}} \left(1-\frac{2M}{R}\right) \ddot{T} +\dot{a}\sqrt{\frac{2M}{R}}+\frac{4M^2a^2}{\left(1-\frac{2M}{R}\right)r_s^3}, \label{Kttpkz}\\
K^{-}_{\theta\theta}&=&R\left[1-\frac{F}{R}+\frac{2M}{R}\right]^{1/2} \label{thd1}, ~~~~
K^{+}_{\theta \theta}= R.\label{Kthpkz}
\end{eqnarray}
The $K_{\tau\tau}$ matching gives us the equation \eqref{Vt1}. The $K_{\theta\theta}$ matching gives us 
the equation: 
\begin{equation}\label{matching_boundary_1}
F(r_{b})\equiv m\, r_{b}^{3}= 2M.
\end{equation}

A similar exercise may also be carried out to join the interior spacetime created due to viscous 
fluid collapse, given by equation \eqref{metprt}, with
the external Schwarzschild spacetime. In that case, the matching gives the
following set of conditions. First, the $K_{\theta\theta}$ matching again gives
\begin{equation}
F(r_{b}, \tau)\equiv m(r_{b}, \tau)\, r_{b}^{3}= 2M,
\end{equation}
This matching, along with the condition from \eqref{F_eqn}, leads to the following relation involving 
the radial pressure and viscosity terms at the boundary
\begin{equation}
p_{r}=\zeta\theta+\,(4/3)\eta\,\sigma.
\end{equation}
%

%
 

\end{document}